\documentclass[sn-mathphys]{sn-jnl}

\usepackage{graphicx}%
\usepackage{multirow}%
\usepackage{amsmath,amssymb,amsfonts}%
\usepackage{amsthm}%
\usepackage{mathrsfs}%
\usepackage[title]{appendix}%
\usepackage{xcolor}%
\usepackage{textcomp}%
\usepackage{manyfoot}%
\usepackage{booktabs}%
\usepackage{algorithm}%
\usepackage{algorithmicx}%
\usepackage{algpseudocode}%
\usepackage{listings}%

\usepackage{comment}
\usepackage{aas_macros}

\def\beq{\begin{equation}}
\def\eeq{\end{equation}}
\def\ba{\begin{eqnarray}}
\def\ea{\end{eqnarray}}
\def\bal{\begin{align}}
\def\eal{\end{align}}
\def\bxi{{\mbox{\boldmath $\xi$}}}

\begin{document}

\title[Giant Planet Tides] {Tidal Dissipation in Giant Planets}

\author[1]{\fnm{Jim} \sur{Fuller}}

\author[2]{\fnm{Tristan} \sur{Guillot}}

\author[3]{\fnm{Stephane} \sur{Mathis}}

\author*[4]{\fnm{Carl} \sur{Murray}}\email{c.d.murray@qmul.ac.uk}


\affil[1]{\orgdiv{TAPIR, Walter Burke Institute for Theoretical Physics}, \orgname{Caltech}, \orgaddress{\street{Mailcode 350-17}, \city{Pasadena}, \postcode{91125}, \state{CA}, \country{USA}}}

\affil[2]{\orgdiv{Observatoire de la C\^ote d'Azur}, \orgname{Universit\'e C\^ote d'Azur}, \orgaddress{\street{CNRS}, \city{Nice}, \postcode{06304}, \country{France}}}

\affil[3]{\orgdiv{Université Paris-Saclay}, \orgname{Université Paris Cité}, \orgaddress{\street{CEA, CNRS, AIM}, \city{Gif-sur-Yvette}, \postcode{F-91191}, \country{France}}}

\affil[4]{\orgdiv{School of Physical and Chemical Sciences}, \orgname{Queen Mary University of London}, \orgaddress{\street{E1 4NS}, \city{London}, \country{United Kingdom}}}


\abstract{Tidal interactions between moons and planets can have major effects on the orbits, spins, and thermal evolution of the moons. In the Saturn system, tidal dissipation in the planet transfers angular momentum from Saturn to the moons, causing them to migrate outwards. The rate of migration is determined by the mechanism of dissipation within the planet, which is closely tied to the planet's uncertain structure. We review current knowledge of giant planet internal structure and evolution, which has improved thanks to data from the \textit{Juno} and \textit{Cassini} missions. We discuss general principles of tidal dissipation, describing both equilibrium and dynamical tides, and how dissipation can occur in a solid core or a fluid envelope. Finally, we discuss the possibility of resonance locking, whereby a moon can lock into resonance with a planetary oscillation mode, producing enhanced tidal migration relative to classical theories, and possibly explaining recent measurements of moon migration rates.}

\maketitle

\section{Introduction}

\begin{figure}
\begin{center}
\includegraphics[scale=0.27]{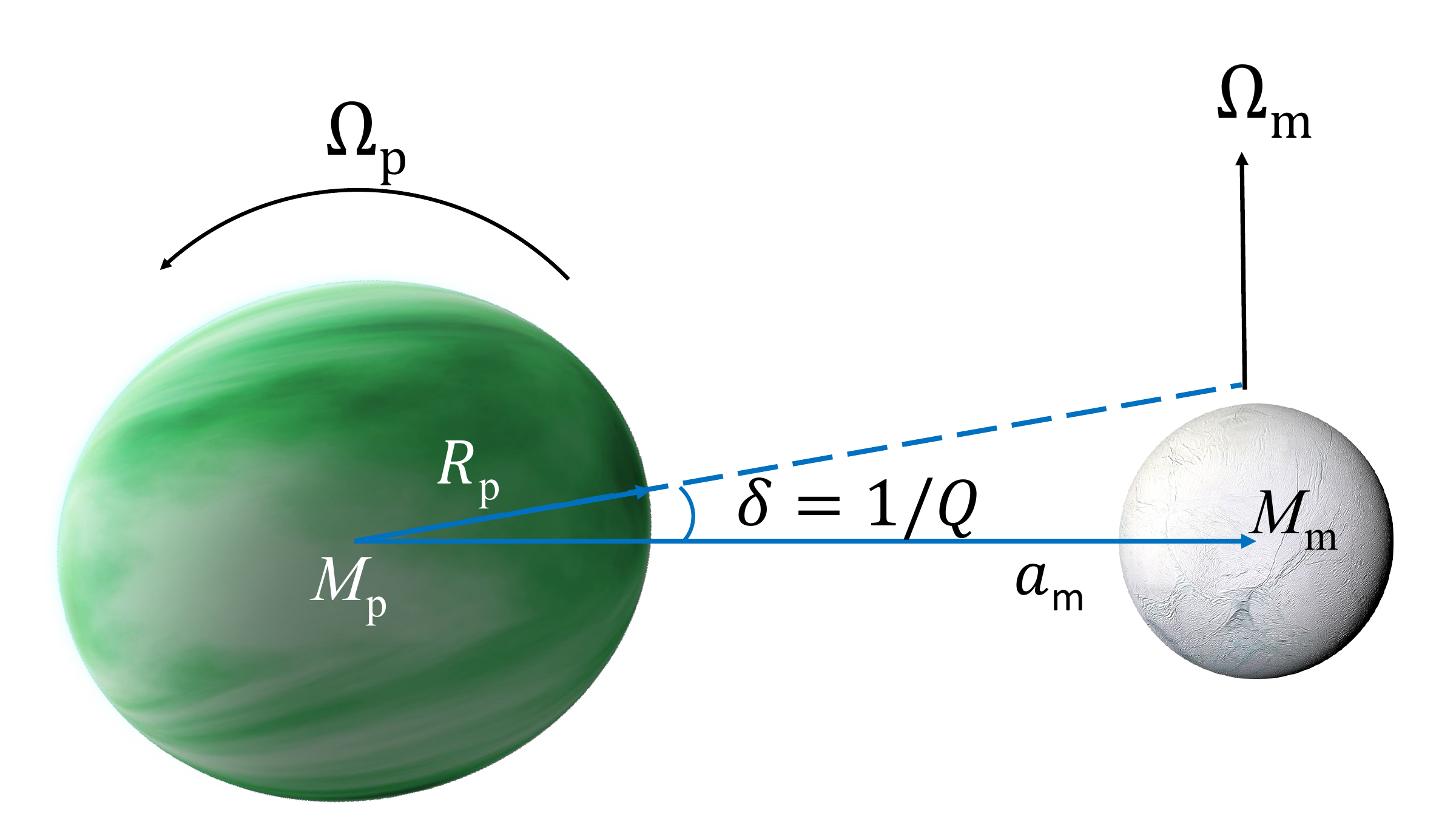}
\end{center} 
\caption{\label{fig:Qfig} 
Cartoon showing a tidally distorted planet of mass $M_{\rm p}$ and radius $R_{\rm p}$, due to a moon of mass $M_{\rm m}$ at semi-major axis $a_{\rm m}$. The planet's spin frequency is $\Omega_{\rm p}$ and the moon's orbital frequency is $\Omega_{\rm m}$. The tidal quality factor of the planet, $Q$, can be visualized as a lag angle of $\delta = 1/Q$ between the tidal bulge and the moon's position. In reality, the effective value of $Q$ can involve more complicated tidal distortion within the planet due to dynamical tides.}
\end{figure}

During the last decade, new discoveries in the Jupiter and Saturn systems have compelled us to revisit the estimates of tidal dissipation in giant gaseous planets of our solar system.  Figure \ref{fig:Qfig} illustrates how dissipation leads to a phase lag between the tidal bulge and the satellite, creating a torque that drags the satellite forward or backward. When the planet rotates faster than the moon orbits ($\Omega_{\rm p} > \Omega_{\rm m}$, as in the case of solar system moon systems), angular momentum is transferred from the planet to the moon, dragging the moon forward and making it migrate outwards. The initial predictions for the tidal quality factor of Jupiter and Saturn were computed by \cite{GoldreichSoter1966}, who solved the evolution equation for the semi-major axis of the moons assuming a frequency-independent tidal dissipation rate, constant over the evolution, in order to place them at their current positions at the age of the solar system. This procedure predicts tidal quality factors of $10^4 \lesssim Q \lesssim 10^6$ for Saturn and Jupiter.

Using high precision astrometry, a series of measurements \citep{Laineyetal2009,Laineyetal2012,Laineyetal2017,Polycarpeetal2018} demonstrated that tidal dissipation in Jupiter and Saturn can be more than an order of magnitude higher than what was computed initially by \cite{GoldreichSoter1966} \citep[see also][]{Sinclair1983}, i.e., a tidal quality factor $Q \lesssim10^3$ instead of $Q \sim10^4$ for Saturn. Moreover, the dissipation in Saturn has a complex variation as a function of the tidal frequency: we observe a weak (and potentially smooth) dependence on the frequency for Enceladus, Tethys and Dione, and a much stronger dissipation for Rhea and Titan. Observations of faster moon migration than expected (and hence stronger dissipation) may require revisions of our understanding of the formation and evolution of giant planets systems \citep{Charnozetal2011}, and of tidal dissipation in planetary interiors.


These results are also of major interest for exoplanetary systems since giant gaseous planets have been the first type of discovered exoplanets. Indeed, the discovery of 51 Pegasi b by \cite{MayorQueloz1995} opened the path to the exploration of giant planets outside our Solar system. The occurrence of giant gaseous planets close to their host stars, the so-called ``hot'' Jupiters, demonstrated that the orbital architecture of our solar system may not be common. It also spawned new challenging questions about the formation, the dynamical evolution, and the stability of exoplanetary systems. In this context, the study of tidal (and magnetic) star-planet interactions between low-mass stars and close giant planets came to the forefront \citep[e.g.,][]{Ahuiretal2021}.

For example, tidal interactions between a host star and orbiting giant exoplanets were proposed as a candidate to explain the ``bloated'' hot Jupiter phenomenon \citep{Bodenheimer+2001, Guillot+Showman2002, Baraffe2005}. The proposed idea is that tidal dissipation can sustain intense internal heating in giant exoplanets, modifying their internal structure with an increase of their radii. Another important but unsolved question is the effects of tides on the orbital distribution of hot Jupiters around their host stars, for which dissipation in both the stars and the planets can play a key role. Let us finally point that these major questions are among the key objectives of ongoing and forthcoming major space missions like CHEOPS, TESS, JWST, PLATO, and ARIEL \citep{Benzetal2017,Rickeretal2015,Lagage2015,Raueretal2014,Tinettietal2018} and ground-based instruments like SPIRou \citep{Moutouetal2015}.

While tidal dissipation is crucial for orbital evolution of both giant planet moon systems and exoplanet systems, the underlying physics is complex. In this chapter, we review our current understanding of giant planet structure, and we discuss how planetary structure and internal dynamics affect various tidal dissipation mechanisms. Comparison between models and measured migration rates are now allowing us to uncover the complex and interrelated histories of giant planet interiors and their moon systems.

\section{Giant Planet Structure}
\label{sec:GPS}

\subsection{Interiors of Jupiter, Saturn, Uranus and Neptune}

In the solar system, four planets have acquired enough hydrogen and helium during the early formation phase of the system to grow rapidly in size and mass and earn the qualification of ``giant". With radii of about 10 times our Earth, Jupiter and Saturn are commonly known as ``gas giants" while Uranus and Neptune, which are about 4 times larger than Earth, are usually called ``ice giants". Both denominations are convenient, but we should be aware of their limitations: most of Jupiter and Saturn's interiors behave as a liquid rather than a gas \citep[they are in fact {\em fluid}, e.g.,][]{Guillot2005}, while the compositions of Uranus and Neptune are highly uncertain: they could well be more rocky than icy \citep{Helled+Fortney2020, Kunitomo+2018}.

An important property of giant planets in our solar system is that they radiate about twice as much energy as they receive from the Sun \citep[][and references therein]{Pearl+Conrath1991}. The difference is due to their progressive contraction and cooling \citep{Hubbard1968}. A notable exception is Uranus, whose intrinsic heat luminosity is about 5 to 10 times lower than the other three giant planets and compatible with zero \citep{Pearl+Conrath1991}. Several hypotheses have been put forward \citep[see][]{Helled+Fortney2020}, but the recent 30\% upward revision of Jupiter's intrinsic luminosity \citep{Li_Liming+2018} calls for a reanalysis of available data, and ultimately, in situ measurements at Uranus \citep[e.g.,][]{Fletcher+2020, Guillot2022}. 

In most of the interiors of these four planets, the high radiative opacities and low heat conductivities imply that any non-negligible intrinsic luminosity should ensure convection \citep{Stevenson1976, Guillot+1994, Guillot+2004}. The atmospheres of all planets, including Uranus, indeed exhibit significant convective activity \citep{Hueso+2020}. At photospheric levels, their temperatures are remarkably uniform in spite of the strongly varying insulation, also a sign of efficient heat transport and hence convection \citep{Ingersoll+Porco1978}. In planets, only very slightly super-adiabatic temperature gradients are required for convection to carry the planetary heat flux, so convective regions are thought to have a nearly adiabatic density profile. These facts motivate the widely used hypothesis that interiors are convective and hence nearly adiabatically stratified.

\begin{figure}
\begin{center}
\includegraphics[scale=0.35]{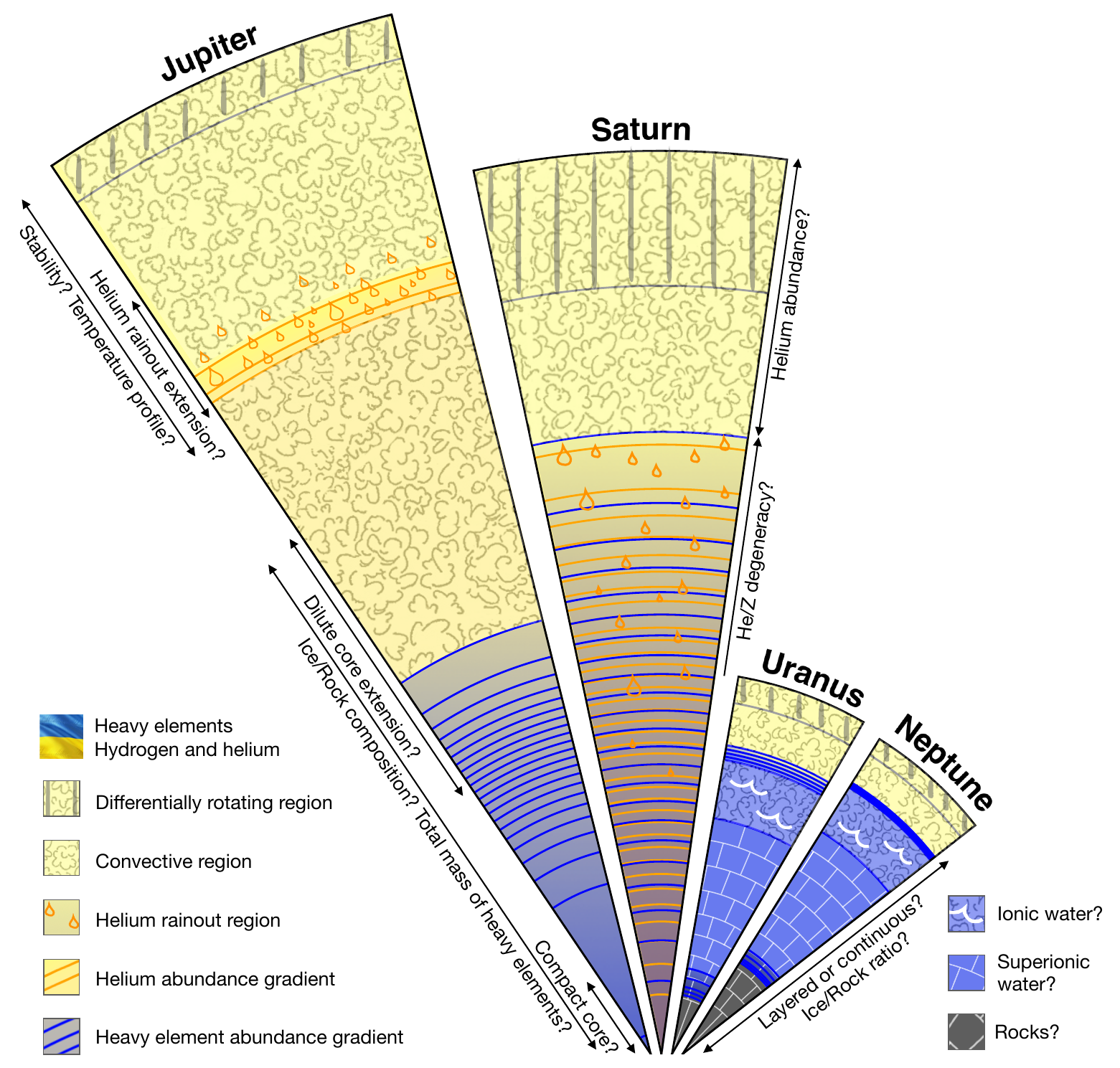}
\end{center} 
\caption{ \label{fig:interiorslices} 
Slices of the internal structures of Jupiter, Saturn, Uranus and Neptune highlighting compositional gradients in helium (orange) and heavy elements (blue). Jupiter and Saturn are characterized by a phase separation of helium in metallic hydrogen starting near Mbar pressures. Juno and Cassini measurements also indicate the presence of a dilute heavy element core in these planets, implying that heavy elements are partially mixed with hydrogen and helium rather forming separate pure layers. In this diagram, each layer corresponds to a 2\% increase of the helium/hydrogen ratio (orange) or of the heavy element mass fraction $Z$ (blue). For Uranus and Neptune, a structure dominated by well-separated layers, including a solid superionic water layer, is presented. However, solutions with compositions that evolve more continuously are also highly plausible [From \citet{Guillot+2023}].}
\end{figure}

The structures of our four giant planets as obtained from interior models using the latest gravity measurements by Juno for Jupiter \citep{Durante+2020}, gravity and seismology measurements by Cassini for Saturn \citep{Iess+2019, Hedman+2019}, and by much less accurate gravity measurements for Uranus \citep{Jacobson2014} and Neptune \citep{Jacobson2009},  are shown in Fig.~\ref{fig:interiorslices}. For Jupiter, the models predict the existence of a fluid envelope separated into a helium-poor upper part and a helium-rich lower part, and a deep dilute core with a higher proportion of heavy elements mixed with hydrogen \citep{Wahl+2017, Debras+Chabrier2019, Miguel+2022, Militzer+2022}.

For Saturn, models predict that the transition from helium-poor to helium-rich region occurs at much larger depths, and combines with a progressive increase in heavy elements \citep{Mankovich+Fuller2021}. Models of Uranus and Neptune have considerable uncertainties \citep{Nettelmann+2013, Helled+2020} and point to a small, 0.5 to 4\,M$_\oplus$ fluid hydrogen-helium envelope overlaying a core made of ices and rocks which could either be layered or mixed, possibly including hydrogen and helium as well. In these models, the envelopes are believed to be largely convective and of uniform composition, except for Jupiter and Saturn, in a region in which helium separates from hydrogen and rains out \citep[e.g.,][]{Schottler+Redmer2018, Brygoo+2021}. The cores, however, may be highly stratified and exhibit large composition gradients with depth (see Section \ref{sec:static}).

\begin{figure}
\begin{center}
\includegraphics[scale=0.22]{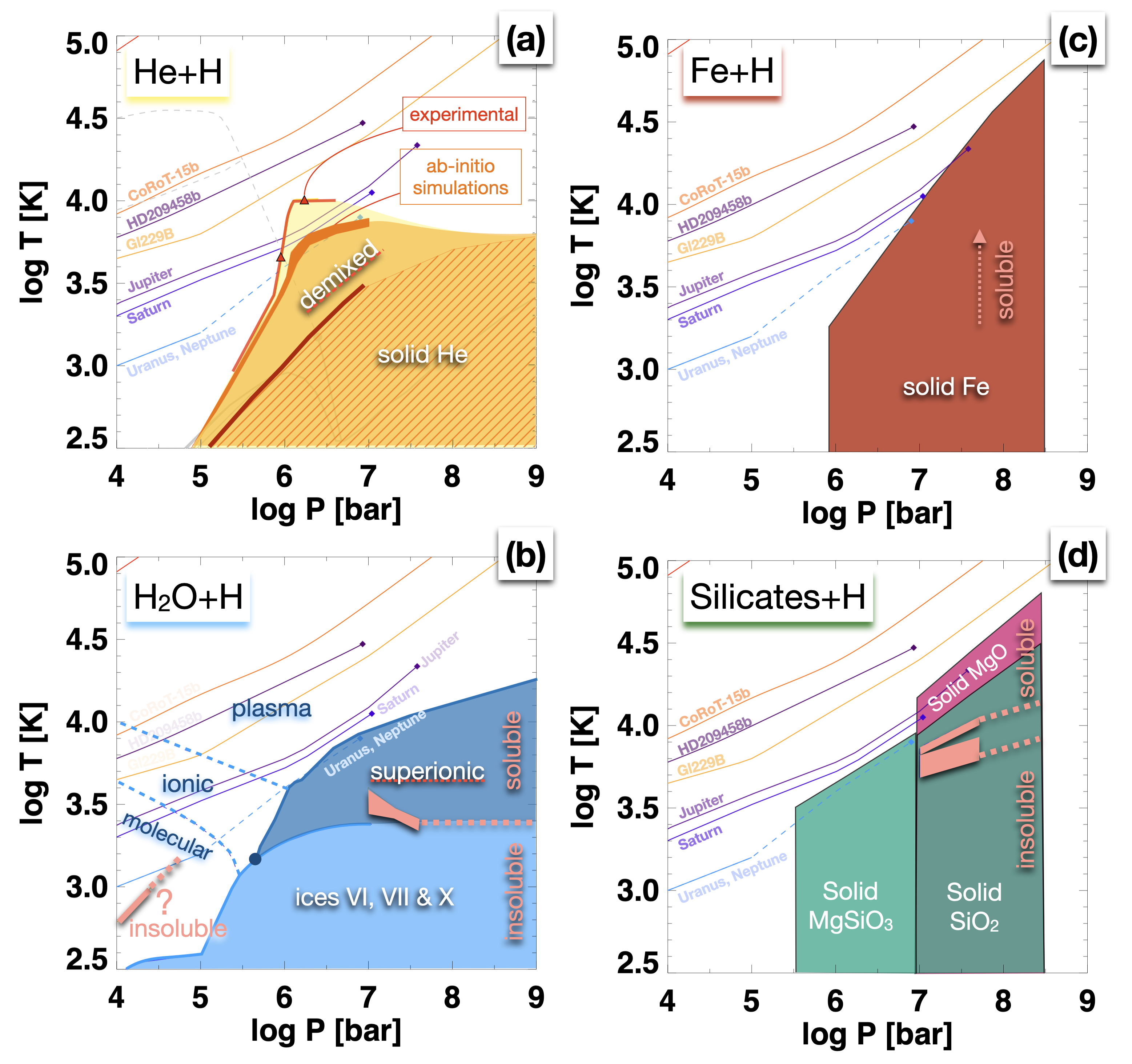}
\end{center} 
\caption{ \label{fig:phases} 
Phase diagrams of key elements with pressure-temperature profiles of relevant astrophysical objects (the Sun, brown dwarfs CoRoT-15b and Gl229~B, giant exoplanet HD~209458~b, and Jupiter, Saturn, Uranus and Neptune). Pressures are in bars ($\rm 1\,bar=10^5\,Pa=10^6\,dyn/cm^{2}$). (a): Location of the hydrogen-helium phase separation leading to helium rain-out in Jupiter and Saturn (yellow). The critical demixing temperatures obtained from ab-initio simulations are shown in orange. Those obtained from high-pressure experiments are shown as two triangles connected by a red curve. (b): Phase diagram of H$_2$O, including the fluid molecular, ionic and plasma phases, the solid ice phases and the superionic phase, as labelled. (c): Region of iron solidification, relevant for the central regions of Jupiter and possibly Saturn. (d): Regions of silicate solidification, relevant in the interiors of Uranus, Neptune, Saturn and Jupiter. Note that MgSiO$_3$ decomposes at high pressures into MgO (which should be solid in Jupiter and Saturn) and SiO$_2$ (which may still be liquid in the deep interior of Jupiter and Saturn).
[Adapted from \citet{Guillot+2023} --- see references therein].
}
\end{figure}

Although hydrogen and helium are definitely fluid, the state of other elements is more difficult to assess. In Jupiter and Saturn, the temperature profile lies above the estimated solidification curves for most elements shown in Fig.~\ref{fig:phases}, except possibly iron and magnesium oxide \citep{Mazevet+2019, Guillot+2023}. Even in this case, these elements are believed to be soluble in metallic hydrogen \citep{Gonzalez-Cataldo+2014}. Whether a solid compact core remains will depend on the availability of energy to mix elements upward \citep{Guillot+2004, Vazan+2018} and on how high-pressure elemental interactions between different species affect phase diagrams.

In Uranus and Neptune, gravitational moments indicate that the hydrogen-helium envelope should transition to a denser fluid \citep{Helled+2020}. For models assuming a three layer structure (hydrogen-helium overlying ices overlying rocks), water in the middle layer forms an ionic fluid which transitions to being superionic at Mbar pressures (\citealt{French+2009, Redmer+2011}, see Fig.~\ref{fig:phases}). Ab-initio calculations indicate that superionic water should behave as a solid \citep{Millot+2019}, with strong consequences for the planetary evolution \citep{Stixrude+2021}. This situation corresponds to the structure shown in Fig.~\ref{fig:interiorslices}. However, in the case of a non-layered structure, the mixing of other elements may affect the picture, perhaps considerably \citep{Guarguaglini+2019}, so that whether Uranus and Neptune are partially solid remains an open question \citep[see][for details]{Guillot+2023}. 

\subsection{Interiors of Giant Exoplanets}

So far, the primary measurements that we can rely on to constrain the global composition of exoplanets are their size and mass, coupled to theoretical expectations drawn from our understanding of solar system giant planets and planetary evolution. The known sample of giant exoplanets is extremely diverse, and includes planets with surprisingly large radii which cannot be reproduced by conventional evolution models as well as very dense planets requiring the presence of tens to hundreds of Earth masses of heavy elements in their interior \citep[e.g.,][]{Guillot+2006, Thorngren+2016}. 

Many of the giant exoplanets known today orbit much closer to their parent star than planets in our solar system. Because of observational biases, some of the planets that can be best characterized (in particular from transit photometry and radial velocimetry) also endure extremely high level of irradiation, orders of magnitude higher than Jupiter. This implies the existence of an outer radiative region that grows as the planet cools and contracts \citep{Guillot+1996}, potentially to kbar depths for mature hot Jupiters \citep{Guillot+Showman2002}. Thus, these planets could be made of an inner convective envelope and an outer radiative shell. 

Complications arise however: a large fraction of exoplanets are bloated beyond what is predicted by standard evolution models, implying that some missing physics is slowing their contraction \citep[see][]{Guillot+2006}. A number of explanations have been proposed \citep[e.g.,][]{Fortney+2021}. Although tidal interactions were initially identified as a possible explanation \citep{Bodenheimer+2001, Guillot+Showman2002, Baraffe2005}, the small eccentricities of hot Jupiters indicate tidal heating could only account for a small fraction of the anomalously large planets \citep{Fortney+2021}. More likely mechanisms such as Ohmic dissipation or downward energy transport \citep[][and references therein]{Batygin2010,Fortney+2021} alter the simple picture of a thick radiative shell. It could be much narrower \citep{Sarkis+2021}, or more generally, several radiative/convective zones could be present.
In the deeper interior, the probable existence of compositional gradients would also affect static stability in the interior, as well as the cooling of the giant planets themselves \citep{Chabrier+Baraffe2007}.

Finally, giant exoplanets which are more irradiated than those in the solar system (i.e., most of the ones which are currently observed with large instruments) should be warmer and therefore even less likely of being partially solid in their interior. This is particularly obvious when considering the case of HD~209458~b in Fig.~\ref{fig:phases}.

\subsection{Static stability in giant planets}
\label{sec:static}

While simple models of giant planets are fully convective and adiabatically stratified, several signs point toward stable (non-adiabatic) stratification due to compositional gradients at large depths. The existence of a largely stable deep interior in Saturn had been previously proposed in order to account for Saturn's high luminosity relative to predictions of adiabatic models \citep[e.g.,][]{leconte:12,leconte:13}.  \citet{fuller:14} also appealed to stable stratification in order to account for the splitting of normal modes as detected in the rings by Cassini \citep{Hedman+Nicholson2013}. The mode splitting is interpreted as the interaction between f-modes present in the whole planet and g-modes which are present only in stably stratified regions.  The presence of a dilute core both in Jupiter and in Saturn (see Fig.~\ref{fig:interiorslices}) seems to indicate that the existence of such a stable region may be the norm rather than the exception. However, for giant planets more massive than Jupiter, more energy is available to mix any primordial core more efficiently \citep[see][]{Guillot+2004}, raising the possibility that these massive giant planets might have fully eroded their core and be fully convective, erasing any primordial composition gradients.

The question of stability also arises in other contexts: in the hydrogen-helium phase separation region (see Fig.~\ref{fig:phases}), the formation of helium-rich droplets may or may not lead to the inhibition of convection \citep{Stevenson+Salpeter1977a, Mankovich+2016}. In the presence of compositional gradients, double-diffusive convection may arise, possibly leading to a small-scale staircase structure of convective regions sandwiched between conductive interfaces \citep{Rosenblum+2011, Leconte+Chabrier2012}. However, double-diffusive staircases may not always form for realistic conditions in giant planets \citep{Rosenblum+2011,Mirouh2012,Fuentes2022}, and they may become unstable over long time scales \citep{Moore2016}.

\begin{figure}
\begin{center}
\includegraphics[scale=0.26]{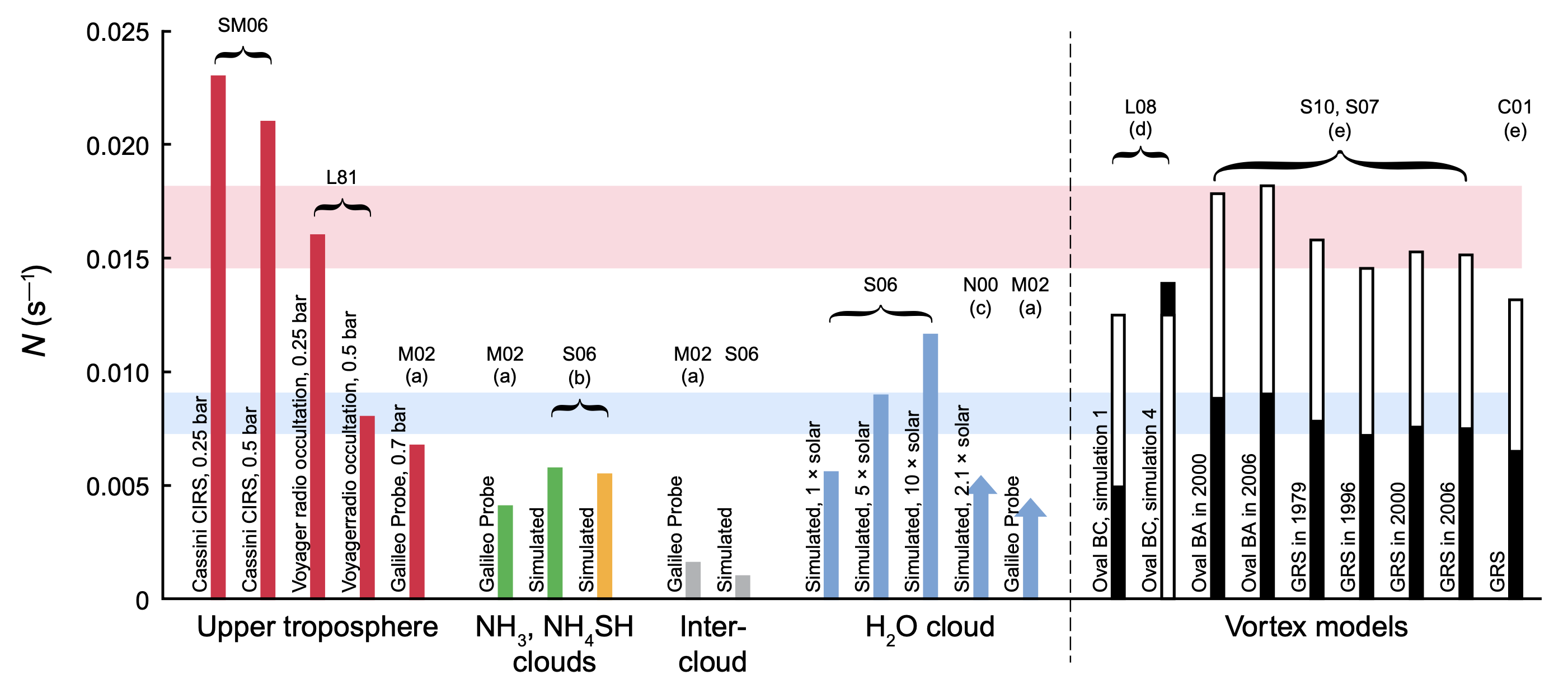}
\end{center} 
\caption{ \label{fig:static_stability} 
Estimates of static stability in Jupiter’s atmosphere as measured from its Br\"unt-V\"ais\"al\"a frequency $N$. Shaded bars (to the left of the dashed line) are from measurements and models of Jupiter’s vertical structure; black and white bars (to the right of the dashed line) are for vortex surroundings as determined by numerical models of the Great Red Spot, and of Jupiter's anticyclones. [From \citealt{Wong+2011}].
}
\end{figure}

In Jupiter, the existence of stable regions deep in the molecular envelope (between 95\% and 80\% of the planet's total radius) have been invoked to account for the inferred zonal wind decay \citep{Christensen+2020} and measured magnetic field properties \citep{Moore+2022, Connerney+2022, Sharan+2022}. But even features in the convective part of the atmosphere require that the atmosphere is at least partially stable. This is shown in Fig.~\ref{fig:static_stability} by measurements of the Br\"unt-V\"ais\"al\"a frequency, $N$, defined by 
\begin{equation}
    N^2 = \frac{g}{T}\left(\frac{dT}{dz} + \Gamma_{\rm ad}\right),
\label{eq:BV}    
\end{equation}
where $g$ is the gravitational acceleration, $T$ temperature, $z$ altitude, $\Gamma_{\rm ad}\equiv -(dT/dz)_S$ is the dry adiabatic lapse rate (at constant entropy $S$), and a perfect gas of uniform composition has been assumed for simplicity \citep{Wong+2011}. Thus, $N^2>0$ only in stable regions. The values of $N$ shown in Fig.~\ref{fig:static_stability} directly measured range from high values of around $0.02\rm\,s^{-1}$ in the radiative upper atmosphere, to about $2\times 10^{-3}\rm\,s^{-1}$ in the troposphere near 10 bars, as measured by the Galileo probe. Intermediate values are required to model the observed vortexes. In these regions, latent heat release due to the condensation of H$_2$O, NH$_4$SH and NH$_3$ can lead to subadiabatic temperature gradients accounting for the observed static stability. However, the penetration of cyclones and anticyclones to great depths as seen by Juno \citep[][and references therein]{Bolton+2021, Guillot+2023} indicates that the stable region is not limited solely to the water condensation zone near 5 bar.

\section{Tidal Dissipation Mechanisms}

\subsection{Tides: general principles}
\label{section:generalprinciple}

\subsubsection{Tidal force and potential}

Gaseous giant planets of our solar system have strong tidal interactions with their numerous moons, with tides raised on the moons by the host planet, and vice versa. Tides are for instance responsible for the volcanism of Io \citep{Johnsonetal1984,OjakangasStevenson1986} and for the heating of Enceladus \citep{Laineyetal2012}. The same is hypothesized to occur between close-in giant exoplanets and their host stars \citep[e.g.][and references therein]{WinnFabrycky2015}. In this section, we focus on tidal interactions with moons, but many of the same processes could happen in gas giant exoplanet systems where the tidal force is produced by the planet's host star.

Giant planets (hereafter the primary) are large, so the gravitational force exerted by the companions (e.g., a moon, hereafter the secondary)  varies non-negligibly between the center and the surface. The resulting differential force is the tidal force (${\bf f}_{\rm T}$), whose dominant quadrupolar component is derived from the tidal potential ($U_{\rm T}$) as:
\begin{equation}
{\bf f}_{\rm T}=-{\boldsymbol\nabla}U_{\rm T}\quad\hbox{with}\quad U_{\rm T}=-\frac{GM_{\rm m}r^2}{d^3}P_{2}\left[\frac{{\bf d}\cdot{\bf r}}{d \, r} \right], 
\end{equation} 
where $G$ is the gravitational constant, $M_{\rm m}$ the mass of the companion (treated here as a point mass), $d$ (${\bf d}$) the distance (vector) between the centres of mass of the primary and the secondary, ${\bf r}$ the position vector in the primary relative to its center of mass, and $P_{2}$ the quadrupolar Legendre polynomial. $U_{\rm T}$ is often expanded as a function of the Keplerian orbital parameters \citep[e.g.,][]{MurrayDermott1999,Kaula1962,MathisLePoncinLafitte2009,Ogilvie2014}. We also define the tidal frequency in the rotating frame of the planet:
\begin{equation}
\omega \equiv l \Omega_{\rm m}-m\Omega_{\rm p}, 
\label{TidalFreq}
\end{equation}
where $\Omega_{\rm m}$ and $\Omega_{\rm p}$ are the angular velocities of the moon's orbit and of the mean rotation of the planet, respectively, and $l$ and $m$ are integers.

\subsubsection{Equilibrium and dynamical tides and their dissipation}

\begin{figure}
\begin{center}
\includegraphics[scale=0.38]{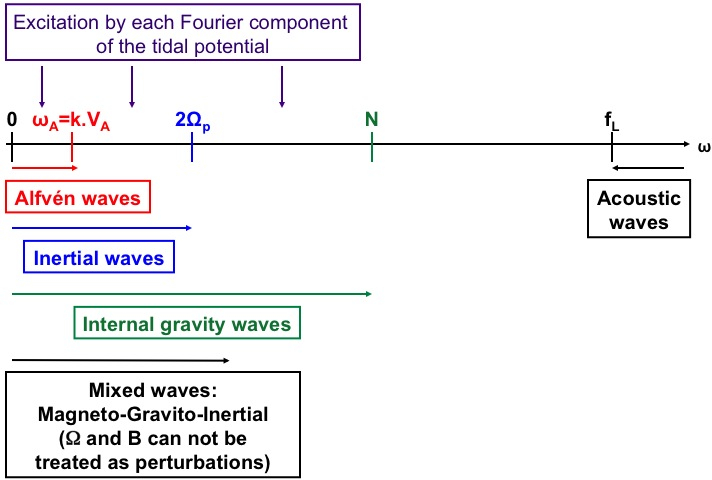}
\end{center}
\caption{Type of waves constituting the dynamical tide in fluid planetary layers. We introduce their characteristic frequencies: the Alfv\'en frequency $\omega_{\rm A}$, where $\vec k$ is the wave vector and ${\vec V}_{\rm A}={\vec B}/\sqrt{\mu_0\rho}$ is the Alfv\'en velocity ($\vec B$ being the magnetic field, $\rho$ the density, $\mu_0$ the vacuum permeability), the inertial frequency $2\Omega_{\rm p}$ ($\Omega_{\rm p}$ being the angular velocity of the planet), the Br\"unt-V\"ais\"al\"a frequency $N$ defined in Eq. \ref{eq:BV}, and the Lamb frequency $f_{\rm L}$.  For tidally excited waves in stably stratified regions of giant planets, we typically expect $\omega_{\rm A} < 2 \Omega_{\rm p} < N < f_{\rm L}$, but this may not always be true. [Adapted from \citealt{Mathis2011}].}
\label{fig:DTfreq}      
\end{figure}

The perturbation of the hydrostatic balance of celestial bodies by tidal forces results in mass redistribution and gravity and pressure perturbations. This mass redistribution creates a tidal bulge approximately in the direction of the companion (\autoref{fig:Qfig}) and a large-scale velocity or an elastic displacement in fluid/solid layers, respectively, the so-called fluid/elastic equilibrium tide \citep[e.g.][]{Zahn1966a,RMZ2012,Love1911,Tobieetal2005,Remusetal2012}. Because the equilibrium tide is not a solution of the full momentum equation, it is completed by the so-called dynamical tide. In fluid bodies, the dynamical tide is composed of inertial, gravity, Alfv\'en, and acoustic waves excited by the tidal force \citep[e.g.][]{Mathisetal2013}. Their restoring forces are the Coriolis acceleration, buoyancy (in convectively stable layers or at the free surface), magnetic tension, and pressure, respectively \citep[e.g.][and Fig. \ref{fig:DTfreq}]{RieutordSpringer}. In solid bodies, the dynamical tide includes tidally-excited shear waves \citep[e.g.][]{Altermanetal1959,Tobieetal2005}.

In an ideal case without any friction in the primary, the equilibrium and dynamical tides would be in phase with the tidal potential, with the resulting tidal bulges (one per Fourier mode of $U_{\rm T}$) being aligned along the line of centres. In this adiabatic situation, the time-averaged tidal torques on the orbit would vanish \citep[e.g.][]{Zahn1966a}. However, the equilibrium and dynamical tides are submitted to dissipation because of viscous or turbulent friction, the diffusion of heat and chemicals, Ohmic heating from magnetic fields, and viscoelastic dissipation in solid regions. Because of the induced losses of energy, the direction of the tidal bulge lags behind the secondary's position (\autoref{fig:Qfig}), resulting in a tidal torque. In addition, the conversion of the kinetic and potential energies of tides into heat impacts the structure and the evolution of the primary. 

In the simplest case of a circular co-planar system with semi-major axis $a_{\rm m}$, the torque applied to the spin of the primary can be approximated as \citep{Zahn2013}
\begin{equation}
\Gamma=-\frac{\left(\Omega_{\rm p}-\Omega_{\rm m}\right)}{t_{\rm friction}}\left(\frac{M_{\rm m}}{M_{\rm p}}\right)^2 M_{\rm p} R_{\rm p}^2 \left(\frac{R_{\rm p}}{a_{\rm m}}\right)^6,
\label{tidaltorque}
\end{equation}
where $M_{\rm p}$ and $R_{\rm p}$ are the mass and the radius of the primary and $t_{\rm friction}$ is the time characterising tidal dissipation. This expression shows the importance of understanding the physics of tidal dissipation from first principles to obtain a robust evaluation of $t_{\rm friction}$ and a prediction of the evolution of the system.

It is common to express $t_{\rm friction}$ as a function of a corresponding tidal quality factor $Q$ \citep{MacDonald1964}, in analogy with the theory of forced oscillators \citep[e.g.][]{Greenberg2009}. A short friction time corresponds to a strong dissipation and a low quality factor (and vice-versa). To evaluate tidal dissipation, we compute the ratio of $Q$ with the quadrupolar Love number $k_2$, defined as the ratio of the primary's perturbed gravitational potential to the tidal potential at the primary's surface, which depends on the primary's mass concentration. The tidal torque is proportional to $k_2/Q$ (which is equal to the imaginary component of $k_2$, Im[$k_2$]), i.e., the component of the tidal bulge's gravitational field that is misaligned with respect to the moon.

\subsubsection{Tidal orbital and rotational evolution}

For moon orbits that are nearly circular and aligned with the planet's rotation (as is the case for most of the solar system's major moons), tidal evolution is relatively simple. In Eq. \ref{tidaltorque}, one can identify the crucial importance of the so-called {co-rotation radius} at which $\Omega_{\rm p}=\Omega_{\rm m}$. If $\Omega_{\rm m} >\Omega_{\rm p}$, then the companion migrates inwards and the primary rotation accelerates. If $\Omega_{\rm m} < \Omega_{\rm p}$, the companion migrates outwards and the primary rotation slows down. The latter is the configuration of the Earth-Moon system where the Moon migrates outwards by $\sim$3.8 centimetres per year. When $\Omega_{\rm m} >\Omega_{\rm p}$, the Darwin instability results if $L_{\rm orb}\le 3L_{\rm p}$, where $L_{\rm orb}$ and $L_{\rm p}$ are the angular momentum of the orbit and of the primary's rotation respectively. In this case, the companion spirals towards the primary until it is tidally disrupted at the Roche limit \citep{Hut1980}. If $L_{\rm orb}\ge 3L_{\rm p}$, the system evolves towards an equilibrium state where the orbit is circular and the primary's rotation is aligned and synchronized with the orbit.

We can see that the time characterising dissipation ($t_{\rm friction}$) allows one to predict orbital migration and circularisation time scales, and spin alignment and synchronisation time scales. Each dissipation mechanism leads to a specific frequency-dependence of $t_{\rm friction}$ \cite[i.e. smooth or resonant; we refer to][for mathematical formalisms]{Mathisetal2013,Ogilvie2014}, with very sensitive frequency-dependence for resonant fluid dynamical tides. This has major consequences for the dynamical evolution of planetary systems \citep{ZahnBouchet1989,WitteSavonije2002,EfroimskyLainey2007,ADLPM2014}.

\subsection{Dissipation in the solid core}

\begin{figure}
\begin{center}
\includegraphics[scale=0.38]{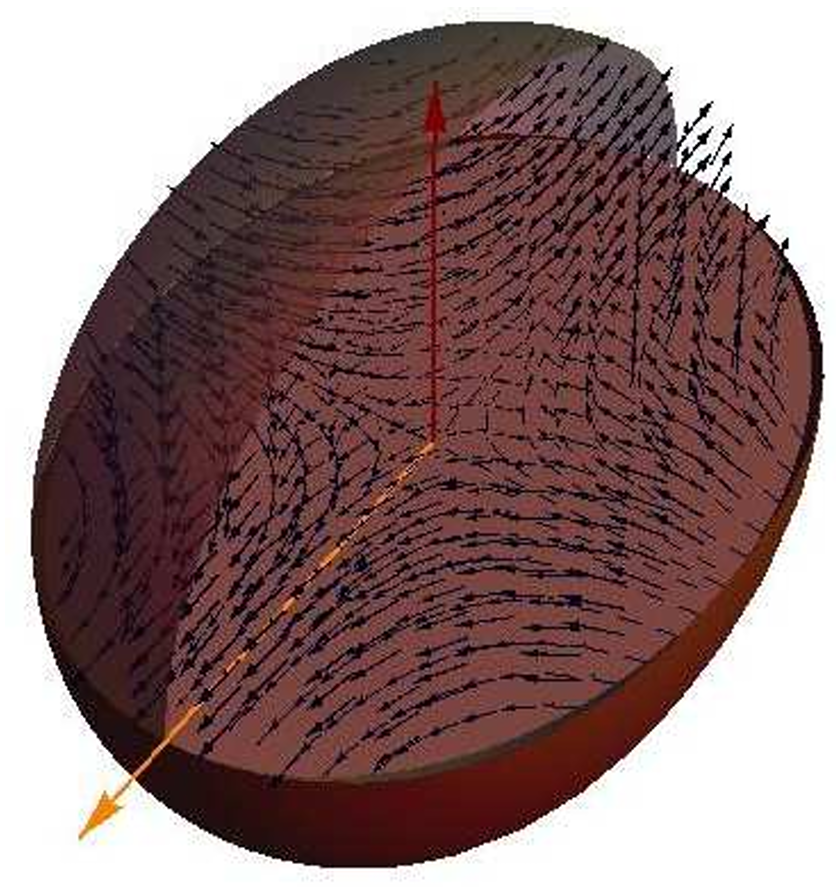}
\includegraphics[scale=0.45]{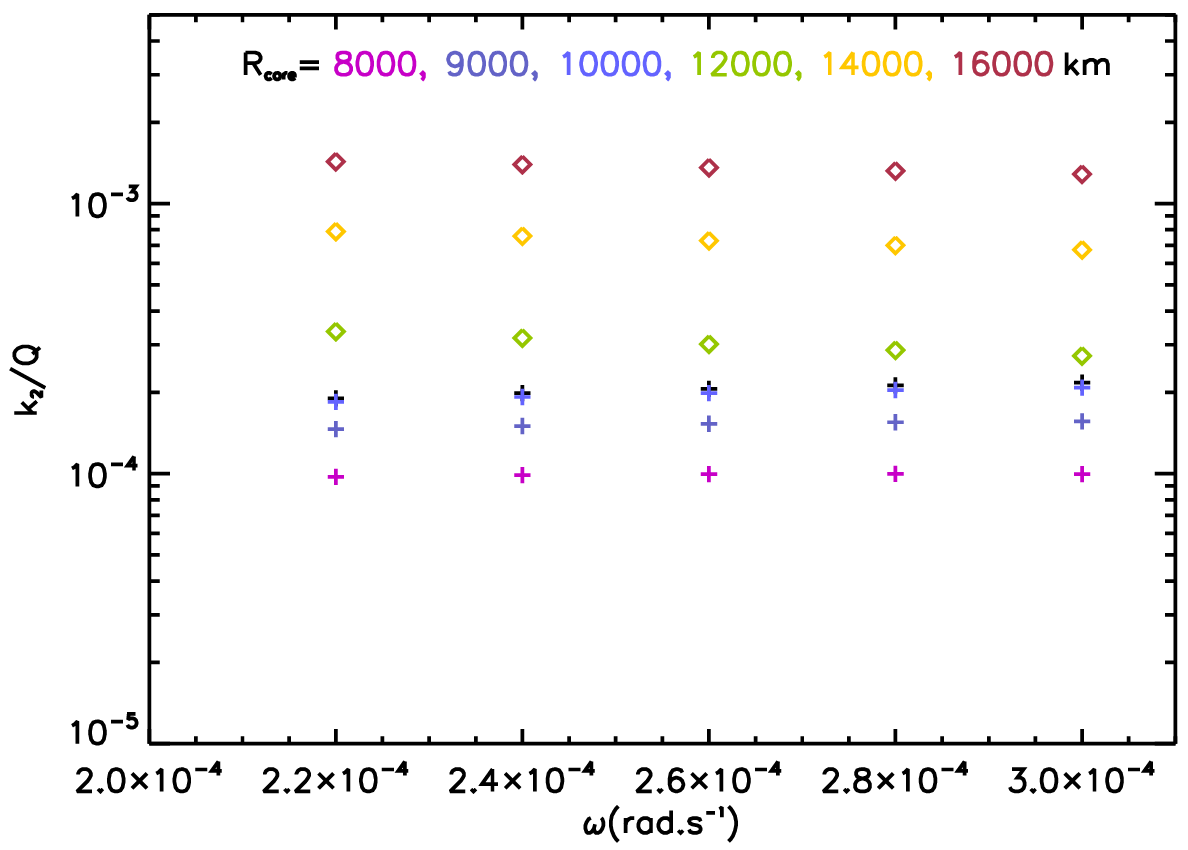}
\end{center}
\caption{{\bf Left:} elastic equilibrium tide in the {rocky/icy core} of a giant planet \citep[taken from][courtesy Astronomy \& Astrophysics]{Remusetal2012}. The red and orange arrows are the rotation axis and the direction of the companion respectively. {\bf Right:} Tidal dissipation in the solid core of Saturn, computed for different values of its radius as a function of the tidal frequency, taking into account its radial density stratification, for a viscosity $\eta=10^{15}$ Pa s \citep[taken from][courtesy Icarus]{Laineyetal2017}. The dissipation is expressed as a function of the second-order Love number $k_2$, which evaluates the amplitude of the hydrostatic tidal deformation, and the tidal quality factor $Q$.}
\label{fig:solid}      
\end{figure}

In the standard core-accretion scenario for the formation of gaseous giant planets, a rocky/icy core is needed to ensure the accretion and the stability of a deep gaseous envelope \citep{Pollacketal1996}. If it solidifies, this core constitutes a potentially strong source of tidal dissipation, because of the friction applied through its rheology on the solid equilibrium tide \citep[e.g.][and Fig. \ref{fig:solid} left panel]{Tobieetal2005,Efroimsky2013,tobie2019}, and a boundary for tidal flows studied in the next section. 

In this framework, \cite{Dermott1979} studied the equilibrium tide in homogeneous solid cores. He showed how its amplitude is boosted thanks to the hydrostatic pressure exerted by a friction-less homogeneous deep gaseous envelope when compared to the case of a sole solid core without any surrounding fluid layer. \cite{Remusetal2012,Remusetal2015} built on this first work and provided a coherent model of the dissipation of solid tides in a simplified bi-layer model with an homogeneous viscoelastic solid core surrounded by a deep homogeneous non-dissipative fluid envelope. Using complex Love numbers and the correspondence principle \citep[e.g.][]{Tobieetal2005}, they computed the elastic deformation of solid layers and the viscous dissipation in its bulk. Because of our lack of knowledge of the rheology of the core of giant planets in their extreme conditions of pressure and temperature, they considered the case of the simplest Maxwell rheology and they explored plausible values of the core's rigidity and viscosity.

These studies showed that friction can simultaneously explain the high amplitude of the tidal dissipation observed for Jupiter and Saturn by \cite{Laineyetal2009} and \cite {Laineyetal2012} and its smooth frequency-dependence in the Saturnian system for Enceladus, Tethys and Dione.  
The moon measurements, made using high-precision ground-based astrometric measurements of the orbital migration, seemed to favour at that time the hypothesis of the dissipation of a solid tide. Indeed, the dissipation of tidal waves propagating in the gaseous envelope would have a resonant frequency-dependence with possible sharp variations of $Q$ by several orders of magnitude \citep{Ogilvie2004,Mathisetal2016}.

The measured values of $Q$ were an order of magnitude smaller than what was predicted assuming a constant value of $Q$ and simultaneous formation of the planets and of their satellites. This opened the possibility of having young moons as proposed in \cite{Charnozetal2011} and in \cite{crida:12}. However,  \cite{ShojiHussmann2017} took into account the frequency-dependence of the tidal dissipation predicted in the case of a visco-elastic core when they solved the equations for the orbital migration of the moons. They showed that a strong dissipation in Saturn does not necessarily imply young moons. \cite{StorchLai2014} also studied this mechanism to provide us an explanation of the formation of hot Jupiters through circularization in a high-eccentricity migration scenario. Finally, \cite{StorchLai2015} improved this bi-layer model by considering a set-up with an homogeneous core surrounded by a polytropic gas envelope.

These works were generalized in \cite{Laineyetal2017} by taking into account the radial profiles of the density and of the incompressibility parameter of the assumed Maxwell rheology \citep[we refer the reader to][for details of the numerical modelling]{Tobieetal2005}. As illustrated in the right panel of Fig. \ref{fig:solid}, the computed viscoelastic dissipation is compatible with the measured dissipation at the frequencies of Enceladus, Tethys and Dione. However, this viscoelastic dissipation, if it is the source of dissipation for these three moons, cannot explain the stronger dissipation by one order of magnitude observed at the frequency of Rhea. Additional dissipation within the fluid envelope must be invoked: we have to focus our efforts on modelling the dissipation of the fluid dynamical tide.

\subsection{Dissipation in the fluid envelope}

A seminal study of the dynamical tide, using a standard tri-layer model of a gaseous giant planet with a deep convective envelope modelled as a polytrope, was achieved by \cite{OgilvieLin2004}. Assuming the presence of a central solid core, they showed how tidally-excited inertial waves can be efficiently dissipated assuming an effective (turbulent) viscous friction. The restoring force governing the dynamics of these waves is the Coriolis acceleration. The dissipation is primary localised in intense shear layers called wave attractors launched at the critical co-latitude $\theta_{\rm c}= \arccos \left(\omega/2\Omega_{\rm p}\right)$ \citep{Ogilvie2005,GoodmanLackner2009,rieutord:10}. It varies by several orders of magnitude with the tidal frequency and the value of the effective viscosity. The resonances in the dissipation frequency spectrum are stronger and sharper for low viscosity.

In this framework, \cite{Mathisetal2016} studied the efficiency of the turbulent friction applied on tidal inertial waves by rapidly rotating convection. Using scaling laws of the rotating mixing-length theory derived by \cite{Stevenson1979}, which have been confirmed in recent numerical simulations \citep{Barkeretal2014,Currieetal2020,Vasiletal2021,Fuentesetal2023,devriesetal2023}, they showed that rapid rotation leads to less efficient turbulent viscous friction applied to tidally-excited inertial modes. The associated sharper and higher resonant peaks in the dissipation frequency-spectrum \citep[e.g.][]{ADMLP2015} may provide a possible explanation for rapid moon migration as observed in the case of Rhea. 

The efficiency of inertial wave dissipation also depends on the structure of the planet. In simple planetary models with a uniform density envelope surrounding a solid core, the frequency-average inertial wave dissipation rate scales approximately as the core radius to the fifth power \citep{ogilvie:13}. For more realistic planetary models with barotropic structures, significant inertial wave dissipation can occur even in the absence of a solid core \citep{ogilvie:13,lazovik:24}. Inertial wave dissipation has also been investigated for highly eccentric orbits applicable to exoplanet systems \citep{Papaloizou:04,ivanov:07,papaloizou:10}. These studies investigate the more complex dynamics that can occur in the case of eccentric/misaligned orbits.

In the case where a solid (denser) core was assumed, it is also important to compare the relative strengths of the dissipation of tidal inertial waves and of the viscoelastic dissipation. This was achieved by \cite{Gueneletal2014}, who considered a simplified bi-layer model with a homogeneous solid core and a deep convective envelope with tidally excited inertial modes. Using the method developed by \cite{Ogilvie2013}, they computed the frequency-averaged values of both types of dissipation. They showed that fluid and solid tidal friction might have comparable strength when assuming plausible values for the core rigidity. This demonstrated the importance of computing dissipation in each fluid (gaseous or liquid) and solid planetary layer. In the case of a fully convective core-less gaseous giant planet, \cite{Wu2005a,Wu2005b} demonstrated that the dissipation of tidal inertial modes will be weakened, as in the case where a solid core is replaced by a denser motionless fluid core \citep{Mathis2015}.

\begin{figure*}
\begin{center}
\includegraphics[scale=0.3]{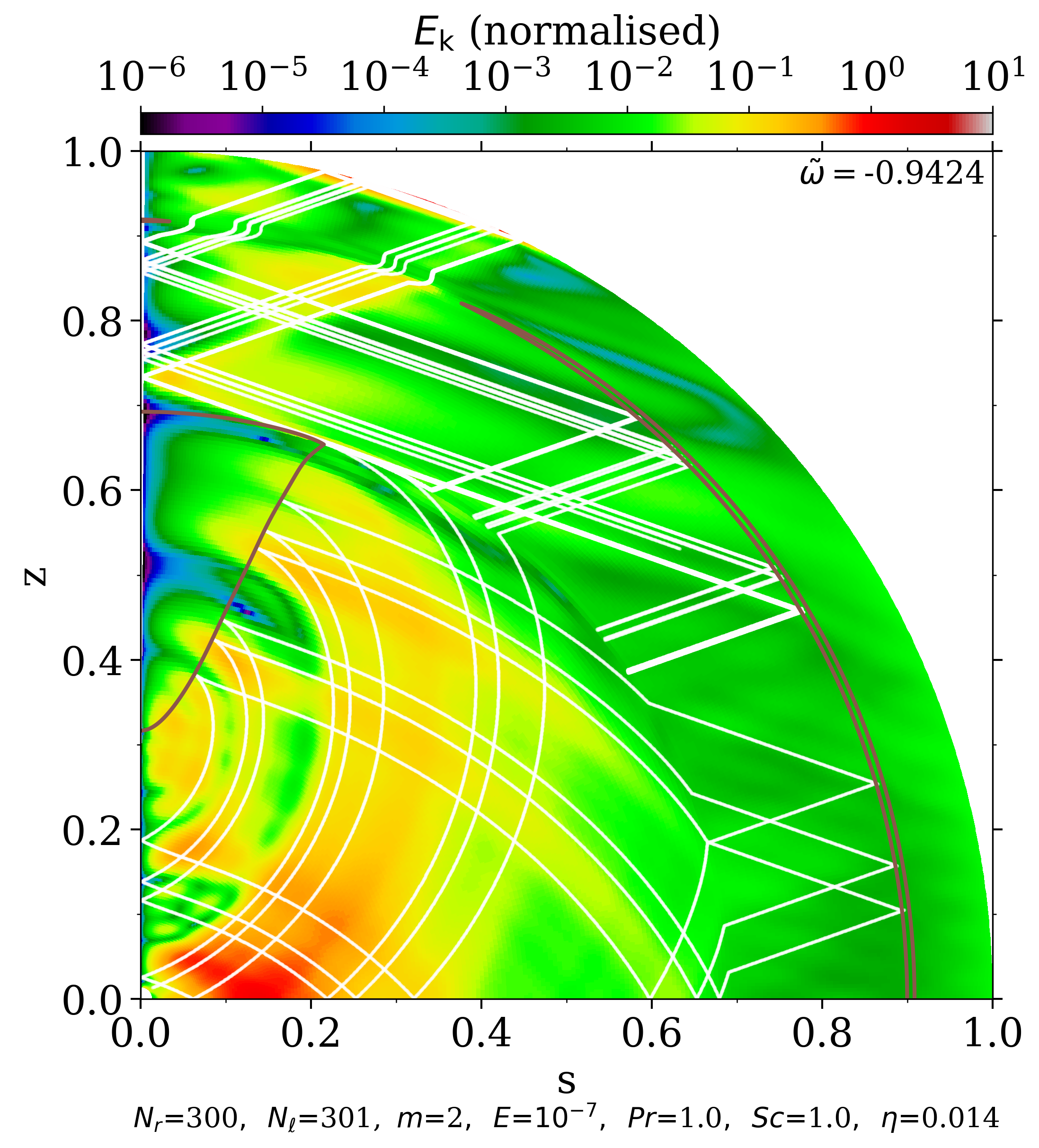}\quad
\includegraphics[scale=0.22]{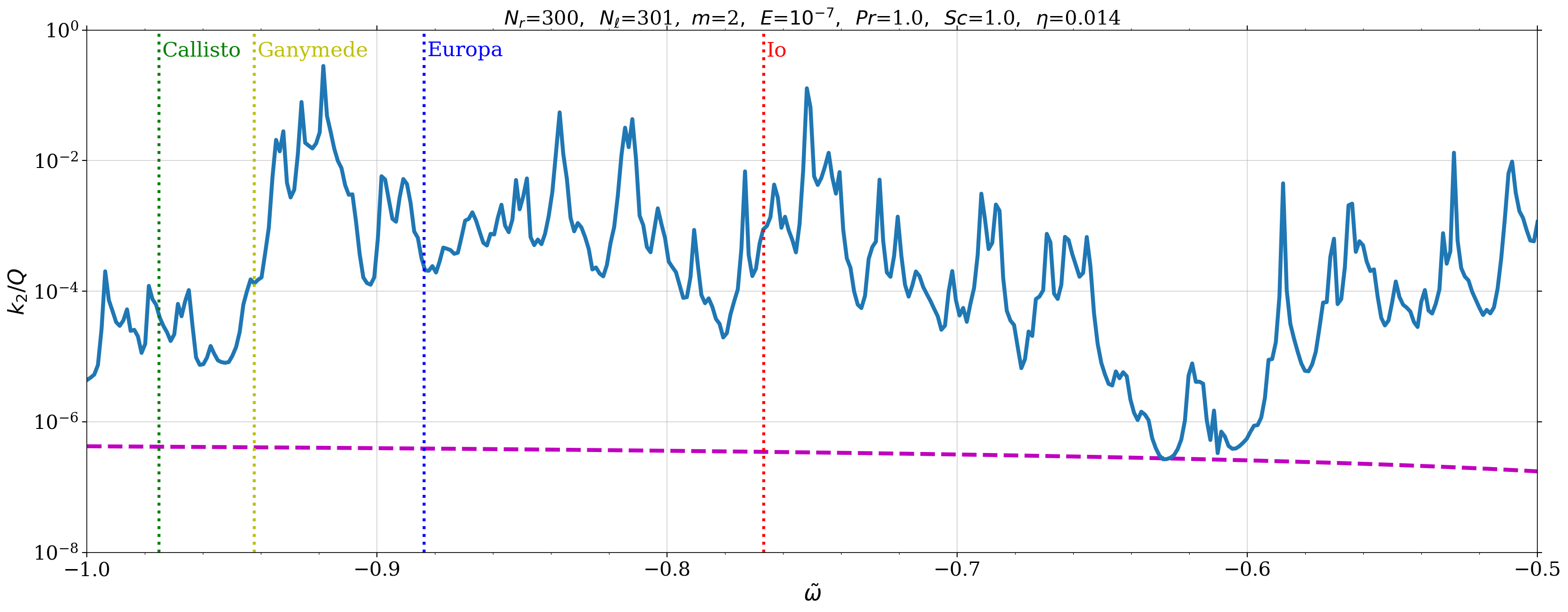}
\end{center}
\caption{{\bf Left:} Kinetic energy density of tidal waves propagating in the interior of Jupiter at the tidal frequency of Ganymede. We use a model of the internal structure of the planet computed by \citealt{Debras+Chabrier2019} with (from core to surface) a dilute stably stratified core, a convective metallic hydrogen shell, a very thin stably stratified layer induced by the immiscibility of Helium into Hydrogen, and an external convective molecular envelope. We recognise the propagation of gravito-inertial waves in the central stable core and the one of inertial waves in convective layers. We assume a solid core with a radius $R_{\rm c}=0.014R_{\rm p}$ with $R_{\rm p}$ the radius of the planet.The ratio of the (turbulent) viscous force to the Coriolis acceleration (the Ekman number $E$) is $10^{-7}$ and viscous, heat and chemical diffusions are equal (i.e. the Prandtl ($P_r$) and the Schmidt ($S_c$) numbers are set to unity) (adapted from Dhouib, et al. 2023, courtesy H. Dhouib). {\bf Right:} The value of $k_2/Q$ as a function of the tidal frequency corresponding to the dissipation of tidal waves as computed in the left panel (blue solid line). The purple line corresponds to the value obtained in the case of tidal inertial waves computed in a sole deep convective envelope surrounding the small core as this was expected before the constraints obtained by the Juno mission.}
\label{fig:fluiddissip}      
\end{figure*}

Our picture of the tidal dissipation mechanism has been greatly changed by the revolution of our understanding of gaseous giant planets interiors thanks to the {\it Juno} and {\it Cassini} space missions as explained in Sec. \ref{sec:GPS}. Indeed, the newly obtained constraints show the importance of possible stably stratified regions within giant planets. As illustrated in Fig. \ref{fig:interiorslices}, state-of-the-art models of the internal structure of giant planets are all constituted by a combination of convective and stably-stratified regions. These convectively stable layers are triggered by heavy element and helium abundance gradients, which stabilise the entropy gradient.

Stably stratified layers allow for dissipation of tidally excited gravito-inertial waves which propagate within these layers. The restoring forces that govern their dynamics are buoyancy and the Coriolis acceleration. \cite{ioannou:93a,ioannou:93b} were pioneers in envisioning the possible importance of these waves to predict tidal dissipation within Jupiter. The development of a new generation of models of the tidal dissipation in giant planets, with the dissipation of tidal inertial waves in convective zones and of tidal gravito-inertial waves in stable layers has been undertaken first in local Cartesian geometry \citep{Andreetal2017,Andreetal2019} and now in global spherical \citep[][Dhouib et al., submitted; Pontin et al., submitted]{Pontinetal2020,Pontinetal2023a,Lin2023,lazovik:24} and spheroidal geometries \citep{Dewberry2023}. These models show how gravito-inertial waves within stably stratified layers enhance the tidal dissipation when compared to the case of a ``standard" fully convective envelope (see Fig. \ref{fig:fluiddissip}). They also show that a stably stratified layer increases the amplitude of inertial wave excitation and associated dissipation within convective regions.

\cite{Fuller2016,Luanetal2018} proposed that the gravito-inertial modes in either stably stratified or convective regions can trap moons into resonances \citep[][]{WitteSavonije2002}. This mechanism is a natural candidate to explain the dissipation rates measured by astrometric observations at the frequencies of Enceladus, Tethys, Dione, and Rhea (see Section \ref{sec:reslock}). However, all of these models still require improvements in the modelling of possible nonlinearities and instabilities affecting tidal waves, of wave-turbulence nonlinear interactions, and of the impact of differential rotation and magnetic fields.

\subsection{Current challenges}

\subsubsection{Nonlinearities and resonances}

As of today, hydrodynamical modelling of the propagation and dissipation of tidal waves in giant planets has been done primarily in the linear regime. As a consequence, many challenging important questions remain open.

First, as in the case of high-amplitude tidal gravity waves in the radiative cores of low-mass stars \citep[e.g.][]{BarkerOgilvie2010}, tidal waves propagating in giant planets may break. This may be the case for gravito-inertial waves propagating towards the center of the planet in absence of a core because of geometric focusing \citep[see][in the case of gravity waves]{Press1981}. Tidal (gravito-)inertial waves propagating near the planetary surface may also break because of the decreasing density as in the case of waves propagating in early-type stars \citep[e.g.][]{Rogersetal2013}. Inertial waves have very large wave numbers at critical latitudes, potentially driving non-linear wave breaking. Even when waves do not break, non-linear wave coupling may increase wave damping rates \citep{Weinberg2012,Essick2016}.

Second, future studies should examine the complex nonlinear wave-turbulence interactions that mediate dissipation. The key questions to answer are: can we model the action of the convective turbulence on tidal inertial waves as an effective turbulent viscous friction, and what is its strength (see, e.g., \citealt{Terquem2021,Barker2021})? How do the attractors predicted in the linear regime evolve in the presence of turbulent convection or in a nonlinear regime? Do they survive or are they destroyed? Are we entering in the so-called wave turbulence regime? How do tidal inertial waves interact with the numerous convective vortices observed at the surface of giant planets with a broad diversity of characteristic scales?

Studies have begun to examine those crucial questions \citep[e.g.][]{JO2014,LeReunetal2017,AstoulBarker2022,Dandoyetal2023,AstoulBarker2023,Terquem2023} but this research area in the specific framework of the prediction of tidal dissipation is still in its infancy. In the case where the modelling of the turbulent friction can be done using an effective eddy viscosity, important progress has been made in predicting its strength and frequency-dependence thanks to systematic studies using direct nonlinear numerical simulations of the interactions between turbulent convection and oscillatory flows \citep{Duguidetal2020a,Duguidetal2020b,VidalBarker2020a,VidalBarker2020b,BarkerAstoul2021}, and more realistic tidal flows \citep{devriesetal2023}.

When tidal deformation is large (which can occur in hot Jupiters but not in solar system planets), global-scale tidal flows can become unstable because of the elliptic instability \citep{Kerswell2002}. In this case, the oscillatory tidal flow non-linearly excites a pair of gravity waves. This triggers the instability, turbulence, and dissipation \citep[for a complete review, we refer the reader to][and references therein]{LeBarsetal2015}. Using hydrodynamical simulations, \cite{BarkerLithwick2013} and \cite{Barker2016} have identified the important role of the differential rotation triggered by the instability, which acts to saturate it. This leads to an intermittent dissipation. In the presence of a weak magnetic field, differential rotation is damped and turbulence is sustained that triggers a stronger dissipation \citep{BarkerLithwick2014}. The elliptic instability may be most important in exoplanet systems where the tidal distortion from the host star can be much larger than that produced by moons.

\subsubsection{Zonal flows and magnetic fields}

The observations provided by {\it Juno} and {\it Cassini} (during its ``grand finale") have revealed that the convective molecular envelopes of Jupiter and Saturn are the seat of the differential rotation observed at their surface \citep{Guillotetal2018,Galantietal2019}. Below $\sim$3000 km for Jupiter and $\sim$8000 km for Saturn, the differential rotation weakens as the electrical conductivity increases with depth. That means that the impact of zonal flows on tidal (gravito-)inertial waves excited in these external layers must be taken into account.

First, differential rotation can modify the cavities where (gravito-)inertial waves propagate and also change their frequencies \citep{Mathis2009,BaruteauRieutord2013,Gueneletal2016,Mirouhetal2016}. Next, critical layers appear where the local Doppler-shifted frequency of the fluid oscillation vanishes. They are of crucial importance since they are the place of strong wave-mean zonal flows interactions that can constitute the dominant component of the tidal torque both in stably-stratified \citep{GoldreichNicholson1989} and in convective regions \citep[e.g.][Baruteau et al., in prep]{Favieretal2014,Gueneletal2016,Astouletal2021,AstoulBarker2022}. Finally, an interesting and necessary perspective will be to compare the relative strengths of zonal flows induced by tidal waves \citep[e.g.][]{Morizeetal2010,Cebronetal2021} and those triggered by the turbulent convection.

Gaseous giant planets are also magnetic celestial bodies \citep{Connerneyetal2018,Duarteetal2018,Doughertyetal2018,Yadavetal2022}. As a consequence, it will be necessary to consider the impact of magnetic fields on tidal waves \citep[e.g.][]{Wei2016,LinOgilvie2018,Wei2018}. Tidal (gravito-)inertial waves become tidal magneto-(gravito-)inertial waves while the Ohmic heating related to their dissipation adds to the viscous and heat diffusion. Finally, it will be important to study carefully the tidally-induced modification of planetary magnetic fields \citep[e.g.][]{CebronHollerbach2014}.

\section{Resonance Locking}
\label{sec:reslock}

As discussed above, the tidal torque due to dynamical tides can easily dominate over equilibrium tides, and is usually very sensitive to the tidal forcing frequency. Unfortunately, our understanding of giant planet structures is not detailed enough to accurately calculate the frequencies of gravito-inertial modes at which tidal dissipation peaks. Hence, we cannot determine whether the observed moons will resonate with planetary oscillation modes. A resonance occurs when the tidal forcing frequency in the inertial frame, $m \Omega_{\rm m}$, is nearly equal to one of the planetary oscillation mode frequencies, $\sigma_\alpha$.
Since the resonances are typically narrow in frequency, it is tempting to conclude that the moons are unlikely to be in resonance, and that tidal dissipation due to dynamical tides should be weak.

\begin{figure}
\begin{center}
\includegraphics[scale=0.38]{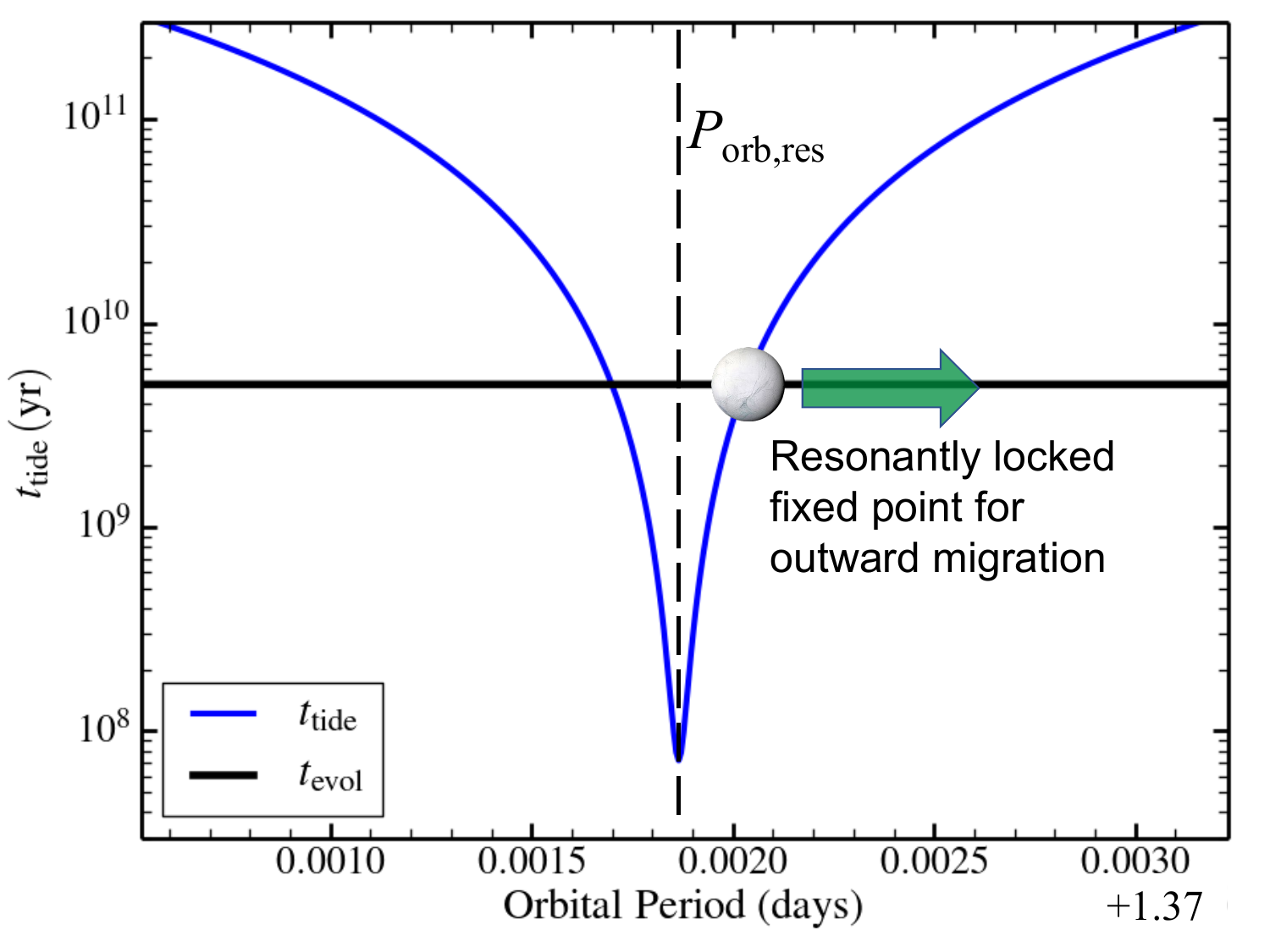}
\end{center} 
\caption{ \label{fig:ResLock} Example of the tidal migration time scale (blue line) as a function of orbital period, near a resonance with a Saturn oscillation mode. A moon with orbital period $P_{\rm orb,res}$ and semi-major axis $a_{\rm res}$ would lie at exact resonance with the oscillation mode (vertical dashed line). The horizontal black line denotes the time scale $t_{\rm evol}$ at which $P_{\rm orb,res}$ moves outward. At the resonantly locked fixed point (where the blue and black lines intersect), the moon would migrate out at the same rate as the resonance location, allowing for a stable resonance lock.}
\end{figure}

However, just as there are physical processes that allow moons to be stably locked into mean-motion resonances with one another, there are also processes that allow moons to be stably locked into resonances with planetary oscillations. This phenomenon is known as ``tidal resonance locking", as was first described by \cite{witte:99} and first considered for moons by \cite{Fuller2016}. If resonance locking can act in the Saturn system, it is much more likely (perhaps even probable) that a moon will be in resonance with one of Saturn's oscillations, allowing it to migrate much faster than expected otherwise. While it is not currently clear whether resonance locking is required to explain the observed moon migration rates, it is an appealing mechanism to increase tidal dissipation rates.

Resonance locking arises from the fact that planets and stars are not static objects, but are evolving in time. Because a planet's internal structure is slowly changing, so too are the oscillation mode frequencies that can become resonantly excited by one of the moons. Since the moons migrate outward, their tidal forcing frequencies $m \Omega_{\rm m}$ decrease with time. Crucially, if the oscillation modes that resonate with the moons also have decreasing frequencies $\sigma_\alpha$ in the inertial frame (corresponding to resonant semi-major axes that move outward in time), then resonance locking can potentially occur, as illustrated in \autoref{fig:ResLock}.

In many cases, the moon migration rates are fast at exact resonance, and slow away from resonance. Just outward of the resonant location, there is a ``fixed point" at which the moon will tidally migrate at the same rate that the resonance location is moving outwards (see \autoref{fig:ResLock}). This fixed point is stable: an inward perturbation will move the moon deeper into resonance, so that it migrates outward faster, and returns to the fixed point. An outward perturbation will move the moon out of resonance, so that it migrates slower, allowing the resonance location to catch up to the moon. Hence, moons can stably ``surf" these resonance locations outwards as the planet evolves, so that their migration rates are determined by the rate at which the planetary structure is evolving. Moreover, planetary evolution may allow the resonant locations to sweep past the moons, such that moons will naturally be caught up in these resonances, regardless of when and where the moons formed.

\subsection{When resonance locking will not occur}

There are several reasons why some moons may not be caught in resonance locks. First, if the damping rates of Saturn's oscillation modes are too high, the resonant migration time scales (blue line in Figure \ref{fig:ResLock}) will be broader and shallower, such that the blue line no longer intercepts the black line, and the resonance locking fixed point no longer exists. When this occurs, the moon will not be able to migrate outward as fast as the resonance location is moving, so resonance locks cannot be sustained. This can also occur if Saturn evolves too quickly (i.e., the black line moves downward in Figure \ref{fig:ResLock}), or if a moon has a large semi-major axis (causing the blue line to shift upward). Hence, we expect resonance locks to eventually break as the moons migrate outwards, though it is not clear when exactly this will occur.

Resonance locks also cannot occur if the resonant locations are moving inward, because then there is no resonance locking fixed point. In this scenario, the moons simply pass through the resonances on short time scales, so we do not expect them to be in resonance today.

To determine whether resonance locking can occur, we must examine the resonance condition, 
\beq
\label{res1}
\sigma_\alpha \simeq m \Omega_{\rm m}  \, .
\eeq 
Differentiating the resonance criterion of equation \ref{res1} with respect to time leads to the locking criterion
\beq
\label{reslock}
 {\dot \sigma}_\alpha \simeq m {\dot \Omega}_{\rm m}  \, .
\eeq
Since the moons migrate outward, $\dot{\Omega}_{\rm m}$ is negative, so the mode frequency must decrease in the inertial frame in order for resonance locking to occur.

In Saturn's rotating frame the mode frequency is 
\begin{equation}
\label{omega}
\omega_\alpha = \sigma_\alpha - m \Omega_{\rm p},    
\end{equation}
which is negative for all of Saturn's moons (using $m\geq2$), such that resonant modes have negative frequencies and are retrograde modes. Resonance locking thus requires
\beq
\label{resomega}
\dot{\omega}_\alpha \simeq  m ( \dot{\Omega}_{\rm m} - \dot{\Omega}_{\rm p} )  \, .
\eeq
For a constant rotation rate of Saturn, or if Saturn's rotation rate increases with time, the mode frequency $\omega_\alpha$ must decrease in the rotating frame, i.e., become more negative and such that $|\omega_\alpha|$ increases. Increasing rotation frequency and $|\omega_\alpha|$ are what we expect if Saturn is slowly contracting, but explicit verification of this requirement has not been demonstrated with evolutionary models of Saturn.

Finally, resonance locking may not occur if non-linear effects (which are neglected in most calculations of tidally excited oscillations) become important. A tidally driven parent mode can non-linearly couple to two daughter modes, as described in detail in works such as \cite{Weinberg2012,Essick2016}. The transfer of energy to daughter modes (and subsequently to granddaughter modes, etc.) effectively acts as damping that may saturate the resonance and prevent the fixed point from existing, as described above. This likely prevents resonance locking from occurring in many Sun-like stars (see \citealt{Ma2021}), but non-linear coupling calculations have not been performed for tidally excited modes of giant planets. While we can be sure that resonance locks will eventually break as moons move to larger semi-major axis, it is not clear when this non-linear saturation will occur.

\subsection{Consequences of Resonance Locking}

If a moon becomes caught in a resonance lock, its tidal migration time scale is dictated by the evolution of the resonant mode frequency, which is determined by the structure of the planet. Hence, the moon's tidal migration time scale is directly linked to the time scale on which the planet's structure changes. For gas giant planets undergoing slow contraction, we expect this time scale to be comparable to (or somewhat larger than) the age of the solar system. Hence, to order of magnitude, we expect moon migration time scales of billions to tens of billions of years. 

Given a model for the evolution of Saturn's internal structure, one can calculate the resonance locking-driven migration rates exactly. We defined the mode frequency evolution time scale $t_{\alpha} = \omega_\alpha/\dot{\omega}_\alpha$, and the planetary spin evolution timescale $t_{\rm p} = \Omega_{\rm p}/\dot{\Omega}_{\rm p}$. Equation \ref{resomega} becomes
\beq
\label{omdot}
\dot{\Omega}_{\rm m} = \frac{\Omega_{\rm p}}{t_{\rm p}} + \frac{\omega_\alpha}{m t_\alpha} \, .
\eeq
Since the moon's orbital frequency changes with semi-major axis as $\dot{a}_{\rm m}/a_{\rm m} = -(2/3) \dot{\Omega}_{\rm m}/\Omega_{\rm m}$, we have
\beq
\label{adot}
\frac{\dot{a}_{\rm m}}{a_{\rm m}} = -\frac{2}{3} \bigg[ \frac{\omega_\alpha}{m \Omega_{\rm m} t_\alpha} + \frac{\Omega_{\rm p}}{\Omega_{\rm m} t_{\rm p}}  \bigg] \, .
\eeq
Since $\omega_\alpha$ is negative, we see that outward migration requires that $\omega_{\alpha}$ be decreasing with time ($t_\alpha$ is positive) or that the planetary spin frequency be decreasing with time ($t_{\rm p}$ is negative).

For moons at large semi-major axis, resonant modes have $\omega_\alpha \simeq - m \Omega_{\rm p}$, in which case we obtain 
\beq
\label{adot2}
\frac{\dot{a}_{\rm m}}{a_{\rm m}} \approx \frac{2}{3} \frac{\Omega_{\rm p}}{\Omega_{\rm m}} \bigg[ \frac{1}{t_\alpha} - \frac{1}{t_{\rm p}}  \bigg] \, .
\eeq
Hence, if $t_\alpha$ is not a strong function of mode frequency, we expect outer moons (which have smaller $\Omega_{\rm m}$) to migrate faster than inner moons. This is obviously a very different prediction from viscoelastic dissipation or constant $Q$ models. Using the definition
\begin{equation}
t_{\rm tide}^{-1} = \frac{\dot{a}_{\rm m}}{a_{\rm m}} \, ,
\end{equation}
we can also write 
\beq
\label{ttide}
t_{\rm tide} \approx \frac{3}{2} \frac{\Omega_{\rm m}}{\Omega_{\rm p}} \bigg[ \frac{1}{t_\alpha} - \frac{1}{t_{\rm p}}  \bigg]^{-1} \, .
\eeq

If resonance locks are driven by planetary spin evolution such that the $t_{\rm p}$ term dominates, outward migration requires that $t_p < 0$, i.e., the planet's spin frequency must be decreasing with time. Hence, if giant planets are gradually spinning up as they contract, resonance locking cannot occur unless mode frequencies evolve such that $t_\alpha < t_{\rm p}$. 

Using the definition of the tidal quality factor $Q$, the moon migration rate is given by
\begin{equation}
    \label{qdef}
    \frac{\dot{a}_{\rm m}}{a_{\rm m}} = \frac{3 k_2}{Q} \frac{M_m}{M_{\rm p}} \left(\frac{R_{\rm p}}{a_{\rm m}}\right)^5 \Omega_{\rm m}
\end{equation}
Equating this with equation \ref{adot} yields the effective quality factor due to a resonance lock, $Q_{\rm RL}$
\beq
\label{q}
Q_{\rm RL} = \frac{9 k_2}{2} \frac{M_{\rm m}}{M_{\rm p}} \bigg(\frac{R}{a}\bigg)^{\!5} \bigg[ \frac{\omega_\alpha}{m \Omega_{\rm m}^2 t_\alpha} - \frac{\Omega_{\rm p}}{\Omega_{\rm m}^2 t_{\rm p}}  \bigg]^{-1} \, .
\eeq
The effective tidal quality factor therefore depends on many parameters of the system and is expected to be different for each moon.

\subsection{Flavors of Resonance Locking}

Depending on the type of oscillation mode that is resonantly locked with a moon, and depending on how a planet is evolving, migration driven by resonance locking could exhibit different behaviors. Let us examine equation \ref{ttide} above.

Let us start with the limit of constant planetary spin such that $t_p \rightarrow \infty$. In this case, $t_{\rm tide} \approx (3 \Omega_{\rm m}/2 \Omega_{\rm p}) t_\alpha$. If the mode evolution time scale $t_\alpha$ is the same for different moons locked in resonance, then this entails that $t_{\rm tide} \propto \Omega_{\rm m}$. Outer moons with smaller orbital frequencies would migrate \textit{faster} than inner moons, as mentioned above. This also entails that pairs of moons would cross mean-motion resonances divergently, if both moons are driven outwards by resonance locking. Obviously this cannot always be the case for the Saturn system, because Tethys and Dione are in mean-motion resonance with Mimas and Enceladus, respectively. It is possible that Mimas and Enceladus are in resonance locks with Saturn, while Tethys and Dione are not.

Another possibility is that outer moons (e.g., Rhea) were previously in a mean-motion resonance with one of the inner moons. If Rhea then encountered a resonance lock with Saturn, it could start migrating outwards faster than the inner moon, escaping the mean-motion resonance and migrating outwards as we see it today. 

However, it is important to remember that the value of $t_\alpha$ is not the same for each oscillation mode. If we consider pure gravity modes trapped in a stably stratified region, then we expect the g mode frequencies to scale as $\omega_\alpha \propto \int N dr/r$. For a contracting planet with increasing $N$, we would expect the value of $t_\alpha$ to be positive (allowing for resonance locking) and similar for each mode. However, if modes are partially trapped in different regions of the planet, their values of $t_\alpha$ could be different. For instance, in models of evolving stars \citep{fuller:17}, the value of $t_\alpha$ changes by a factor of several (and can even change sign) because some g modes are trapped near the core of the star, while others are trapped in the envelope. Detailed models of Saturn and its evolving structure are needed to quantify the expected values of $t_\alpha$.


Another flavor of resonance locking discussed in \cite{Lainey2020} is with predominantly inertial ``modes." It is well known (e.g., \citealt{ogilvie:04}) that inertial waves confined to a spherical shell can be reflected onto paths at critical latitudes where their wavenumber diverges, internal shear layers are created, and large amounts of dissipation are produced. This creates a stronger tidal torque at certain tidal forcing frequencies. While these frequencies are not normal oscillation mode frequencies, recent work \citep{lin:21} shows that this pattern is related to resonances with an underlying spectrum of pure global-scale inertial modes. In any case, an evolving planetary structure and/or spin rate would cause these ``mode" frequencies to evolve with time, possibly allowing for resonance locking. 

Unfortunately, the associated values of $t_\alpha$ (and even their sign) for inertial ``modes" are unclear, making it difficult to predict the associated trends in resonance locking migration rates. \cite{Lainey2020} posited that inertial ``mode" frequencies may evolve such that $t_\alpha \propto \Omega_{\rm m}^{-1}$. If we again let $t_{\rm p} \rightarrow \infty$, this would mean that $t_{\rm tide}$ would be nearly the same for each moon caught in a resonance lock, consistent with observations of Saturn's moons. However, this possibility needs to be investigated with detailed calculations of inertial wave dissipation in a sequence of evolutionary models, such that the realistic values of $t_\alpha$ can be estimated.

\subsection{Long-term Orbital Evolution}

The long-term orbital evolution resulting from resonance locking is very difficult to predict. It depends on the long-term evolution of Saturn's structure and oscillation modes, whereas even the current structure is poorly understood. Moreover, the relevant oscillation modes have frequencies (in Saturn's rotating frame) $|\omega_\alpha| < 2 \Omega_{\rm p}$ and hence lie in the gravito-inertial regime. Even computing the relevant oscillation modes (frequencies, eigenfunctions, and damping rates) for a single planetary model is challenging due to the non-separability of the oscillation equations and the presence of critical latitudes where inertial waves develop extremely short wavelengths. Quantitative predictions await improved planetary structure models, evolution models, and mode calculations. 

Nonetheless, we can predict some general properties of tidal migration driven by resonance locking, and how it differs from constant $Q$ or viscoelastic models. As mentioned above, the resonance locking time scale is proportional to the planetary evolution time scale, regardless of the planet's mass and orbital period. Hence, whereas constant $Q$ models predict that higher mass moons migrate otward faster, the resonance locking migration timescale is independent of mass. Most importantly, whereas constant $Q$ models predict a very strong dependence on semi-major axis (with migration time scale proportional to $a_{\rm m}^{13/2}$, equation \ref{qdef}), the resonance lock migration time scale is only weakly dependent on semi-major axis. Hence, resonance locking typically predicts much faster migration for outer moons (and correspondingly low effective $Q$ values) relative to constant $Q$ models.

\begin{figure}
\begin{center}
\includegraphics[scale=0.5]{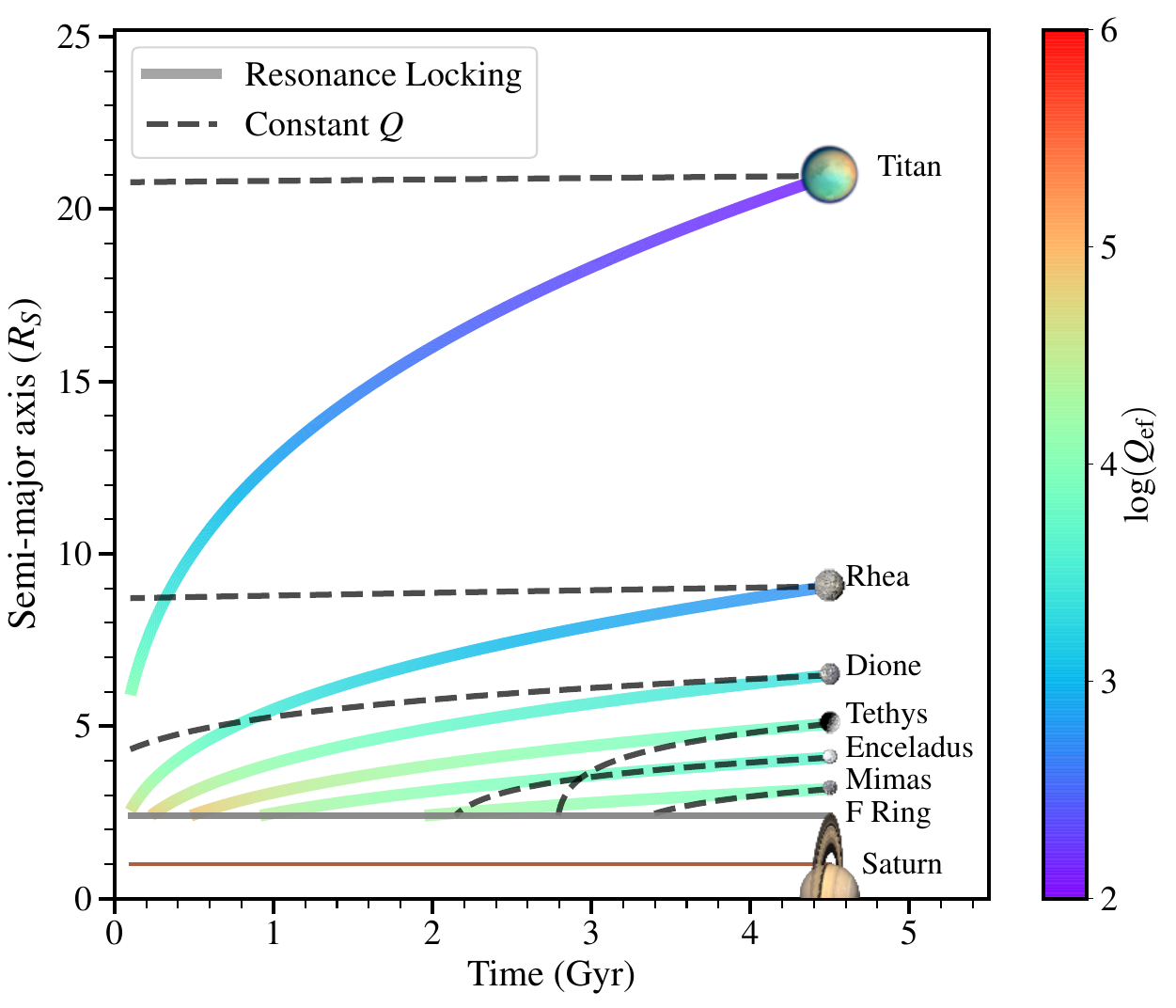}
\end{center} 
\caption{ \label{fig:OrbEvol} An example of the differing orbital evolution between resonance locking (colored lines) and constant $Q$ models (dashed lines). For the latter, $Q$ values for each moon of Saturn are taken from \citealt{Lainey2020} and assumed to be constant in time. For resonance locking, we use $t_{\rm tide}=3 t$, which matches measured moon migration rates fairly well, and the color of the lines corresponds to the effective quality factor (equation \ref{q}). Neither evolution accounts for mean-motion resonances, which will significantly affect the orbital evolution of the inner moons. [From \citealt{Lainey2020}]}
\end{figure}

\autoref{fig:OrbEvol} shows an example of the qualitative differences between migration driven by resonance locking vs. migration driven by constant $Q$ models. The biggest difference is that with resonance locking, outer moons migrate relatively faster, and inner moons migrate slower. In Saturn, the consequence is that Rhea and Titan may have migrated by a large fraction of their current semi-major axes, whereas constant $Q$ models predict almost no migration for these moons. Additionally, Mimas, Enceladus, and Tethys migrate slower and hence may be much older than inferred from constant $Q$ models. 

However, detailed predictions are plagued by the uncertainties mentioned above. While present day migration rates and hence $t_\alpha$ values can be inferred, it is not clear how those values have evolved over time. \autoref{fig:OrbEvol} assumes that $t_{\rm tide} \sim t$, where $t$ is the age of Saturn, but the constant of proportionality has not been predicted from theoretical models. Current moon migration rates require $t_{\rm tide} \sim 3 t$ \citep{Lainey2020} In reality, the form of $t_{\rm tide}(t)$ and $t_\alpha(t)$ is determined by the very uncertain evolution of Saturn's structure and oscillation mode frequencies. Even the sign of $t_\alpha$ has not been calculated with realistic models. Extrapolation from the present day to small values of $t$ becomes increasingly uncertain. 

Another major uncertainty is whether moons pass through mean-motion resonances, and what happens when they do so. If two moons are each caught in a resonance lock, their relative migration rates are determined by the value of $t_\alpha$ for each resonantly locked mode, which could be quite different from one another. Hence it is certainly possible for either the inner or outer moon to migrate faster, i.e., the migration may be either convergent or divergent. In the latter case (the outer moon migrates faster), the system will simply pass through the mean-motion resonance, and each moon will presumably remain resonantly locked with its respective planetary oscillation. 

In the case of convergent migration, the moons may get locked into mean motion resonance, but the details of this process are highly uncertain. At present, nobody has ever computed an orbital integration for this case, including the changing tidal torque, which is an extremely sensitive function of orbital period near resonance (\autoref{fig:ResLock}). Hence, the mean-motion resonance could potentially disrupt the resonance lock by pushing the moon out of resonance. 

Alternatively, the system may be able to adjust so that the inner moon stably locks into mean motion resonance with the outer moon, driving it outwards. In this case, the inner moon would need to be pushed deeper into resonance with the planetary oscillation, so that the tidal torque on it can be stronger to account for the extra energy and angular momentum needed to drive out the outer moon. If this can occur, the entire system would be driven outwards at the same resonance locking time scale that the inner moon had before the mean-motion resonance was established.

It is very unlikely that both an inner and outer moon could simultaneously be in resonance locks with planetary oscillations, and in mean-motion resonance with each other. The mean-motion resonances occur near integer ratios of the orbital frequencies, but there is no reason to suspect planetary oscillation mode frequencies to obey such a relationship. Even if they did at one point in time, it is unlikely that each oscillation mode frequency would evolve together to sustain this integer ratio for an extended period of time. 

Consequently, if the migration of Saturn's inner moons are driven by resonance locking, it is most likely that the inner moons of mean-motion pairs (i.e., Mimas and Enceladus) are caught in resonance locks while the outer moons (Tethys and Dione) are not. Hence, we would expect to measure a strong tidal torque (small planet $Q$) on Mimas and Enceladus, and a weak tidal torque (large planet $Q$) on Tethys and Dione. The measured values \citep{Lainey2020} may indeed be larger for those outer moons, but current measurement uncertainties are very large so it is difficult to claim anything with confidence. Continued monitoring of moon migration rates will help distinguish between different tidal dissipation mechanisms and hence differentiate between possible evolutionary histories.

\section{Conclusions}

Our understanding of tidal dissipation in giant planets and its consequences for moon migration remain incomplete, but they have improved greatly over the last decade. A major advance has been in our picture of giant planet structure. Both Jupiter and Saturn show evidence for a large and ``diffuse" core, likely containing a smooth and gradual compositional change from ice/rock in the core to the gaseous envelope. In Jupiter, this diffuse core helps explain gravity data from \textit{Juno}, while in Saturn it is revealed in part by ring seismology data from \textit{Cassini}. A diffuse core is also predicted by modern planetary formation models.

Stably stratified regions between the core and envelope have major implications for tidal dissipation in giant planets. They allow for the excitation of gravity waves in these regions, which can contribute greatly to tidal dissipation. Additionally, thick stably stratified shells increase the excitation rate of inertial waves in the overlying convective envelope, enhancing their contribution to tidal dissipation. Recent models accounting for these effects may explain the rapid moon migration rates in Saturn that have been measured using long-baseline astrometry.

Finally, the continually evolving structures of giant planets may feedback onto tidal migration through a resonance locking process. As a planet's structure evolves (e.g., due to cooling, helium rain, compositional settling, or core erosion), the frequencies of gravito-inertial modes evolve as well. This may allow moons to lock into resonance with these modes, such that the energy dissipation rate is enhanced over long periods of time. This process could be necessary to explain the rapid migration of Saturn's outer moons Rhea and Titan.

Although recent progress is promising, there is plenty of room for improvement in future work. More sophisticated planetary formation and evolution models matched to thermal/gravity/seismology data can be developed to provide more realistic planetary internal structures. These will allow for more reliable calculations of tidal energy dissipation rates, including the feedback between the evolving planetary structure and expanding orbits of the moons. Such work will also shed light on the physics of ice giant systems, exoplanet systems, tidal heating of moons, and even the possibility of life in the oceans of Enceladus and Europa.

\section*{Availability of data and materials}

No new data has been used in this work. Data and analysis scripts can be found in referenced materials. 

\section*{Competing interests}

Not applicable.

\section*{Funding}

We gratefully acknowledge support from the International Space Science Institute. S. M. acknowledges support from PNP (CNRS/INSU) and PLATO CNES grant at CEA/DAp.  

\section*{Authors' contributions}

Authors Fuller, Guillot, and Mathis contributed equally to the writing of this manuscript. Author Murray contributed expertise on moon orbital evolution. 

\section*{Acknowledgments}

This work was inspired by a workshop hosted by the International Space Science Institute in Bern, Switzerland.

\bibliography{SatBib,Bib_AstrofluidMathis}


\begin{thebibliography}{192}
\ifx \bisbn   \undefined \def \bisbn  #1{ISBN #1}\fi
\ifx \binits  \undefined \def \binits#1{#1}\fi
\ifx \bauthor  \undefined \def \bauthor#1{#1}\fi
\ifx \batitle  \undefined \def \batitle#1{#1}\fi
\ifx \bjtitle  \undefined \def \bjtitle#1{#1}\fi
\ifx \bvolume  \undefined \def \bvolume#1{\textbf{#1}}\fi
\ifx \byear  \undefined \def \byear#1{#1}\fi
\ifx \bissue  \undefined \def \bissue#1{#1}\fi
\ifx \bfpage  \undefined \def \bfpage#1{#1}\fi
\ifx \blpage  \undefined \def \blpage #1{#1}\fi
\ifx \burl  \undefined \def \burl#1{\textsf{#1}}\fi
\ifx \doiurl  \undefined \def \doiurl#1{\url{https://doi.org/#1}}\fi
\ifx \betal  \undefined \def \betal{\textit{et al.}}\fi
\ifx \binstitute  \undefined \def \binstitute#1{#1}\fi
\ifx \binstitutionaled  \undefined \def \binstitutionaled#1{#1}\fi
\ifx \bctitle  \undefined \def \bctitle#1{#1}\fi
\ifx \beditor  \undefined \def \beditor#1{#1}\fi
\ifx \bpublisher  \undefined \def \bpublisher#1{#1}\fi
\ifx \bbtitle  \undefined \def \bbtitle#1{#1}\fi
\ifx \bedition  \undefined \def \bedition#1{#1}\fi
\ifx \bseriesno  \undefined \def \bseriesno#1{#1}\fi
\ifx \blocation  \undefined \def \blocation#1{#1}\fi
\ifx \bsertitle  \undefined \def \bsertitle#1{#1}\fi
\ifx \bsnm \undefined \def \bsnm#1{#1}\fi
\ifx \bsuffix \undefined \def \bsuffix#1{#1}\fi
\ifx \bparticle \undefined \def \bparticle#1{#1}\fi
\ifx \barticle \undefined \def \barticle#1{#1}\fi
\bibcommenthead
\ifx \bconfdate \undefined \def \bconfdate #1{#1}\fi
\ifx \botherref \undefined \def \botherref #1{#1}\fi
\ifx \url \undefined \def \url#1{\textsf{#1}}\fi
\ifx \bchapter \undefined \def \bchapter#1{#1}\fi
\ifx \bbook \undefined \def \bbook#1{#1}\fi
\ifx \bcomment \undefined \def \bcomment#1{#1}\fi
\ifx \oauthor \undefined \def \oauthor#1{#1}\fi
\ifx \citeauthoryear \undefined \def \citeauthoryear#1{#1}\fi
\ifx \endbibitem  \undefined \def \endbibitem {}\fi
\ifx \bconflocation  \undefined \def \bconflocation#1{#1}\fi
\ifx \arxivurl  \undefined \def \arxivurl#1{\textsf{#1}}\fi
\csname PreBibitemsHook\endcsname

\bibitem[\protect\citeauthoryear{{Goldreich} and
  {Soter}}{1966}]{GoldreichSoter1966}
\begin{barticle}
\bauthor{\bsnm{{Goldreich}}, \binits{P.}},
\bauthor{\bsnm{{Soter}}, \binits{S.}}:
\batitle{{Q in the Solar System}}.
\bjtitle{Icarus}
\bvolume{5},
\bfpage{375}--\blpage{389}
(\byear{1966})
\doiurl{10.1016/0019-1035(66)90051-0}
\end{barticle}
\endbibitem

\bibitem[\protect\citeauthoryear{{Lainey} et~al.}{2009}]{Laineyetal2009}
\begin{barticle}
\bauthor{\bsnm{{Lainey}}, \binits{V.}},
\bauthor{\bsnm{{Arlot}}, \binits{J.-E.}},
\bauthor{\bsnm{{Karatekin}}, \binits{{\"O}.}},
\bauthor{\bsnm{{van Hoolst}}, \binits{T.}}:
\batitle{{Strong tidal dissipation in Io and Jupiter from astrometric
  observations}}.
\bjtitle{\nat}
\bvolume{459},
\bfpage{957}--\blpage{959}
(\byear{2009})
\doiurl{10.1038/nature08108}
\end{barticle}
\endbibitem

\bibitem[\protect\citeauthoryear{{Lainey} et~al.}{2012}]{Laineyetal2012}
\begin{barticle}
\bauthor{\bsnm{{Lainey}}, \binits{V.}},
\bauthor{\bsnm{{Karatekin}}, \binits{{\"O}.}},
\bauthor{\bsnm{{Desmars}}, \binits{J.}},
\bauthor{\bsnm{{Charnoz}}, \binits{S.}},
\bauthor{\bsnm{{Arlot}}, \binits{J.-E.}},
\bauthor{\bsnm{{Emelyanov}}, \binits{N.}},
\bauthor{\bsnm{{Le Poncin-Lafitte}}, \binits{C.}},
\bauthor{\bsnm{{Mathis}}, \binits{S.}},
\bauthor{\bsnm{{Remus}}, \binits{F.}},
\bauthor{\bsnm{{Tobie}}, \binits{G.}},
\bauthor{\bsnm{{Zahn}}, \binits{J.-P.}}:
\batitle{{Strong Tidal Dissipation in Saturn and Constraints on Enceladus'
  Thermal State from Astrometry}}.
\bjtitle{\apj}
\bvolume{752},
\bfpage{14}
(\byear{2012})
\doiurl{10.1088/0004-637X/752/1/14}
{\href{https://arxiv.org/abs/1204.0895}{{arXiv:1204.0895}}}
{[astro-ph.EP]}
\end{barticle}
\endbibitem

\bibitem[\protect\citeauthoryear{{Lainey} et~al.}{2017}]{Laineyetal2017}
\begin{barticle}
\bauthor{\bsnm{{Lainey}}, \binits{V.}},
\bauthor{\bsnm{{Jacobson}}, \binits{R.A.}},
\bauthor{\bsnm{{Tajeddine}}, \binits{R.}},
\bauthor{\bsnm{{Cooper}}, \binits{N.J.}},
\bauthor{\bsnm{{Murray}}, \binits{C.}},
\bauthor{\bsnm{{Robert}}, \binits{V.}},
\bauthor{\bsnm{{Tobie}}, \binits{G.}},
\bauthor{\bsnm{{Guillot}}, \binits{T.}},
\bauthor{\bsnm{{Mathis}}, \binits{S.}},
\bauthor{\bsnm{{Remus}}, \binits{F.}},
\bauthor{\bsnm{{Desmars}}, \binits{J.}},
\bauthor{\bsnm{{Arlot}}, \binits{J.-E.}},
\bauthor{\bsnm{{De Cuyper}}, \binits{J.-P.}},
\bauthor{\bsnm{{Dehant}}, \binits{V.}},
\bauthor{\bsnm{{Pascu}}, \binits{D.}},
\bauthor{\bsnm{{Thuillot}}, \binits{W.}},
\bauthor{\bsnm{{Le Poncin-Lafitte}}, \binits{C.}},
\bauthor{\bsnm{{Zahn}}, \binits{J.-P.}}:
\batitle{{New constraints on Saturn's interior from Cassini astrometric data}}.
\bjtitle{Icarus}
\bvolume{281},
\bfpage{286}--\blpage{296}
(\byear{2017})
\doiurl{10.1016/j.icarus.2016.07.014}
{\href{https://arxiv.org/abs/1510.05870}{{arXiv:1510.05870}}}
{[astro-ph.EP]}
\end{barticle}
\endbibitem

\bibitem[\protect\citeauthoryear{{Polycarpe} et~al.}{2018}]{Polycarpeetal2018}
\begin{barticle}
\bauthor{\bsnm{{Polycarpe}}, \binits{W.}},
\bauthor{\bsnm{{Saillenfest}}, \binits{M.}},
\bauthor{\bsnm{{Lainey}}, \binits{V.}},
\bauthor{\bsnm{{Vienne}}, \binits{A.}},
\bauthor{\bsnm{{Noyelles}}, \binits{B.}},
\bauthor{\bsnm{{Rambaux}}, \binits{N.}}:
\batitle{{Strong tidal energy dissipation in Saturn at Titan's frequency as an
  explanation for Iapetus orbit}}.
\bjtitle{\aap}
\bvolume{619},
\bfpage{133}
(\byear{2018})
\doiurl{10.1051/0004-6361/201833930}
{\href{https://arxiv.org/abs/1809.11065}{{arXiv:1809.11065}}}
{[astro-ph.EP]}
\end{barticle}
\endbibitem

\bibitem[\protect\citeauthoryear{{Sinclair}}{1983}]{Sinclair1983}
\begin{bchapter}
\bauthor{\bsnm{{Sinclair}}, \binits{A.T.}}:
\bctitle{{A re-consideration of the evolution hypothesis of the origin of the
  resonances among Saturn's satellites}}.
In: \beditor{\bsnm{{Markellos}}, \binits{V.V.}},
\beditor{\bsnm{{Kozai}}, \binits{Y.}} (eds.)
\bbtitle{IAU Colloq. 74: Dynamical Trapping and Evolution in the Solar System}.
\bsertitle{Astrophysics and Space Science Library},
vol. \bseriesno{106},
pp. \bfpage{19}--\blpage{25}
(\byear{1983}).
\doiurl{10.1007/978-94-009-7214-8_2}
\end{bchapter}
\endbibitem

\bibitem[\protect\citeauthoryear{{Charnoz} et~al.}{2011}]{Charnozetal2011}
\begin{barticle}
\bauthor{\bsnm{{Charnoz}}, \binits{S.}},
\bauthor{\bsnm{{Crida}}, \binits{A.}},
\bauthor{\bsnm{{Castillo-Rogez}}, \binits{J.C.}},
\bauthor{\bsnm{{Lainey}}, \binits{V.}},
\bauthor{\bsnm{{Dones}}, \binits{L.}},
\bauthor{\bsnm{{Karatekin}}, \binits{{\"O}.}},
\bauthor{\bsnm{{Tobie}}, \binits{G.}},
\bauthor{\bsnm{{Mathis}}, \binits{S.}},
\bauthor{\bsnm{{Le Poncin-Lafitte}}, \binits{C.}},
\bauthor{\bsnm{{Salmon}}, \binits{J.}}:
\batitle{{Accretion of Saturn's mid-sized moons during the viscous spreading of
  young massive rings: Solving the paradox of silicate-poor rings versus
  silicate-rich moons}}.
\bjtitle{Icarus}
\bvolume{216},
\bfpage{535}--\blpage{550}
(\byear{2011})
\doiurl{10.1016/j.icarus.2011.09.017}
{\href{https://arxiv.org/abs/1109.3360}{{arXiv:1109.3360}}}
{[astro-ph.EP]}
\end{barticle}
\endbibitem

\bibitem[\protect\citeauthoryear{{Mayor} and {Queloz}}{1995}]{MayorQueloz1995}
\begin{barticle}
\bauthor{\bsnm{{Mayor}}, \binits{M.}},
\bauthor{\bsnm{{Queloz}}, \binits{D.}}:
\batitle{{A Jupiter-mass companion to a solar-type star}}.
\bjtitle{\nat}
\bvolume{378},
\bfpage{355}--\blpage{359}
(\byear{1995})
\doiurl{10.1038/378355a0}
\end{barticle}
\endbibitem

\bibitem[\protect\citeauthoryear{{Ahuir} et~al.}{2021}]{Ahuiretal2021}
\begin{barticle}
\bauthor{\bsnm{{Ahuir}}, \binits{J.}},
\bauthor{\bsnm{{Strugarek}}, \binits{A.}},
\bauthor{\bsnm{{Brun}}, \binits{A.-S.}},
\bauthor{\bsnm{{Mathis}}, \binits{S.}}:
\batitle{{Magnetic and tidal migration of close-in planets. Influence of
  secular evolution on their population}}.
\bjtitle{\aap}
\bvolume{650},
\bfpage{126}
(\byear{2021})
\doiurl{10.1051/0004-6361/202040173}
{\href{https://arxiv.org/abs/2104.01004}{{arXiv:2104.01004}}}
{[astro-ph.EP]}
\end{barticle}
\endbibitem

\bibitem[\protect\citeauthoryear{{Bodenheimer} et~al.}{2001}]{Bodenheimer+2001}
\begin{barticle}
\bauthor{\bsnm{{Bodenheimer}}, \binits{P.}},
\bauthor{\bsnm{{Lin}}, \binits{D.N.C.}},
\bauthor{\bsnm{{Mardling}}, \binits{R.A.}}:
\batitle{{On the Tidal Inflation of Short-Period Extrasolar Planets}}.
\bjtitle{\apj}
\bvolume{548}(\bissue{1}),
\bfpage{466}--\blpage{472}
(\byear{2001})
\doiurl{10.1086/318667}
\end{barticle}
\endbibitem

\bibitem[\protect\citeauthoryear{{Guillot} and
  {Showman}}{2002}]{Guillot+Showman2002}
\begin{barticle}
\bauthor{\bsnm{{Guillot}}, \binits{T.}},
\bauthor{\bsnm{{Showman}}, \binits{A.P.}}:
\batitle{{Evolution of ``51 Pegasus b-like'' planets}}.
\bjtitle{\aap}
\bvolume{385},
\bfpage{156}--\blpage{165}
(\byear{2002})
\doiurl{10.1051/0004-6361:20011624}
{\href{https://arxiv.org/abs/astro-ph/0202234}{{arXiv:astro-ph/0202234}}}
{[astro-ph]}
\end{barticle}
\endbibitem

\bibitem[\protect\citeauthoryear{{Baraffe}}{2005}]{Baraffe2005}
\begin{barticle}
\bauthor{\bsnm{{Baraffe}}, \binits{I.}}:
\batitle{{Structure and Evolution of Giant Planets}}.
\bjtitle{\ssr}
\bvolume{116},
\bfpage{67}--\blpage{76}
(\byear{2005})
\doiurl{10.1007/s11214-005-1948-0}
\end{barticle}
\endbibitem

\bibitem[\protect\citeauthoryear{{Benz} et~al.}{2017}]{Benzetal2017}
\begin{bbook}
\bauthor{\bsnm{{Benz}}, \binits{W.}},
\bauthor{\bsnm{{Ehrenreich}}, \binits{D.}},
\bauthor{\bsnm{{Isaak}}, \binits{K.}}:
\bbtitle{{CHEOPS: CHaracterizing ExOPlanets Satellite}},
p. \bfpage{84}
(\byear{2017}).
\doiurl{10.1007/978-3-319-30648-3_84-1}
\end{bbook}
\endbibitem

\bibitem[\protect\citeauthoryear{{Ricker} et~al.}{2015}]{Rickeretal2015}
\begin{barticle}
\bauthor{\bsnm{{Ricker}}, \binits{G.R.}},
\bauthor{\bsnm{{Winn}}, \binits{J.N.}},
\bauthor{\bsnm{{Vanderspek}}, \binits{R.}},
\bauthor{\bsnm{{Latham}}, \binits{D.W.}},
\bauthor{\bsnm{{Bakos}}, \binits{G.{\'A}.}},
\bauthor{\bsnm{{Bean}}, \binits{J.L.}},
\bauthor{\bsnm{{Berta-Thompson}}, \binits{Z.K.}},
\bauthor{\bsnm{{Brown}}, \binits{T.M.}},
\bauthor{\bsnm{{Buchhave}}, \binits{L.}},
\bauthor{\bsnm{{Butler}}, \binits{N.R.}},
\bauthor{\bsnm{{Butler}}, \binits{R.P.}},
\bauthor{\bsnm{{Chaplin}}, \binits{W.J.}},
\bauthor{\bsnm{{Charbonneau}}, \binits{D.}},
\bauthor{\bsnm{{Christensen-Dalsgaard}}, \binits{J.}},
\bauthor{\bsnm{{Clampin}}, \binits{M.}},
\bauthor{\bsnm{{Deming}}, \binits{D.}},
\bauthor{\bsnm{{Doty}}, \binits{J.}},
\bauthor{\bsnm{{De Lee}}, \binits{N.}},
\bauthor{\bsnm{{Dressing}}, \binits{C.}},
\bauthor{\bsnm{{Dunham}}, \binits{E.W.}},
\bauthor{\bsnm{{Endl}}, \binits{M.}},
\bauthor{\bsnm{{Fressin}}, \binits{F.}},
\bauthor{\bsnm{{Ge}}, \binits{J.}},
\bauthor{\bsnm{{Henning}}, \binits{T.}},
\bauthor{\bsnm{{Holman}}, \binits{M.J.}},
\bauthor{\bsnm{{Howard}}, \binits{A.W.}},
\bauthor{\bsnm{{Ida}}, \binits{S.}},
\bauthor{\bsnm{{Jenkins}}, \binits{J.M.}},
\bauthor{\bsnm{{Jernigan}}, \binits{G.}},
\bauthor{\bsnm{{Johnson}}, \binits{J.A.}},
\bauthor{\bsnm{{Kaltenegger}}, \binits{L.}},
\bauthor{\bsnm{{Kawai}}, \binits{N.}},
\bauthor{\bsnm{{Kjeldsen}}, \binits{H.}},
\bauthor{\bsnm{{Laughlin}}, \binits{G.}},
\bauthor{\bsnm{{Levine}}, \binits{A.M.}},
\bauthor{\bsnm{{Lin}}, \binits{D.}},
\bauthor{\bsnm{{Lissauer}}, \binits{J.J.}},
\bauthor{\bsnm{{MacQueen}}, \binits{P.}},
\bauthor{\bsnm{{Marcy}}, \binits{G.}},
\bauthor{\bsnm{{McCullough}}, \binits{P.R.}},
\bauthor{\bsnm{{Morton}}, \binits{T.D.}},
\bauthor{\bsnm{{Narita}}, \binits{N.}},
\bauthor{\bsnm{{Paegert}}, \binits{M.}},
\bauthor{\bsnm{{Palle}}, \binits{E.}},
\bauthor{\bsnm{{Pepe}}, \binits{F.}},
\bauthor{\bsnm{{Pepper}}, \binits{J.}},
\bauthor{\bsnm{{Quirrenbach}}, \binits{A.}},
\bauthor{\bsnm{{Rinehart}}, \binits{S.A.}},
\bauthor{\bsnm{{Sasselov}}, \binits{D.}},
\bauthor{\bsnm{{Sato}}, \binits{B.}},
\bauthor{\bsnm{{Seager}}, \binits{S.}},
\bauthor{\bsnm{{Sozzetti}}, \binits{A.}},
\bauthor{\bsnm{{Stassun}}, \binits{K.G.}},
\bauthor{\bsnm{{Sullivan}}, \binits{P.}},
\bauthor{\bsnm{{Szentgyorgyi}}, \binits{A.}},
\bauthor{\bsnm{{Torres}}, \binits{G.}},
\bauthor{\bsnm{{Udry}}, \binits{S.}},
\bauthor{\bsnm{{Villasenor}}, \binits{J.}}:
\batitle{{Transiting Exoplanet Survey Satellite (TESS)}}.
\bjtitle{Journal of Astronomical Telescopes, Instruments, and Systems}
\bvolume{1}(\bissue{1}),
\bfpage{014003}
(\byear{2015})
\doiurl{10.1117/1.JATIS.1.1.014003}
\end{barticle}
\endbibitem

\bibitem[\protect\citeauthoryear{{Lagage}}{2015}]{Lagage2015}
\begin{barticle}
\bauthor{\bsnm{{Lagage}}, \binits{P.-O.}}:
\batitle{{Exoplanets characterisation with the JWST and particularly MIRI}}.
\bjtitle{European Planetary Science Congress 2015}
\bvolume{10},
\bfpage{2015}--\blpage{757}
(\byear{2015})
\end{barticle}
\endbibitem

\bibitem[\protect\citeauthoryear{{Rauer} et~al.}{2014}]{Raueretal2014}
\begin{barticle}
\bauthor{\bsnm{{Rauer}}, \binits{H.}},
\bauthor{\bsnm{{Catala}}, \binits{C.}},
\bauthor{\bsnm{{Aerts}}, \binits{C.}},
\bauthor{\bsnm{{Appourchaux}}, \binits{T.}},
\bauthor{\bsnm{{Benz}}, \binits{W.}},
\bauthor{\bsnm{{Brandeker}}, \binits{A.}},
\bauthor{\bsnm{{Christensen-Dalsgaard}}, \binits{J.}},
\bauthor{\bsnm{{Deleuil}}, \binits{M.}},
\bauthor{\bsnm{{Gizon}}, \binits{L.}},
\bauthor{\bsnm{{Goupil}}, \binits{M.-J.}},
\bauthor{\bsnm{{G{\"u}del}}, \binits{M.}},
\bauthor{\bsnm{{Janot-Pacheco}}, \binits{E.}},
\bauthor{\bsnm{{Mas-Hesse}}, \binits{M.}},
\bauthor{\bsnm{{Pagano}}, \binits{I.}},
\bauthor{\bsnm{{Piotto}}, \binits{G.}},
\bauthor{\bsnm{{Pollacco}}, \binits{D.}},
\bauthor{\bsnm{{Santos}}, \binits{{\. C}.}},
\bauthor{\bsnm{{Smith}}, \binits{A.}},
\bauthor{\bsnm{{Su{\'a}rez}}, \binits{J.-C.}},
\bauthor{\bsnm{{Szab{\'o}}}, \binits{R.}},
\bauthor{\bsnm{{Udry}}, \binits{S.}},
\bauthor{\bsnm{{Adibekyan}}, \binits{V.}},
\bauthor{\bsnm{{Alibert}}, \binits{Y.}},
\bauthor{\bsnm{{Almenara}}, \binits{J.-M.}},
\bauthor{\bsnm{{Amaro-Seoane}}, \binits{P.}},
\bauthor{\bsnm{{Eiff}}, \binits{M.A.-v.}},
\bauthor{\bsnm{{Asplund}}, \binits{M.}},
\bauthor{\bsnm{{Antonello}}, \binits{E.}},
\bauthor{\bsnm{{Barnes}}, \binits{S.}},
\bauthor{\bsnm{{Baudin}}, \binits{F.}},
\bauthor{\bsnm{{Belkacem}}, \binits{K.}},
\bauthor{\bsnm{{Bergemann}}, \binits{M.}},
\bauthor{\bsnm{{Bihain}}, \binits{G.}},
\bauthor{\bsnm{{Birch}}, \binits{A.C.}},
\bauthor{\bsnm{{Bonfils}}, \binits{X.}},
\bauthor{\bsnm{{Boisse}}, \binits{I.}},
\bauthor{\bsnm{{Bonomo}}, \binits{A.S.}},
\bauthor{\bsnm{{Borsa}}, \binits{F.}},
\bauthor{\bsnm{{Brand{\~a}o}}, \binits{I.M.}},
\bauthor{\bsnm{{Brocato}}, \binits{E.}},
\bauthor{\bsnm{{Brun}}, \binits{S.}},
\bauthor{\bsnm{{Burleigh}}, \binits{M.}},
\bauthor{\bsnm{{Burston}}, \binits{R.}},
\bauthor{\bsnm{{Cabrera}}, \binits{J.}},
\bauthor{\bsnm{{Cassisi}}, \binits{S.}},
\bauthor{\bsnm{{Chaplin}}, \binits{W.}},
\bauthor{\bsnm{{Charpinet}}, \binits{S.}},
\bauthor{\bsnm{{Chiappini}}, \binits{C.}},
\bauthor{\bsnm{{Church}}, \binits{R.P.}},
\bauthor{\bsnm{{Csizmadia}}, \binits{S.}},
\bauthor{\bsnm{{Cunha}}, \binits{M.}},
\bauthor{\bsnm{{Damasso}}, \binits{M.}},
\bauthor{\bsnm{{Davies}}, \binits{M.B.}},
\bauthor{\bsnm{{Deeg}}, \binits{H.J.}},
\bauthor{\bsnm{{D{\'{\i}}az}}, \binits{R.F.}},
\bauthor{\bsnm{{Dreizler}}, \binits{S.}},
\bauthor{\bsnm{{Dreyer}}, \binits{C.}},
\bauthor{\bsnm{{Eggenberger}}, \binits{P.}},
\bauthor{\bsnm{{Ehrenreich}}, \binits{D.}},
\bauthor{\bsnm{{Eigm{\"u}ller}}, \binits{P.}},
\bauthor{\bsnm{{Erikson}}, \binits{A.}},
\bauthor{\bsnm{{Farmer}}, \binits{R.}},
\bauthor{\bsnm{{Feltzing}}, \binits{S.}},
\bauthor{\bsnm{{de Oliveira Fialho}}, \binits{F.}},
\bauthor{\bsnm{{Figueira}}, \binits{P.}},
\bauthor{\bsnm{{Forveille}}, \binits{T.}},
\bauthor{\bsnm{{Fridlund}}, \binits{M.}},
\bauthor{\bsnm{{Garc{\'{\i}}a}}, \binits{R.A.}},
\bauthor{\bsnm{{Giommi}}, \binits{P.}},
\bauthor{\bsnm{{Giuffrida}}, \binits{G.}},
\bauthor{\bsnm{{Godolt}}, \binits{M.}},
\bauthor{\bsnm{{Gomes da Silva}}, \binits{J.}},
\bauthor{\bsnm{{Granzer}}, \binits{T.}},
\bauthor{\bsnm{{Grenfell}}, \binits{J.L.}},
\bauthor{\bsnm{{Grotsch-Noels}}, \binits{A.}},
\bauthor{\bsnm{{G{\"u}nther}}, \binits{E.}},
\bauthor{\bsnm{{Haswell}}, \binits{C.A.}},
\bauthor{\bsnm{{Hatzes}}, \binits{A.P.}},
\bauthor{\bsnm{{H{\'e}brard}}, \binits{G.}},
\bauthor{\bsnm{{Hekker}}, \binits{S.}},
\bauthor{\bsnm{{Helled}}, \binits{R.}},
\bauthor{\bsnm{{Heng}}, \binits{K.}},
\bauthor{\bsnm{{Jenkins}}, \binits{J.M.}},
\bauthor{\bsnm{{Johansen}}, \binits{A.}},
\bauthor{\bsnm{{Khodachenko}}, \binits{M.L.}},
\bauthor{\bsnm{{Kislyakova}}, \binits{K.G.}},
\bauthor{\bsnm{{Kley}}, \binits{W.}},
\bauthor{\bsnm{{Kolb}}, \binits{U.}},
\bauthor{\bsnm{{Krivova}}, \binits{N.}},
\bauthor{\bsnm{{Kupka}}, \binits{F.}},
\bauthor{\bsnm{{Lammer}}, \binits{H.}},
\bauthor{\bsnm{{Lanza}}, \binits{A.F.}},
\bauthor{\bsnm{{Lebreton}}, \binits{Y.}},
\bauthor{\bsnm{{Magrin}}, \binits{D.}},
\bauthor{\bsnm{{Marcos-Arenal}}, \binits{P.}},
\bauthor{\bsnm{{Marrese}}, \binits{P.M.}},
\bauthor{\bsnm{{Marques}}, \binits{J.P.}},
\bauthor{\bsnm{{Martins}}, \binits{J.}},
\bauthor{\bsnm{{Mathis}}, \binits{S.}},
\bauthor{\bsnm{{Mathur}}, \binits{S.}},
\bauthor{\bsnm{{Messina}}, \binits{S.}},
\bauthor{\bsnm{{Miglio}}, \binits{A.}},
\bauthor{\bsnm{{Montalban}}, \binits{J.}},
\bauthor{\bsnm{{Montalto}}, \binits{M.}},
\bauthor{\bsnm{{Monteiro}}, \binits{M.J.P.F.G.}},
\bauthor{\bsnm{{Moradi}}, \binits{H.}},
\bauthor{\bsnm{{Moravveji}}, \binits{E.}},
\bauthor{\bsnm{{Mordasini}}, \binits{C.}},
\bauthor{\bsnm{{Morel}}, \binits{T.}},
\bauthor{\bsnm{{Mortier}}, \binits{A.}},
\bauthor{\bsnm{{Nascimbeni}}, \binits{V.}},
\bauthor{\bsnm{{Nelson}}, \binits{R.P.}},
\bauthor{\bsnm{{Nielsen}}, \binits{M.B.}},
\bauthor{\bsnm{{Noack}}, \binits{L.}},
\bauthor{\bsnm{{Norton}}, \binits{A.J.}},
\bauthor{\bsnm{{Ofir}}, \binits{A.}},
\bauthor{\bsnm{{Oshagh}}, \binits{M.}},
\bauthor{\bsnm{{Ouazzani}}, \binits{R.-M.}},
\bauthor{\bsnm{{P{\'a}pics}}, \binits{P.}},
\bauthor{\bsnm{{Parro}}, \binits{V.C.}},
\bauthor{\bsnm{{Petit}}, \binits{P.}},
\bauthor{\bsnm{{Plez}}, \binits{B.}},
\bauthor{\bsnm{{Poretti}}, \binits{E.}},
\bauthor{\bsnm{{Quirrenbach}}, \binits{A.}},
\bauthor{\bsnm{{Ragazzoni}}, \binits{R.}},
\bauthor{\bsnm{{Raimondo}}, \binits{G.}},
\bauthor{\bsnm{{Rainer}}, \binits{M.}},
\bauthor{\bsnm{{Reese}}, \binits{D.R.}},
\bauthor{\bsnm{{Redmer}}, \binits{R.}},
\bauthor{\bsnm{{Reffert}}, \binits{S.}},
\bauthor{\bsnm{{Rojas-Ayala}}, \binits{B.}},
\bauthor{\bsnm{{Roxburgh}}, \binits{I.W.}},
\bauthor{\bsnm{{Salmon}}, \binits{S.}},
\bauthor{\bsnm{{Santerne}}, \binits{A.}},
\bauthor{\bsnm{{Schneider}}, \binits{J.}},
\bauthor{\bsnm{{Schou}}, \binits{J.}},
\bauthor{\bsnm{{Schuh}}, \binits{S.}},
\bauthor{\bsnm{{Schunker}}, \binits{H.}},
\bauthor{\bsnm{{Silva-Valio}}, \binits{A.}},
\bauthor{\bsnm{{Silvotti}}, \binits{R.}},
\bauthor{\bsnm{{Skillen}}, \binits{I.}},
\bauthor{\bsnm{{Snellen}}, \binits{I.}},
\bauthor{\bsnm{{Sohl}}, \binits{F.}},
\bauthor{\bsnm{{Sousa}}, \binits{S.G.}},
\bauthor{\bsnm{{Sozzetti}}, \binits{A.}},
\bauthor{\bsnm{{Stello}}, \binits{D.}},
\bauthor{\bsnm{{Strassmeier}}, \binits{K.G.}},
\bauthor{\bsnm{{{\v S}vanda}}, \binits{M.}},
\bauthor{\bsnm{{Szab{\'o}}}, \binits{G.M.}},
\bauthor{\bsnm{{Tkachenko}}, \binits{A.}},
\bauthor{\bsnm{{Valencia}}, \binits{D.}},
\bauthor{\bsnm{{Van Grootel}}, \binits{V.}},
\bauthor{\bsnm{{Vauclair}}, \binits{S.D.}},
\bauthor{\bsnm{{Ventura}}, \binits{P.}},
\bauthor{\bsnm{{Wagner}}, \binits{F.W.}},
\bauthor{\bsnm{{Walton}}, \binits{N.A.}},
\bauthor{\bsnm{{Weingrill}}, \binits{J.}},
\bauthor{\bsnm{{Werner}}, \binits{S.C.}},
\bauthor{\bsnm{{Wheatley}}, \binits{P.J.}},
\bauthor{\bsnm{{Zwintz}}, \binits{K.}}:
\batitle{{The PLATO 2.0 mission}}.
\bjtitle{Experimental Astronomy}
\bvolume{38},
\bfpage{249}--\blpage{330}
(\byear{2014})
\doiurl{10.1007/s10686-014-9383-4}
{\href{https://arxiv.org/abs/1310.0696}{{arXiv:1310.0696}}}
{[astro-ph.EP]}
\end{barticle}
\endbibitem

\bibitem[\protect\citeauthoryear{{Tinetti} et~al.}{2018}]{Tinettietal2018}
\begin{barticle}
\bauthor{\bsnm{{Tinetti}}, \binits{G.}},
\bauthor{\bsnm{{Drossart}}, \binits{P.}},
\bauthor{\bsnm{{Eccleston}}, \binits{P.}},
\bauthor{\bsnm{{Hartogh}}, \binits{P.}},
\bauthor{\bsnm{{Heske}}, \binits{A.}},
\bauthor{\bsnm{{Leconte}}, \binits{J.}},
\bauthor{\bsnm{{Micela}}, \binits{G.}},
\bauthor{\bsnm{{Ollivier}}, \binits{M.}},
\bauthor{\bsnm{{Pilbratt}}, \binits{G.}},
\bauthor{\bsnm{{Puig}}, \binits{L.}},
\bauthor{\bsnm{al.}}:
\batitle{{A chemical survey of exoplanets with ARIEL}}.
\bjtitle{Experimental Astronomy}
\bvolume{46},
\bfpage{135}--\blpage{209}
(\byear{2018})
\doiurl{10.1007/s10686-018-9598-x}
\end{barticle}
\endbibitem

\bibitem[\protect\citeauthoryear{{Moutou} et~al.}{2015}]{Moutouetal2015}
\begin{bchapter}
\bauthor{\bsnm{{Moutou}}, \binits{C.}},
\bauthor{\bsnm{{Boisse}}, \binits{I.}},
\bauthor{\bsnm{{H{\'e}brard}}, \binits{G.}},
\bauthor{\bsnm{{H{\'e}brard}}, \binits{E.}},
\bauthor{\bsnm{{Donati}}, \binits{J.-F.}},
\bauthor{\bsnm{{Delfosse}}, \binits{X.}},
\bauthor{\bsnm{{Kouach}}, \binits{D.}}:
\bctitle{{SPIRou: a spectropolarimeter for the CFHT}}.
In: \beditor{\bsnm{{Martins}}, \binits{F.}},
\beditor{\bsnm{{Boissier}}, \binits{S.}},
\beditor{\bsnm{{Buat}}, \binits{V.}},
\beditor{\bsnm{{Cambr{\'e}sy}}, \binits{L.}},
\beditor{\bsnm{{Petit}}, \binits{P.}} (eds.)
\bbtitle{SF2A-2015: Proceedings of the Annual Meeting of the French Society of
  \aap},
pp. \bfpage{205}--\blpage{212}
(\byear{2015})
\end{bchapter}
\endbibitem

\bibitem[\protect\citeauthoryear{{Guillot}}{2005}]{Guillot2005}
\begin{barticle}
\bauthor{\bsnm{{Guillot}}, \binits{T.}}:
\batitle{{THE INTERIORS OF GIANT PLANETS: Models and Outstanding Questions}}.
\bjtitle{Annual Review of Earth and Planetary Sciences}
\bvolume{33},
\bfpage{493}--\blpage{530}
(\byear{2005})
\doiurl{10.1146/annurev.earth.32.101802.120325}
{\href{https://arxiv.org/abs/astro-ph/0502068}{{arXiv:astro-ph/0502068}}}
{[astro-ph]}
\end{barticle}
\endbibitem

\bibitem[\protect\citeauthoryear{{Helled} and
  {Fortney}}{2020}]{Helled+Fortney2020}
\begin{barticle}
\bauthor{\bsnm{{Helled}}, \binits{R.}},
\bauthor{\bsnm{{Fortney}}, \binits{J.J.}}:
\batitle{{The interiors of Uranus and Neptune: current understanding and open
  questions}}.
\bjtitle{Philosophical Transactions of the Royal Society of London Series A}
\bvolume{378}(\bissue{2187}),
\bfpage{20190474}
(\byear{2020})
\doiurl{10.1098/rsta.2019.0474}
{\href{https://arxiv.org/abs/2007.10783}{{arXiv:2007.10783}}}
{[astro-ph.EP]}
\end{barticle}
\endbibitem

\bibitem[\protect\citeauthoryear{{Kunitomo} et~al.}{2018}]{Kunitomo+2018}
\begin{barticle}
\bauthor{\bsnm{{Kunitomo}}, \binits{M.}},
\bauthor{\bsnm{{Guillot}}, \binits{T.}},
\bauthor{\bsnm{{Ida}}, \binits{S.}},
\bauthor{\bsnm{{Takeuchi}}, \binits{T.}}:
\batitle{{Revisiting the pre-main-sequence evolution of stars. II. Consequences
  of planet formation on stellar surface composition}}.
\bjtitle{\aap}
\bvolume{618},
\bfpage{132}
(\byear{2018})
\doiurl{10.1051/0004-6361/201833127}
{\href{https://arxiv.org/abs/1808.07396}{{arXiv:1808.07396}}}
{[astro-ph.SR]}
\end{barticle}
\endbibitem

\bibitem[\protect\citeauthoryear{{Pearl} and
  {Conrath}}{1991}]{Pearl+Conrath1991}
\begin{barticle}
\bauthor{\bsnm{{Pearl}}, \binits{J.C.}},
\bauthor{\bsnm{{Conrath}}, \binits{B.J.}}:
\batitle{{The albedo, effective temperature, and energy balance of Neptune, as
  determined from Voyager data}}.
\bjtitle{\jgr}
\bvolume{96},
\bfpage{18921}--\blpage{18930}
(\byear{1991})
\doiurl{10.1029/91JA01087}
\end{barticle}
\endbibitem

\bibitem[\protect\citeauthoryear{{Hubbard}}{1968}]{Hubbard1968}
\begin{barticle}
\bauthor{\bsnm{{Hubbard}}, \binits{W.B.}}:
\batitle{{Thermal structure of Jupiter}}.
\bjtitle{\apj}
\bvolume{152},
\bfpage{745}--\blpage{754}
(\byear{1968})
\doiurl{10.1086/149591}
\end{barticle}
\endbibitem

\bibitem[\protect\citeauthoryear{Li et~al.}{2018}]{Li_Liming+2018}
\begin{barticle}
\bauthor{\bsnm{Li}, \binits{L.}},
\bauthor{\bsnm{Jiang}, \binits{X.}},
\bauthor{\bsnm{West}, \binits{R.A.}},
\bauthor{\bsnm{Gierasch}, \binits{P.J.}},
\bauthor{\bsnm{Perez-Hoyos}, \binits{S.}},
\bauthor{\bsnm{Sanchez-Lavega}, \binits{A.}},
\bauthor{\bsnm{Fletcher}, \binits{L.N.}},
\bauthor{\bsnm{Fortney}, \binits{J.J.}},
\bauthor{\bsnm{Knowles}, \binits{B.}},
\bauthor{\bsnm{Porco}, \binits{C.C.}},
\bauthor{\bsnm{Baines}, \binits{K.H.}},
\bauthor{\bsnm{Fry}, \binits{P.M.}},
\bauthor{\bsnm{Mallama}, \binits{A.}},
\bauthor{\bsnm{Achterberg}, \binits{R.K.}},
\bauthor{\bsnm{Simon}, \binits{A.A.}},
\bauthor{\bsnm{Nixon}, \binits{C.A.}},
\bauthor{\bsnm{Orton}, \binits{G.S.}},
\bauthor{\bsnm{Dyudina}, \binits{U.A.}},
\bauthor{\bsnm{Ewald}, \binits{S.P.}},
\bauthor{\bsnm{Schmude}, \binits{R.W.}}:
\batitle{Less absorbed solar energy and more internal heat for jupiter}.
\bjtitle{Nature Communications}
\bvolume{9}(\bissue{1}),
\bfpage{3709}
(\byear{2018})
\doiurl{10.1038/s41467-018-06107-2}
\end{barticle}
\endbibitem

\bibitem[\protect\citeauthoryear{{Fletcher} et~al.}{2020}]{Fletcher+2020}
\begin{barticle}
\bauthor{\bsnm{{Fletcher}}, \binits{L.N.}},
\bauthor{\bsnm{{Helled}}, \binits{R.}},
\bauthor{\bsnm{{Roussos}}, \binits{E.}},
\bauthor{\bsnm{{Jones}}, \binits{G.}},
\bauthor{\bsnm{{Charnoz}}, \binits{S.}},
\bauthor{\bsnm{{Andr{\'e}}}, \binits{N.}},
\bauthor{\bsnm{{Andrews}}, \binits{D.}},
\bauthor{\bsnm{{Bannister}}, \binits{M.}},
\bauthor{\bsnm{{Bunce}}, \binits{E.}},
\bauthor{\bsnm{{Cavali{\'e}}}, \binits{T.}},
\bauthor{\bsnm{{Ferri}}, \binits{F.}},
\bauthor{\bsnm{{Fortney}}, \binits{J.}},
\bauthor{\bsnm{{Grassi}}, \binits{D.}},
\bauthor{\bsnm{{Griton}}, \binits{L.}},
\bauthor{\bsnm{{Hartogh}}, \binits{P.}},
\bauthor{\bsnm{{Hueso}}, \binits{R.}},
\bauthor{\bsnm{{Kaspi}}, \binits{Y.}},
\bauthor{\bsnm{{Lamy}}, \binits{L.}},
\bauthor{\bsnm{{Masters}}, \binits{A.}},
\bauthor{\bsnm{{Melin}}, \binits{H.}},
\bauthor{\bsnm{{Moses}}, \binits{J.}},
\bauthor{\bsnm{{Mousis}}, \binits{O.}},
\bauthor{\bsnm{{Nettleman}}, \binits{N.}},
\bauthor{\bsnm{{Plainaki}}, \binits{C.}},
\bauthor{\bsnm{{Schmidt}}, \binits{J.}},
\bauthor{\bsnm{{Simon}}, \binits{A.}},
\bauthor{\bsnm{{Tobie}}, \binits{G.}},
\bauthor{\bsnm{{Tortora}}, \binits{P.}},
\bauthor{\bsnm{{Tosi}}, \binits{F.}},
\bauthor{\bsnm{{Turrini}}, \binits{D.}}:
\batitle{{Ice Giant Systems: The scientific potential of orbital missions to
  Uranus and Neptune}}.
\bjtitle{\planss}
\bvolume{191},
\bfpage{105030}
(\byear{2020})
\doiurl{10.1016/j.pss.2020.105030}
{\href{https://arxiv.org/abs/1907.02963}{{arXiv:1907.02963}}}
{[astro-ph.EP]}
\end{barticle}
\endbibitem

\bibitem[\protect\citeauthoryear{{Guillot}}{2022}]{Guillot2022}
\begin{barticle}
\bauthor{\bsnm{{Guillot}}, \binits{T.}}:
\batitle{{Uranus and Neptune are key to understand planets with hydrogen
  atmospheres}}.
\bjtitle{Experimental Astronomy}
\bvolume{54}(\bissue{2-3}),
\bfpage{1027}--\blpage{1049}
(\byear{2022})
\doiurl{10.1007/s10686-021-09812-x}
{\href{https://arxiv.org/abs/1908.02092}{{arXiv:1908.02092}}}
{[astro-ph.EP]}
\end{barticle}
\endbibitem

\bibitem[\protect\citeauthoryear{{Stevenson}}{1976}]{Stevenson1976}
\begin{botherref}
\oauthor{\bsnm{{Stevenson}}, \binits{D.J.}}:
{The Interior of Jupiter.}
PhD thesis,
Cornell University, New York
(January 1976)
\end{botherref}
\endbibitem

\bibitem[\protect\citeauthoryear{{Guillot} et~al.}{1994}]{Guillot+1994}
\begin{barticle}
\bauthor{\bsnm{{Guillot}}, \binits{T.}},
\bauthor{\bsnm{{Gautier}}, \binits{D.}},
\bauthor{\bsnm{{Chabrier}}, \binits{G.}},
\bauthor{\bsnm{{Mosser}}, \binits{B.}}:
\batitle{{Are the Giant Planets Fully Convective?}}
\bjtitle{Icarus}
\bvolume{112}(\bissue{2}),
\bfpage{337}--\blpage{353}
(\byear{1994})
\doiurl{10.1006/icar.1994.1188}
\end{barticle}
\endbibitem

\bibitem[\protect\citeauthoryear{{Guillot} et~al.}{2004}]{Guillot+2004}
\begin{bchapter}
\bauthor{\bsnm{{Guillot}}, \binits{T.}},
\bauthor{\bsnm{{Stevenson}}, \binits{D.J.}},
\bauthor{\bsnm{{Hubbard}}, \binits{W.B.}},
\bauthor{\bsnm{{Saumon}}, \binits{D.}}:
\bctitle{{The interior of Jupiter}}.
In: \beditor{\bsnm{{Bagenal}}, \binits{F.}},
\beditor{\bsnm{{Dowling}}, \binits{T.E.}},
\beditor{\bsnm{{McKinnon}}, \binits{W.B.}} (eds.)
\bbtitle{Jupiter. The Planet, Satellites and Magnetosphere}
vol. \bseriesno{1},
pp. \bfpage{35}--\blpage{57}
(\byear{2004})
\end{bchapter}
\endbibitem

\bibitem[\protect\citeauthoryear{{Hueso} et~al.}{2020}]{Hueso+2020}
\begin{barticle}
\bauthor{\bsnm{{Hueso}}, \binits{R.}},
\bauthor{\bsnm{{Guillot}}, \binits{T.}},
\bauthor{\bsnm{{S{\'a}nchez-Lavega}}, \binits{A.}}:
\batitle{{Convective storms and atmospheric vertical structure in Uranus and
  Neptune}}.
\bjtitle{Philosophical Transactions of the Royal Society of London Series A}
\bvolume{378}(\bissue{2187}),
\bfpage{20190476}
(\byear{2020})
\doiurl{10.1098/rsta.2019.0476}
{\href{https://arxiv.org/abs/2111.15494}{{arXiv:2111.15494}}}
{[astro-ph.EP]}
\end{barticle}
\endbibitem

\bibitem[\protect\citeauthoryear{{Ingersoll} and
  {Porco}}{1978}]{Ingersoll+Porco1978}
\begin{barticle}
\bauthor{\bsnm{{Ingersoll}}, \binits{A.P.}},
\bauthor{\bsnm{{Porco}}, \binits{C.C.}}:
\batitle{{Solar heating and internal heat flow on Jupiter}}.
\bjtitle{Icarus}
\bvolume{35}(\bissue{1}),
\bfpage{27}--\blpage{43}
(\byear{1978})
\end{barticle}
\endbibitem

\bibitem[\protect\citeauthoryear{{Guillot} et~al.}{2022}]{Guillot+2023}
\begin{botherref}
\oauthor{\bsnm{{Guillot}}, \binits{T.}},
\oauthor{\bsnm{{Fletcher}}, \binits{L.N.}},
\oauthor{\bsnm{{Helled}}, \binits{R.}},
\oauthor{\bsnm{{Ikoma}}, \binits{M.}},
\oauthor{\bsnm{{Line}}, \binits{M.R.}},
\oauthor{\bsnm{{Parmentier}}, \binits{V.}}:
{Giant Planets from the Inside-Out}.
arXiv e-prints,
2205--04100
(2022)
{\href{https://arxiv.org/abs/2205.04100}{{arXiv:2205.04100}}}
{[astro-ph.EP]}
\end{botherref}
\endbibitem

\bibitem[\protect\citeauthoryear{{Durante} et~al.}{2020}]{Durante+2020}
\begin{barticle}
\bauthor{\bsnm{{Durante}}, \binits{D.}},
\bauthor{\bsnm{{Parisi}}, \binits{M.}},
\bauthor{\bsnm{{Serra}}, \binits{D.}},
\bauthor{\bsnm{{Zannoni}}, \binits{M.}},
\bauthor{\bsnm{{Notaro}}, \binits{V.}},
\bauthor{\bsnm{{Racioppa}}, \binits{P.}},
\bauthor{\bsnm{{Buccino}}, \binits{D.R.}},
\bauthor{\bsnm{{Lari}}, \binits{G.}},
\bauthor{\bsnm{{Gomez Casajus}}, \binits{L.}},
\bauthor{\bsnm{{Iess}}, \binits{L.}},
\bauthor{\bsnm{{Folkner}}, \binits{W.M.}},
\bauthor{\bsnm{{Tommei}}, \binits{G.}},
\bauthor{\bsnm{{Tortora}}, \binits{P.}},
\bauthor{\bsnm{{Bolton}}, \binits{S.J.}}:
\batitle{{Jupiter's Gravity Field Halfway Through the Juno Mission}}.
\bjtitle{\grl}
\bvolume{47}(\bissue{4}),
\bfpage{86572}
(\byear{2020})
\doiurl{10.1029/2019GL086572}
\end{barticle}
\endbibitem

\bibitem[\protect\citeauthoryear{{Iess} et~al.}{2019}]{Iess+2019}
\begin{barticle}
\bauthor{\bsnm{{Iess}}, \binits{L.}},
\bauthor{\bsnm{{Militzer}}, \binits{B.}},
\bauthor{\bsnm{{Kaspi}}, \binits{Y.}},
\bauthor{\bsnm{{Nicholson}}, \binits{P.}},
\bauthor{\bsnm{{Durante}}, \binits{D.}},
\bauthor{\bsnm{{Racioppa}}, \binits{P.}},
\bauthor{\bsnm{{Anabtawi}}, \binits{A.}},
\bauthor{\bsnm{{Galanti}}, \binits{E.}},
\bauthor{\bsnm{{Hubbard}}, \binits{W.}},
\bauthor{\bsnm{{Mariani}}, \binits{M.J.}},
\bauthor{\bsnm{{Tortora}}, \binits{P.}},
\bauthor{\bsnm{{Wahl}}, \binits{S.}},
\bauthor{\bsnm{{Zannoni}}, \binits{M.}}:
\batitle{{Measurement and implications of Saturn's gravity field and ring
  mass}}.
\bjtitle{Science}
\bvolume{364}(\bissue{6445}),
\bfpage{2965}
(\byear{2019})
\doiurl{10.1126/science.aat2965}
\end{barticle}
\endbibitem

\bibitem[\protect\citeauthoryear{{Hedman} et~al.}{2019}]{Hedman+2019}
\begin{barticle}
\bauthor{\bsnm{{Hedman}}, \binits{M.M.}},
\bauthor{\bsnm{{Nicholson}}, \binits{P.D.}},
\bauthor{\bsnm{{French}}, \binits{R.G.}}:
\batitle{{Kronoseismology. IV. Six Previously Unidentified Waves in
  Saturn{\textquoteright}s Middle C Ring}}.
\bjtitle{\aj}
\bvolume{157}(\bissue{1}),
\bfpage{18}
(\byear{2019})
\doiurl{10.3847/1538-3881/aaf0a6}
{\href{https://arxiv.org/abs/1811.04796}{{arXiv:1811.04796}}}
{[astro-ph.EP]}
\end{barticle}
\endbibitem

\bibitem[\protect\citeauthoryear{{Jacobson}}{2014}]{Jacobson2014}
\begin{barticle}
\bauthor{\bsnm{{Jacobson}}, \binits{R.A.}}:
\batitle{{The Orbits of the Uranian Satellites and Rings, the Gravity Field of
  the Uranian System, and the Orientation of the Pole of Uranus}}.
\bjtitle{\aj}
\bvolume{148}(\bissue{5}),
\bfpage{76}
(\byear{2014})
\doiurl{10.1088/0004-6256/148/5/76}
\end{barticle}
\endbibitem

\bibitem[\protect\citeauthoryear{{Jacobson}}{2009}]{Jacobson2009}
\begin{barticle}
\bauthor{\bsnm{{Jacobson}}, \binits{R.A.}}:
\batitle{{The Orbits of the Neptunian Satellites and the Orientation of the
  Pole of Neptune}}.
\bjtitle{\aj}
\bvolume{137}(\bissue{5}),
\bfpage{4322}--\blpage{4329}
(\byear{2009})
\doiurl{10.1088/0004-6256/137/5/4322}
\end{barticle}
\endbibitem

\bibitem[\protect\citeauthoryear{{Wahl} et~al.}{2017}]{Wahl+2017}
\begin{barticle}
\bauthor{\bsnm{{Wahl}}, \binits{S.M.}},
\bauthor{\bsnm{{Hubbard}}, \binits{W.B.}},
\bauthor{\bsnm{{Militzer}}, \binits{B.}},
\bauthor{\bsnm{{Guillot}}, \binits{T.}},
\bauthor{\bsnm{{Miguel}}, \binits{Y.}},
\bauthor{\bsnm{{Movshovitz}}, \binits{N.}},
\bauthor{\bsnm{{Kaspi}}, \binits{Y.}},
\bauthor{\bsnm{{Helled}}, \binits{R.}},
\bauthor{\bsnm{{Reese}}, \binits{D.}},
\bauthor{\bsnm{{Galanti}}, \binits{E.}},
\bauthor{\bsnm{{Levin}}, \binits{S.}},
\bauthor{\bsnm{{Connerney}}, \binits{J.E.}},
\bauthor{\bsnm{{Bolton}}, \binits{S.J.}}:
\batitle{{Comparing Jupiter interior structure models to Juno gravity
  measurements and the role of a dilute core}}.
\bjtitle{\grl}
\bvolume{44}(\bissue{10}),
\bfpage{4649}--\blpage{4659}
(\byear{2017})
\doiurl{10.1002/2017GL073160}
{\href{https://arxiv.org/abs/1707.01997}{{arXiv:1707.01997}}}
{[astro-ph.EP]}
\end{barticle}
\endbibitem

\bibitem[\protect\citeauthoryear{{Debras} and
  {Chabrier}}{2019}]{Debras+Chabrier2019}
\begin{barticle}
\bauthor{\bsnm{{Debras}}, \binits{F.}},
\bauthor{\bsnm{{Chabrier}}, \binits{G.}}:
\batitle{{New Models of Jupiter in the Context of Juno and Galileo}}.
\bjtitle{\apj}
\bvolume{872}(\bissue{1}),
\bfpage{100}
(\byear{2019})
\doiurl{10.3847/1538-4357/aaff65}
{\href{https://arxiv.org/abs/1901.05697}{{arXiv:1901.05697}}}
{[astro-ph.EP]}
\end{barticle}
\endbibitem

\bibitem[\protect\citeauthoryear{{Miguel} et~al.}{2022}]{Miguel+2022}
\begin{barticle}
\bauthor{\bsnm{{Miguel}}, \binits{Y.}},
\bauthor{\bsnm{{Bazot}}, \binits{M.}},
\bauthor{\bsnm{{Guillot}}, \binits{T.}},
\bauthor{\bsnm{{Howard}}, \binits{S.}},
\bauthor{\bsnm{{Galanti}}, \binits{E.}},
\bauthor{\bsnm{{Kaspi}}, \binits{Y.}},
\bauthor{\bsnm{{Hubbard}}, \binits{W.B.}},
\bauthor{\bsnm{{Militzer}}, \binits{B.}},
\bauthor{\bsnm{{Helled}}, \binits{R.}},
\bauthor{\bsnm{{Atreya}}, \binits{S.K.}},
\bauthor{\bsnm{{Connerney}}, \binits{J.E.P.}},
\bauthor{\bsnm{{Durante}}, \binits{D.}},
\bauthor{\bsnm{{Kulowski}}, \binits{L.}},
\bauthor{\bsnm{{Lunine}}, \binits{J.I.}},
\bauthor{\bsnm{{Stevenson}}, \binits{D.}},
\bauthor{\bsnm{{Bolton}}, \binits{S.}}:
\batitle{{Jupiter's inhomogeneous envelope}}.
\bjtitle{\aap}
\bvolume{662},
\bfpage{18}
(\byear{2022})
\doiurl{10.1051/0004-6361/202243207}
{\href{https://arxiv.org/abs/2203.01866}{{arXiv:2203.01866}}}
{[astro-ph.EP]}
\end{barticle}
\endbibitem

\bibitem[\protect\citeauthoryear{{Militzer} et~al.}{2022}]{Militzer+2022}
\begin{barticle}
\bauthor{\bsnm{{Militzer}}, \binits{B.}},
\bauthor{\bsnm{{Hubbard}}, \binits{W.B.}},
\bauthor{\bsnm{{Wahl}}, \binits{S.}},
\bauthor{\bsnm{{Lunine}}, \binits{J.I.}},
\bauthor{\bsnm{{Galanti}}, \binits{E.}},
\bauthor{\bsnm{{Kaspi}}, \binits{Y.}},
\bauthor{\bsnm{{Miguel}}, \binits{Y.}},
\bauthor{\bsnm{{Guillot}}, \binits{T.}},
\bauthor{\bsnm{{Moore}}, \binits{K.M.}},
\bauthor{\bsnm{{Parisi}}, \binits{M.}},
\bauthor{\bsnm{{Connerney}}, \binits{J.E.P.}},
\bauthor{\bsnm{{Helled}}, \binits{R.}},
\bauthor{\bsnm{{Cao}}, \binits{H.}},
\bauthor{\bsnm{{Mankovich}}, \binits{C.}},
\bauthor{\bsnm{{Stevenson}}, \binits{D.J.}},
\bauthor{\bsnm{{Park}}, \binits{R.S.}},
\bauthor{\bsnm{{Wong}}, \binits{M.}},
\bauthor{\bsnm{{Atreya}}, \binits{S.K.}},
\bauthor{\bsnm{{Anderson}}, \binits{J.}},
\bauthor{\bsnm{{Bolton}}, \binits{S.J.}}:
\batitle{{Juno Spacecraft Measurements of Jupiter's Gravity Imply a Dilute
  Core}}.
\bjtitle{Planetary Science Journal}
\bvolume{3}(\bissue{8}),
\bfpage{185}
(\byear{2022})
\doiurl{10.3847/PSJ/ac7ec8}
\end{barticle}
\endbibitem

\bibitem[\protect\citeauthoryear{{Mankovich} and
  {Fuller}}{2021}]{Mankovich+Fuller2021}
\begin{barticle}
\bauthor{\bsnm{{Mankovich}}, \binits{C.R.}},
\bauthor{\bsnm{{Fuller}}, \binits{J.}}:
\batitle{{A diffuse core in Saturn revealed by ring seismology}}.
\bjtitle{Nature Astronomy}
(\byear{2021})
\doiurl{10.1038/s41550-021-01448-3}
{\href{https://arxiv.org/abs/2104.13385}{{arXiv:2104.13385}}}
{[astro-ph.EP]}
\end{barticle}
\endbibitem

\bibitem[\protect\citeauthoryear{{Nettelmann} et~al.}{2013}]{Nettelmann+2013}
\begin{barticle}
\bauthor{\bsnm{{Nettelmann}}, \binits{N.}},
\bauthor{\bsnm{{Helled}}, \binits{R.}},
\bauthor{\bsnm{{Fortney}}, \binits{J.J.}},
\bauthor{\bsnm{{Redmer}}, \binits{R.}}:
\batitle{{New indication for a dichotomy in the interior structure of Uranus
  and Neptune from the application of modified shape and rotation data}}.
\bjtitle{\planss}
\bvolume{77},
\bfpage{143}--\blpage{151}
(\byear{2013})
\doiurl{10.1016/j.pss.2012.06.019}
{\href{https://arxiv.org/abs/1207.2309}{{arXiv:1207.2309}}}
{[astro-ph.EP]}
\end{barticle}
\endbibitem

\bibitem[\protect\citeauthoryear{{Helled} et~al.}{2020}]{Helled+2020}
\begin{barticle}
\bauthor{\bsnm{{Helled}}, \binits{R.}},
\bauthor{\bsnm{{Nettelmann}}, \binits{N.}},
\bauthor{\bsnm{{Guillot}}, \binits{T.}}:
\batitle{{Uranus and Neptune: Origin, Evolution and Internal Structure}}.
\bjtitle{\ssr}
\bvolume{216}(\bissue{3}),
\bfpage{38}
(\byear{2020})
\doiurl{10.1007/s11214-020-00660-3}
{\href{https://arxiv.org/abs/1909.04891}{{arXiv:1909.04891}}}
{[astro-ph.EP]}
\end{barticle}
\endbibitem

\bibitem[\protect\citeauthoryear{{Sch{\"o}ttler} and
  {Redmer}}{2018}]{Schottler+Redmer2018}
\begin{barticle}
\bauthor{\bsnm{{Sch{\"o}ttler}}, \binits{M.}},
\bauthor{\bsnm{{Redmer}}, \binits{R.}}:
\batitle{{Ab Initio Calculation of the Miscibility Diagram for Hydrogen-Helium
  Mixtures}}.
\bjtitle{\prl}
\bvolume{120}(\bissue{11}),
\bfpage{115703}
(\byear{2018})
\doiurl{10.1103/PhysRevLett.120.115703}
\end{barticle}
\endbibitem

\bibitem[\protect\citeauthoryear{{Brygoo} et~al.}{2021}]{Brygoo+2021}
\begin{barticle}
\bauthor{\bsnm{{Brygoo}}, \binits{S.}},
\bauthor{\bsnm{{Loubeyre}}, \binits{P.}},
\bauthor{\bsnm{{Millot}}, \binits{M.}},
\bauthor{\bsnm{{Rygg}}, \binits{J.R.}},
\bauthor{\bsnm{{Celliers}}, \binits{P.M.}},
\bauthor{\bsnm{{Eggert}}, \binits{J.H.}},
\bauthor{\bsnm{{Jeanloz}}, \binits{R.}},
\bauthor{\bsnm{{Collins}}, \binits{G.W.}}:
\batitle{{Evidence of hydrogen‒helium immiscibility at Jupiter-interior
  conditions}}.
\bjtitle{\nat}
\bvolume{593}(\bissue{7860}),
\bfpage{517}--\blpage{521}
(\byear{2021})
\doiurl{10.1038/s41586-021-03516-0}
\end{barticle}
\endbibitem

\bibitem[\protect\citeauthoryear{{Mazevet} et~al.}{2019}]{Mazevet+2019}
\begin{barticle}
\bauthor{\bsnm{{Mazevet}}, \binits{S.}},
\bauthor{\bsnm{{Musella}}, \binits{R.}},
\bauthor{\bsnm{{Guyot}}, \binits{F.}}:
\batitle{{The fate of planetary cores in giant and ice-giant planets}}.
\bjtitle{\aap}
\bvolume{631},
\bfpage{4}
(\byear{2019})
\doiurl{10.1051/0004-6361/201936288}
{\href{https://arxiv.org/abs/1909.07640}{{arXiv:1909.07640}}}
{[astro-ph.EP]}
\end{barticle}
\endbibitem

\bibitem[\protect\citeauthoryear{{Gonz{\'a}lez-Cataldo}
  et~al.}{2014}]{Gonzalez-Cataldo+2014}
\begin{barticle}
\bauthor{\bsnm{{Gonz{\'a}lez-Cataldo}}, \binits{F.}},
\bauthor{\bsnm{{Wilson}}, \binits{H.F.}},
\bauthor{\bsnm{{Militzer}}, \binits{B.}}:
\batitle{{Ab Initio Free Energy Calculations of the Solubility of Silica in
  Metallic Hydrogen and Application to Giant Planet Cores}}.
\bjtitle{\apj}
\bvolume{787}(\bissue{1}),
\bfpage{79}
(\byear{2014})
\doiurl{10.1088/0004-637X/787/1/79}
\end{barticle}
\endbibitem

\bibitem[\protect\citeauthoryear{{Vazan} et~al.}{2018}]{Vazan+2018}
\begin{barticle}
\bauthor{\bsnm{{Vazan}}, \binits{A.}},
\bauthor{\bsnm{{Helled}}, \binits{R.}},
\bauthor{\bsnm{{Guillot}}, \binits{T.}}:
\batitle{{Jupiter's evolution with primordial composition gradients}}.
\bjtitle{\aap}
\bvolume{610},
\bfpage{14}
(\byear{2018})
\doiurl{10.1051/0004-6361/201732522}
{\href{https://arxiv.org/abs/1801.08149}{{arXiv:1801.08149}}}
{[astro-ph.EP]}
\end{barticle}
\endbibitem

\bibitem[\protect\citeauthoryear{{French} et~al.}{2009}]{French+2009}
\begin{barticle}
\bauthor{\bsnm{{French}}, \binits{M.}},
\bauthor{\bsnm{{Mattsson}}, \binits{T.R.}},
\bauthor{\bsnm{{Nettelmann}}, \binits{N.}},
\bauthor{\bsnm{{Redmer}}, \binits{R.}}:
\batitle{{Equation of state and phase diagram of water at ultrahigh pressures
  as in planetary interiors}}.
\bjtitle{\prb}
\bvolume{79}(\bissue{5}),
\bfpage{054107}
(\byear{2009})
\doiurl{10.1103/PhysRevB.79.054107}
\end{barticle}
\endbibitem

\bibitem[\protect\citeauthoryear{{Redmer} et~al.}{2011}]{Redmer+2011}
\begin{barticle}
\bauthor{\bsnm{{Redmer}}, \binits{R.}},
\bauthor{\bsnm{{Mattsson}}, \binits{T.R.}},
\bauthor{\bsnm{{Nettelmann}}, \binits{N.}},
\bauthor{\bsnm{{French}}, \binits{M.}}:
\batitle{{The phase diagram of water and the magnetic fields of Uranus and
  Neptune}}.
\bjtitle{Icarus}
\bvolume{211}(\bissue{1}),
\bfpage{798}--\blpage{803}
(\byear{2011})
\doiurl{10.1016/j.icarus.2010.08.008}
\end{barticle}
\endbibitem

\bibitem[\protect\citeauthoryear{{Millot} et~al.}{2019}]{Millot+2019}
\begin{barticle}
\bauthor{\bsnm{{Millot}}, \binits{M.}},
\bauthor{\bsnm{{Coppari}}, \binits{F.}},
\bauthor{\bsnm{{Rygg}}, \binits{J.R.}},
\bauthor{\bsnm{{Correa Barrios}}, \binits{A.}},
\bauthor{\bsnm{{Hamel}}, \binits{S.}},
\bauthor{\bsnm{{Swift}}, \binits{D.C.}},
\bauthor{\bsnm{{Eggert}}, \binits{J.H.}}:
\batitle{{Nanosecond X-ray diffraction of shock-compressed superionic water
  ice}}.
\bjtitle{\nat}
\bvolume{569}(\bissue{7755}),
\bfpage{251}--\blpage{255}
(\byear{2019})
\doiurl{10.1038/s41586-019-1114-6}
\end{barticle}
\endbibitem

\bibitem[\protect\citeauthoryear{{Stixrude} et~al.}{2021}]{Stixrude+2021}
\begin{barticle}
\bauthor{\bsnm{{Stixrude}}, \binits{L.}},
\bauthor{\bsnm{{Baroni}}, \binits{S.}},
\bauthor{\bsnm{{Grasselli}}, \binits{F.}}:
\batitle{{Thermal and Tidal Evolution of Uranus with a Growing Frozen Core}}.
\bjtitle{Planetary Science Journal}
\bvolume{2}(\bissue{6}),
\bfpage{222}
(\byear{2021})
\doiurl{10.3847/PSJ/ac2a47}
\end{barticle}
\endbibitem

\bibitem[\protect\citeauthoryear{{Guarguaglini}
  et~al.}{2019}]{Guarguaglini+2019}
\begin{barticle}
\bauthor{\bsnm{{Guarguaglini}}, \binits{M.}},
\bauthor{\bsnm{{Hernandez}}, \binits{J.-A.}},
\bauthor{\bsnm{{Okuchi}}, \binits{T.}},
\bauthor{\bsnm{{Barroso}}, \binits{P.}},
\bauthor{\bsnm{{Benuzzi-Mounaix}}, \binits{A.}},
\bauthor{\bsnm{{Bethkenhagen}}, \binits{M.}},
\bauthor{\bsnm{{Bolis}}, \binits{R.}},
\bauthor{\bsnm{{Brambrink}}, \binits{E.}},
\bauthor{\bsnm{{French}}, \binits{M.}},
\bauthor{\bsnm{{Fujimoto}}, \binits{Y.}},
\bauthor{\bsnm{{Kodama}}, \binits{R.}},
\bauthor{\bsnm{{Koenig}}, \binits{M.}},
\bauthor{\bsnm{{Lefevre}}, \binits{F.}},
\bauthor{\bsnm{{Miyanishi}}, \binits{K.}},
\bauthor{\bsnm{{Ozaki}}, \binits{N.}},
\bauthor{\bsnm{{Redmer}}, \binits{R.}},
\bauthor{\bsnm{{Sano}}, \binits{T.}},
\bauthor{\bsnm{{Umeda}}, \binits{Y.}},
\bauthor{\bsnm{{Vinci}}, \binits{T.}},
\bauthor{\bsnm{{Ravasio}}, \binits{A.}}:
\batitle{{Laser-driven shock compression of ``synthetic planetary mixtures'' of
  water, ethanol, and ammonia}}.
\bjtitle{Scientific Reports}
\bvolume{9},
\bfpage{10155}
(\byear{2019})
\doiurl{10.1038/s41598-019-46561-6}
\end{barticle}
\endbibitem

\bibitem[\protect\citeauthoryear{{Guillot} et~al.}{2006}]{Guillot+2006}
\begin{barticle}
\bauthor{\bsnm{{Guillot}}, \binits{T.}},
\bauthor{\bsnm{{Santos}}, \binits{N.C.}},
\bauthor{\bsnm{{Pont}}, \binits{F.}},
\bauthor{\bsnm{{Iro}}, \binits{N.}},
\bauthor{\bsnm{{Melo}}, \binits{C.}},
\bauthor{\bsnm{{Ribas}}, \binits{I.}}:
\batitle{{A correlation between the heavy element content of transiting
  extrasolar planets and the metallicity of their parent stars}}.
\bjtitle{\aap}
\bvolume{453}(\bissue{2}),
\bfpage{21}--\blpage{24}
(\byear{2006})
\doiurl{10.1051/0004-6361:20065476}
{\href{https://arxiv.org/abs/astro-ph/0605751}{{arXiv:astro-ph/0605751}}}
{[astro-ph]}
\end{barticle}
\endbibitem

\bibitem[\protect\citeauthoryear{{Thorngren} et~al.}{2016}]{Thorngren+2016}
\begin{barticle}
\bauthor{\bsnm{{Thorngren}}, \binits{D.P.}},
\bauthor{\bsnm{{Fortney}}, \binits{J.J.}},
\bauthor{\bsnm{{Murray-Clay}}, \binits{R.A.}},
\bauthor{\bsnm{{Lopez}}, \binits{E.D.}}:
\batitle{{The Mass-Metallicity Relation for Giant Planets}}.
\bjtitle{\apj}
\bvolume{831}(\bissue{1}),
\bfpage{64}
(\byear{2016})
\doiurl{10.3847/0004-637X/831/1/64}
{\href{https://arxiv.org/abs/1511.07854}{{arXiv:1511.07854}}}
{[astro-ph.EP]}
\end{barticle}
\endbibitem

\bibitem[\protect\citeauthoryear{{Guillot} et~al.}{1996}]{Guillot+1996}
\begin{barticle}
\bauthor{\bsnm{{Guillot}}, \binits{T.}},
\bauthor{\bsnm{{Burrows}}, \binits{A.}},
\bauthor{\bsnm{{Hubbard}}, \binits{W.B.}},
\bauthor{\bsnm{{Lunine}}, \binits{J.I.}},
\bauthor{\bsnm{{Saumon}}, \binits{D.}}:
\batitle{{Giant Planets at Small Orbital Distances}}.
\bjtitle{\apjl}
\bvolume{459},
\bfpage{35}
(\byear{1996})
\doiurl{10.1086/309935}
{\href{https://arxiv.org/abs/astro-ph/9511109}{{arXiv:astro-ph/9511109}}}
{[astro-ph]}
\end{barticle}
\endbibitem

\bibitem[\protect\citeauthoryear{{Fortney} et~al.}{2021}]{Fortney+2021}
\begin{barticle}
\bauthor{\bsnm{{Fortney}}, \binits{J.J.}},
\bauthor{\bsnm{{Dawson}}, \binits{R.I.}},
\bauthor{\bsnm{{Komacek}}, \binits{T.D.}}:
\batitle{{Hot Jupiters: Origins, Structure, Atmospheres}}.
\bjtitle{Journal of Geophysical Research (Planets)}
\bvolume{126}(\bissue{3}),
\bfpage{06629}
(\byear{2021})
\doiurl{10.1029/2020JE006629}
{\href{https://arxiv.org/abs/2102.05064}{{arXiv:2102.05064}}}
{[astro-ph.EP]}
\end{barticle}
\endbibitem

\bibitem[\protect\citeauthoryear{{Batygin} and {Stevenson}}{2010}]{Batygin2010}
\begin{barticle}
\bauthor{\bsnm{{Batygin}}, \binits{K.}},
\bauthor{\bsnm{{Stevenson}}, \binits{D.J.}}:
\batitle{{Inflating Hot Jupiters with Ohmic Dissipation}}.
\bjtitle{\apjl}
\bvolume{714}(\bissue{2}),
\bfpage{238}--\blpage{243}
(\byear{2010})
\doiurl{10.1088/2041-8205/714/2/L238}
{\href{https://arxiv.org/abs/1002.3650}{{arXiv:1002.3650}}}
{[astro-ph.EP]}
\end{barticle}
\endbibitem

\bibitem[\protect\citeauthoryear{{Sarkis} et~al.}{2021}]{Sarkis+2021}
\begin{barticle}
\bauthor{\bsnm{{Sarkis}}, \binits{P.}},
\bauthor{\bsnm{{Mordasini}}, \binits{C.}},
\bauthor{\bsnm{{Henning}}, \binits{T.}},
\bauthor{\bsnm{{Marleau}}, \binits{G.D.}},
\bauthor{\bsnm{{Molli{\`e}re}}, \binits{P.}}:
\batitle{{Evidence of three mechanisms explaining the radius anomaly of hot
  Jupiters}}.
\bjtitle{\aap}
\bvolume{645},
\bfpage{79}
(\byear{2021})
\doiurl{10.1051/0004-6361/202038361}
{\href{https://arxiv.org/abs/2009.04291}{{arXiv:2009.04291}}}
{[astro-ph.EP]}
\end{barticle}
\endbibitem

\bibitem[\protect\citeauthoryear{{Chabrier} and
  {Baraffe}}{2007}]{Chabrier+Baraffe2007}
\begin{barticle}
\bauthor{\bsnm{{Chabrier}}, \binits{G.}},
\bauthor{\bsnm{{Baraffe}}, \binits{I.}}:
\batitle{{Heat Transport in Giant (Exo)planets: A New Perspective}}.
\bjtitle{\apjl}
\bvolume{661}(\bissue{1}),
\bfpage{81}--\blpage{84}
(\byear{2007})
\doiurl{10.1086/518473}
{\href{https://arxiv.org/abs/astro-ph/0703755}{{arXiv:astro-ph/0703755}}}
{[astro-ph]}
\end{barticle}
\endbibitem

\bibitem[\protect\citeauthoryear{{Leconte} and {Chabrier}}{2012}]{leconte:12}
\begin{barticle}
\bauthor{\bsnm{{Leconte}}, \binits{J.}},
\bauthor{\bsnm{{Chabrier}}, \binits{G.}}:
\batitle{{A new vision of giant planet interiors: Impact of double diffusive
  convection}}.
\bjtitle{\aap}
\bvolume{540},
\bfpage{20}
(\byear{2012})
\doiurl{10.1051/0004-6361/201117595}
{\href{https://arxiv.org/abs/1201.4483}{{arXiv:1201.4483}}}
{[astro-ph.EP]}
\end{barticle}
\endbibitem

\bibitem[\protect\citeauthoryear{{Leconte} and {Chabrier}}{2013}]{leconte:13}
\begin{barticle}
\bauthor{\bsnm{{Leconte}}, \binits{J.}},
\bauthor{\bsnm{{Chabrier}}, \binits{G.}}:
\batitle{{Layered convection as the origin of Saturn's luminosity anomaly}}.
\bjtitle{Nature Geoscience}
\bvolume{6},
\bfpage{347}--\blpage{350}
(\byear{2013})
\doiurl{10.1038/ngeo1791}
{\href{https://arxiv.org/abs/1304.6184}{{arXiv:1304.6184}}}
{[astro-ph.EP]}
\end{barticle}
\endbibitem

\bibitem[\protect\citeauthoryear{{Fuller}}{2014}]{fuller:14}
\begin{barticle}
\bauthor{\bsnm{{Fuller}}, \binits{J.}}:
\batitle{{Saturn ring seismology: Evidence for stable stratification in the
  deep interior of Saturn}}.
\bjtitle{Icarus}
\bvolume{242},
\bfpage{283}--\blpage{296}
(\byear{2014})
\doiurl{10.1016/j.icarus.2014.08.006}
{\href{https://arxiv.org/abs/1406.3343}{{arXiv:1406.3343}}}
{[astro-ph.EP]}
\end{barticle}
\endbibitem

\bibitem[\protect\citeauthoryear{{Hedman} and
  {Nicholson}}{2013}]{Hedman+Nicholson2013}
\begin{barticle}
\bauthor{\bsnm{{Hedman}}, \binits{M.M.}},
\bauthor{\bsnm{{Nicholson}}, \binits{P.D.}}:
\batitle{{Kronoseismology: Using Density Waves in Saturn's C Ring to Probe the
  Planet's Interior}}.
\bjtitle{\aj}
\bvolume{146}(\bissue{1}),
\bfpage{12}
(\byear{2013})
\doiurl{10.1088/0004-6256/146/1/12}
{\href{https://arxiv.org/abs/1304.3735}{{arXiv:1304.3735}}}
{[astro-ph.EP]}
\end{barticle}
\endbibitem

\bibitem[\protect\citeauthoryear{{Stevenson} and
  {Salpeter}}{1977}]{Stevenson+Salpeter1977a}
\begin{barticle}
\bauthor{\bsnm{{Stevenson}}, \binits{D.J.}},
\bauthor{\bsnm{{Salpeter}}, \binits{E.E.}}:
\batitle{{The phase diagram and transport properties for hydrogen-helium fluid
  planets}}.
\bjtitle{\apjs}
\bvolume{35},
\bfpage{221}--\blpage{237}
(\byear{1977})
\end{barticle}
\endbibitem

\bibitem[\protect\citeauthoryear{{Mankovich} et~al.}{2016}]{Mankovich+2016}
\begin{barticle}
\bauthor{\bsnm{{Mankovich}}, \binits{C.}},
\bauthor{\bsnm{{Fortney}}, \binits{J.J.}},
\bauthor{\bsnm{{Moore}}, \binits{K.L.}}:
\batitle{{Bayesian Evolution Models for Jupiter with Helium Rain and
  Double-diffusive Convection}}.
\bjtitle{\apj}
\bvolume{832}(\bissue{2}),
\bfpage{113}
(\byear{2016})
\doiurl{10.3847/0004-637X/832/2/113}
{\href{https://arxiv.org/abs/1609.09070}{{arXiv:1609.09070}}}
{[astro-ph.EP]}
\end{barticle}
\endbibitem

\bibitem[\protect\citeauthoryear{{Rosenblum} et~al.}{2011}]{Rosenblum+2011}
\begin{barticle}
\bauthor{\bsnm{{Rosenblum}}, \binits{E.}},
\bauthor{\bsnm{{Garaud}}, \binits{P.}},
\bauthor{\bsnm{{Traxler}}, \binits{A.}},
\bauthor{\bsnm{{Stellmach}}, \binits{S.}}:
\batitle{{Turbulent Mixing and Layer Formation in Double-diffusive Convection:
  Three-dimensional Numerical Simulations and Theory}}.
\bjtitle{\apj}
\bvolume{731}(\bissue{1}),
\bfpage{66}
(\byear{2011})
\doiurl{10.1088/0004-637X/731/1/66}
{\href{https://arxiv.org/abs/1012.0617}{{arXiv:1012.0617}}}
{[astro-ph.SR]}
\end{barticle}
\endbibitem

\bibitem[\protect\citeauthoryear{{Leconte} and
  {Chabrier}}{2012}]{Leconte+Chabrier2012}
\begin{barticle}
\bauthor{\bsnm{{Leconte}}, \binits{J.}},
\bauthor{\bsnm{{Chabrier}}, \binits{G.}}:
\batitle{{A new vision of giant planet interiors: Impact of double diffusive
  convection}}.
\bjtitle{\aap}
\bvolume{540},
\bfpage{20}
(\byear{2012})
\doiurl{10.1051/0004-6361/201117595}
{\href{https://arxiv.org/abs/1201.4483}{{arXiv:1201.4483}}}
{[astro-ph.EP]}
\end{barticle}
\endbibitem

\bibitem[\protect\citeauthoryear{{Mirouh} et~al.}{2012}]{Mirouh2012}
\begin{barticle}
\bauthor{\bsnm{{Mirouh}}, \binits{G.M.}},
\bauthor{\bsnm{{Garaud}}, \binits{P.}},
\bauthor{\bsnm{{Stellmach}}, \binits{S.}},
\bauthor{\bsnm{{Traxler}}, \binits{A.L.}},
\bauthor{\bsnm{{Wood}}, \binits{T.S.}}:
\batitle{{A New Model for Mixing by Double-diffusive Convection
  (Semi-convection). I. The Conditions for Layer Formation}}.
\bjtitle{\apj}
\bvolume{750}(\bissue{1}),
\bfpage{61}
(\byear{2012})
\doiurl{10.1088/0004-637X/750/1/61}
{\href{https://arxiv.org/abs/1112.4819}{{arXiv:1112.4819}}}
{[astro-ph.SR]}
\end{barticle}
\endbibitem

\bibitem[\protect\citeauthoryear{{Fuentes} et~al.}{2022}]{Fuentes2022}
\begin{barticle}
\bauthor{\bsnm{{Fuentes}}, \binits{J.R.}},
\bauthor{\bsnm{{Cumming}}, \binits{A.}},
\bauthor{\bsnm{{Anders}}, \binits{E.H.}}:
\batitle{{Layer formation in a stably stratified fluid cooled from above:
  Towards an analog for Jupiter and other gas giants}}.
\bjtitle{Physical Review Fluids}
\bvolume{7}(\bissue{12}),
\bfpage{124501}
(\byear{2022})
\doiurl{10.1103/PhysRevFluids.7.124501}
{\href{https://arxiv.org/abs/2204.12643}{{arXiv:2204.12643}}}
{[astro-ph.EP]}
\end{barticle}
\endbibitem

\bibitem[\protect\citeauthoryear{{Moore} and {Garaud}}{2016}]{Moore2016}
\begin{barticle}
\bauthor{\bsnm{{Moore}}, \binits{K.}},
\bauthor{\bsnm{{Garaud}}, \binits{P.}}:
\batitle{{Main Sequence Evolution with Layered Semiconvection}}.
\bjtitle{\apj}
\bvolume{817}(\bissue{1}),
\bfpage{54}
(\byear{2016})
\doiurl{10.3847/0004-637X/817/1/54}
{\href{https://arxiv.org/abs/1506.01034}{{arXiv:1506.01034}}}
{[astro-ph.SR]}
\end{barticle}
\endbibitem

\bibitem[\protect\citeauthoryear{{Wong} et~al.}{2011}]{Wong+2011}
\begin{barticle}
\bauthor{\bsnm{{Wong}}, \binits{M.H.}},
\bauthor{\bsnm{{de Pater}}, \binits{I.}},
\bauthor{\bsnm{{Asay-Davis}}, \binits{X.}},
\bauthor{\bsnm{{Marcus}}, \binits{P.S.}},
\bauthor{\bsnm{{Go}}, \binits{C.Y.}}:
\batitle{{Vertical structure of Jupiter's Oval BA before and after it reddened:
  What changed?}}
\bjtitle{Icarus}
\bvolume{215}(\bissue{1}),
\bfpage{211}--\blpage{225}
(\byear{2011})
\doiurl{10.1016/j.icarus.2011.06.032}
\end{barticle}
\endbibitem

\bibitem[\protect\citeauthoryear{{Christensen} et~al.}{2020}]{Christensen+2020}
\begin{barticle}
\bauthor{\bsnm{{Christensen}}, \binits{U.R.}},
\bauthor{\bsnm{{Wicht}}, \binits{J.}},
\bauthor{\bsnm{{Dietrich}}, \binits{W.}}:
\batitle{{Mechanisms for Limiting the Depth of Zonal Winds in the Gas Giant
  Planets}}.
\bjtitle{\apj}
\bvolume{890}(\bissue{1}),
\bfpage{61}
(\byear{2020})
\doiurl{10.3847/1538-4357/ab698c}
\end{barticle}
\endbibitem

\bibitem[\protect\citeauthoryear{{Moore} et~al.}{2022}]{Moore+2022}
\begin{barticle}
\bauthor{\bsnm{{Moore}}, \binits{K.M.}},
\bauthor{\bsnm{{Barik}}, \binits{A.}},
\bauthor{\bsnm{{Stanley}}, \binits{S.}},
\bauthor{\bsnm{{Stevenson}}, \binits{D.J.}},
\bauthor{\bsnm{{Nettelmann}}, \binits{N.}},
\bauthor{\bsnm{{Helled}}, \binits{R.}},
\bauthor{\bsnm{{Guillot}}, \binits{T.}},
\bauthor{\bsnm{{Militzer}}, \binits{B.}},
\bauthor{\bsnm{{Bolton}}, \binits{S.}}:
\batitle{{Dynamo Simulations of Jupiter's Magnetic Field: The Role of Stable
  Stratification and a Dilute Core}}.
\bjtitle{Journal of Geophysical Research (Planets)}
\bvolume{127}(\bissue{11}),
\bfpage{2022}--\blpage{007479}
(\byear{2022})
\doiurl{10.1029/2022JE007479}
\end{barticle}
\endbibitem

\bibitem[\protect\citeauthoryear{{Connerney} et~al.}{2022}]{Connerney+2022}
\begin{barticle}
\bauthor{\bsnm{{Connerney}}, \binits{J.E.P.}},
\bauthor{\bsnm{{Timmins}}, \binits{S.}},
\bauthor{\bsnm{{Oliversen}}, \binits{R.J.}},
\bauthor{\bsnm{{Espley}}, \binits{J.R.}},
\bauthor{\bsnm{{Joergensen}}, \binits{J.L.}},
\bauthor{\bsnm{{Kotsiaros}}, \binits{S.}},
\bauthor{\bsnm{{Joergensen}}, \binits{P.S.}},
\bauthor{\bsnm{{Merayo}}, \binits{J.M.G.}},
\bauthor{\bsnm{{Herceg}}, \binits{M.}},
\bauthor{\bsnm{{Bloxham}}, \binits{J.}},
\bauthor{\bsnm{{Moore}}, \binits{K.M.}},
\bauthor{\bsnm{{Mura}}, \binits{A.}},
\bauthor{\bsnm{{Moirano}}, \binits{A.}},
\bauthor{\bsnm{{Bolton}}, \binits{S.J.}},
\bauthor{\bsnm{{Levin}}, \binits{S.M.}}:
\batitle{{A New Model of Jupiter's Magnetic Field at the Completion of Juno's
  Prime Mission}}.
\bjtitle{Journal of Geophysical Research (Planets)}
\bvolume{127}(\bissue{2}),
\bfpage{07055}
(\byear{2022})
\doiurl{10.1029/2021JE007055}
\end{barticle}
\endbibitem

\bibitem[\protect\citeauthoryear{{Sharan} et~al.}{2022}]{Sharan+2022}
\begin{barticle}
\bauthor{\bsnm{{Sharan}}, \binits{S.}},
\bauthor{\bsnm{{Langlais}}, \binits{B.}},
\bauthor{\bsnm{{Amit}}, \binits{H.}},
\bauthor{\bsnm{{Th{\'e}bault}}, \binits{E.}},
\bauthor{\bsnm{{Pinceloup}}, \binits{M.}},
\bauthor{\bsnm{{Verhoeven}}, \binits{O.}}:
\batitle{{The Internal Structure and Dynamics of Jupiter Unveiled by a
  High-Resolution Magnetic Field and Secular Variation Model}}.
\bjtitle{\grl}
\bvolume{49}(\bissue{15}),
\bfpage{98839}
(\byear{2022})
\doiurl{10.1029/2022GL098839}
\end{barticle}
\endbibitem

\bibitem[\protect\citeauthoryear{{Bolton} et~al.}{2021}]{Bolton+2021}
\begin{barticle}
\bauthor{\bsnm{{Bolton}}, \binits{S.J.}},
\bauthor{\bsnm{{Levin}}, \binits{S.M.}},
\bauthor{\bsnm{{Guillot}}, \binits{T.}},
\bauthor{\bsnm{{Li}}, \binits{C.}},
\bauthor{\bsnm{{Kaspi}}, \binits{Y.}},
\bauthor{\bsnm{{Orton}}, \binits{G.}},
\bauthor{\bsnm{{Wong}}, \binits{M.H.}},
\bauthor{\bsnm{{Oyafuso}}, \binits{F.}},
\bauthor{\bsnm{{Allison}}, \binits{M.}},
\bauthor{\bsnm{{Arballo}}, \binits{J.}},
\bauthor{\bsnm{{Atreya}}, \binits{S.}},
\bauthor{\bsnm{{Becker}}, \binits{H.N.}},
\bauthor{\bsnm{{Bloxham}}, \binits{J.}},
\bauthor{\bsnm{{Brown}}, \binits{S.T.}},
\bauthor{\bsnm{{Fletcher}}, \binits{L.N.}},
\bauthor{\bsnm{{Galanti}}, \binits{E.}},
\bauthor{\bsnm{{Gulkis}}, \binits{S.}},
\bauthor{\bsnm{{Janssen}}, \binits{M.}},
\bauthor{\bsnm{{Ingersoll}}, \binits{A.}},
\bauthor{\bsnm{{Lunine}}, \binits{J.L.}},
\bauthor{\bsnm{{Misra}}, \binits{S.}},
\bauthor{\bsnm{{Steffes}}, \binits{P.}},
\bauthor{\bsnm{{Stevenson}}, \binits{D.}},
\bauthor{\bsnm{{Waite}}, \binits{J.H.}},
\bauthor{\bsnm{{Yadav}}, \binits{R.K.}},
\bauthor{\bsnm{{Zhang}}, \binits{Z.}}:
\batitle{{Microwave observations reveal the deep extent and structure of
  Jupiter{\textquoteright}s atmospheric vortices}}.
\bjtitle{Science}
\bvolume{374}(\bissue{6570}),
\bfpage{968}--\blpage{972}
(\byear{2021})
\doiurl{10.1126/science.abf1015}
\end{barticle}
\endbibitem

\bibitem[\protect\citeauthoryear{{Johnson} et~al.}{1984}]{Johnsonetal1984}
\begin{barticle}
\bauthor{\bsnm{{Johnson}}, \binits{T.V.}},
\bauthor{\bsnm{{Morrison}}, \binits{D.}},
\bauthor{\bsnm{{Matson}}, \binits{D.L.}},
\bauthor{\bsnm{{Veeder}}, \binits{G.J.}},
\bauthor{\bsnm{{Brown}}, \binits{R.H.}},
\bauthor{\bsnm{{Nelson}}, \binits{R.M.}}:
\batitle{{Volcanic hotspots on Io - Stability and longitudinal distribution}}.
\bjtitle{Science}
\bvolume{226},
\bfpage{134}--\blpage{137}
(\byear{1984})
\doiurl{10.1126/science.226.4671.134}
\end{barticle}
\endbibitem

\bibitem[\protect\citeauthoryear{{Ojakangas} and
  {Stevenson}}{1986}]{OjakangasStevenson1986}
\begin{barticle}
\bauthor{\bsnm{{Ojakangas}}, \binits{G.W.}},
\bauthor{\bsnm{{Stevenson}}, \binits{D.J.}}:
\batitle{{Episodic volcanism of tidally heated satellites with application to
  Io}}.
\bjtitle{Icarus}
\bvolume{66}(\bissue{2}),
\bfpage{341}--\blpage{358}
(\byear{1986})
\doiurl{10.1016/0019-1035(86)90163-6}
\end{barticle}
\endbibitem

\bibitem[\protect\citeauthoryear{{Winn} and
  {Fabrycky}}{2015}]{WinnFabrycky2015}
\begin{barticle}
\bauthor{\bsnm{{Winn}}, \binits{J.N.}},
\bauthor{\bsnm{{Fabrycky}}, \binits{D.C.}}:
\batitle{{The Occurrence and Architecture of Exoplanetary Systems}}.
\bjtitle{\araa}
\bvolume{53},
\bfpage{409}--\blpage{447}
(\byear{2015})
\doiurl{10.1146/annurev-astro-082214-122246}
{\href{https://arxiv.org/abs/1410.4199}{{arXiv:1410.4199}}}
{[astro-ph.EP]}
\end{barticle}
\endbibitem

\bibitem[\protect\citeauthoryear{{Murray} and
  {Dermott}}{1999}]{MurrayDermott1999}
\begin{bbook}
\bauthor{\bsnm{{Murray}}, \binits{C.D.}},
\bauthor{\bsnm{{Dermott}}, \binits{S.F.}}:
\bbtitle{{Solar System Dynamics, Cambridge University Press}},
(\byear{1999})
\end{bbook}
\endbibitem

\bibitem[\protect\citeauthoryear{{Kaula}}{1962}]{Kaula1962}
\begin{barticle}
\bauthor{\bsnm{{Kaula}}, \binits{W.M.}}:
\batitle{{Development of the lunar and solar disturbing functions for a close
  satellite}}.
\bjtitle{\aj}
\bvolume{67},
\bfpage{300}
(\byear{1962})
\doiurl{10.1086/108729}
\end{barticle}
\endbibitem

\bibitem[\protect\citeauthoryear{{Mathis} and {Le
  Poncin-Lafitte}}{2009}]{MathisLePoncinLafitte2009}
\begin{barticle}
\bauthor{\bsnm{{Mathis}}, \binits{S.}},
\bauthor{\bsnm{{Le Poncin-Lafitte}}, \binits{C.}}:
\batitle{{Tidal dynamics of extended bodies in planetary systems and multiple
  stars}}.
\bjtitle{\aap}
\bvolume{497},
\bfpage{889}--\blpage{910}
(\byear{2009})
\doiurl{10.1051/0004-6361/20079054}
\end{barticle}
\endbibitem

\bibitem[\protect\citeauthoryear{{Ogilvie}}{2014}]{Ogilvie2014}
\begin{barticle}
\bauthor{\bsnm{{Ogilvie}}, \binits{G.I.}}:
\batitle{{Tidal Dissipation in Stars and Giant Planets}}.
\bjtitle{\araa}
\bvolume{52},
\bfpage{171}--\blpage{210}
(\byear{2014})
\doiurl{10.1146/annurev-astro-081913-035941}
{\href{https://arxiv.org/abs/1406.2207}{{arXiv:1406.2207}}}
{[astro-ph.SR]}
\end{barticle}
\endbibitem

\bibitem[\protect\citeauthoryear{{Zahn}}{1966}]{Zahn1966a}
\begin{barticle}
\bauthor{\bsnm{{Zahn}}, \binits{J.P.}}:
\batitle{{Les mar{\'e}es dans une {\'e}toile double serr{\'e}e}}.
\bjtitle{Annales d'Astrophysique}
\bvolume{29},
\bfpage{313}
(\byear{1966})
\end{barticle}
\endbibitem

\bibitem[\protect\citeauthoryear{{Remus} et~al.}{2012}]{RMZ2012}
\begin{barticle}
\bauthor{\bsnm{{Remus}}, \binits{F.}},
\bauthor{\bsnm{{Mathis}}, \binits{S.}},
\bauthor{\bsnm{{Zahn}}, \binits{J.-P.}}:
\batitle{{The equilibrium tide in stars and giant planets. I. The coplanar
  case}}.
\bjtitle{\aap}
\bvolume{544},
\bfpage{132}
(\byear{2012})
\doiurl{10.1051/0004-6361/201118160}
{\href{https://arxiv.org/abs/1205.3536}{{arXiv:1205.3536}}}
{[astro-ph.SR]}
\end{barticle}
\endbibitem

\bibitem[\protect\citeauthoryear{{Love}}{1911}]{Love1911}
\begin{bbook}
\bauthor{\bsnm{{Love}}, \binits{A.E.H.}}:
\bbtitle{{Some Problems of Geodynamics}},
(\byear{1911})
\end{bbook}
\endbibitem

\bibitem[\protect\citeauthoryear{{Tobie} et~al.}{2005}]{Tobieetal2005}
\begin{barticle}
\bauthor{\bsnm{{Tobie}}, \binits{G.}},
\bauthor{\bsnm{{Grasset}}, \binits{O.}},
\bauthor{\bsnm{{Lunine}}, \binits{J.I.}},
\bauthor{\bsnm{{Mocquet}}, \binits{A.}},
\bauthor{\bsnm{{Sotin}}, \binits{C.}}:
\batitle{{Titan's internal structure inferred from a coupled thermal-orbital
  model}}.
\bjtitle{Icarus}
\bvolume{175},
\bfpage{496}--\blpage{502}
(\byear{2005})
\doiurl{10.1016/j.icarus.2004.12.007}
\end{barticle}
\endbibitem

\bibitem[\protect\citeauthoryear{{Remus} et~al.}{2012}]{Remusetal2012}
\begin{barticle}
\bauthor{\bsnm{{Remus}}, \binits{F.}},
\bauthor{\bsnm{{Mathis}}, \binits{S.}},
\bauthor{\bsnm{{Zahn}}, \binits{J.-P.}},
\bauthor{\bsnm{{Lainey}}, \binits{V.}}:
\batitle{{Anelastic tidal dissipation in multi-layer planets}}.
\bjtitle{\aap}
\bvolume{541},
\bfpage{165}
(\byear{2012})
\doiurl{10.1051/0004-6361/201118595}
{\href{https://arxiv.org/abs/1204.1468}{{arXiv:1204.1468}}}
{[astro-ph.EP]}
\end{barticle}
\endbibitem

\bibitem[\protect\citeauthoryear{{Mathis} et~al.}{2013}]{Mathisetal2013}
\begin{bchapter}
\bauthor{\bsnm{{Mathis}}, \binits{S.}},
\bauthor{\bsnm{{Le Poncin-Lafitte}}, \binits{C.}},
\bauthor{\bsnm{{Remus}}, \binits{F.}}:
\bctitle{{Tides in Planetary Systems}}.
In: \beditor{\bsnm{{Souchay}}, \binits{J.}},
\beditor{\bsnm{{Mathis}}, \binits{S.}},
\beditor{\bsnm{{Tokieda}}, \binits{T.}} (eds.)
\bbtitle{Lecture Notes in Physics, Berlin Springer Verlag}.
\bsertitle{Lecture Notes in Physics, Berlin Springer Verlag},
vol. \bseriesno{861},
p. \bfpage{255}
(\byear{2013}).
\doiurl{10.1007/978-3-642-32961-6_7}
\end{bchapter}
\endbibitem

\bibitem[\protect\citeauthoryear{{Rieutord}}{2015}]{RieutordSpringer}
\begin{bbook}
\bauthor{\bsnm{{Rieutord}}, \binits{M.}}:
\bbtitle{{Fluid Dynamics: An Introduction, Springer-Verlag}},
(\byear{2015}).
\doiurl{10.1007/978-3-319-09351-2}
\end{bbook}
\endbibitem

\bibitem[\protect\citeauthoryear{{Alterman} et~al.}{1959}]{Altermanetal1959}
\begin{barticle}
\bauthor{\bsnm{{Alterman}}, \binits{Z.}},
\bauthor{\bsnm{{Jarosch}}, \binits{H.}},
\bauthor{\bsnm{{Pekeris}}, \binits{C.L.}}:
\batitle{{Oscillations of the Earth}}.
\bjtitle{Proceedings of the Royal Society of London Series A}
\bvolume{252},
\bfpage{80}--\blpage{95}
(\byear{1959})
\doiurl{10.1098/rspa.1959.0138}
\end{barticle}
\endbibitem

\bibitem[\protect\citeauthoryear{{Mathis} and {de Brye}}{2011}]{Mathis2011}
\begin{barticle}
\bauthor{\bsnm{{Mathis}}, \binits{S.}},
\bauthor{\bsnm{{de Brye}}, \binits{N.}}:
\batitle{{Low-frequency internal waves in magnetized rotating stellar radiation
  zones. I. Wave structure modification by a toroidal field}}.
\bjtitle{\aap}
\bvolume{526},
\bfpage{65}
(\byear{2011})
\doiurl{10.1051/0004-6361/201015571}
\end{barticle}
\endbibitem

\bibitem[\protect\citeauthoryear{{Zahn}}{2013}]{Zahn2013}
\begin{bchapter}
\bauthor{\bsnm{{Zahn}}, \binits{J.-P.}}:
\bctitle{{Stellar Tides}}.
In: \beditor{\bsnm{{Souchay}}, \binits{J.}},
\beditor{\bsnm{{Mathis}}, \binits{S.}},
\beditor{\bsnm{{Tokieda}}, \binits{T.}} (eds.)
\bbtitle{Lecture Notes in Physics, Berlin Springer Verlag}.
\bsertitle{Lecture Notes in Physics, Berlin Springer Verlag},
vol. \bseriesno{861},
p. \bfpage{301}
(\byear{2013}).
\doiurl{10.1007/978-3-642-32961-6_8}
\end{bchapter}
\endbibitem

\bibitem[\protect\citeauthoryear{{MacDonald}}{1964}]{MacDonald1964}
\begin{barticle}
\bauthor{\bsnm{{MacDonald}}, \binits{G.J.F.}}:
\batitle{{Tidal Friction}}.
\bjtitle{Reviews of Geophysics and Space Physics}
\bvolume{2},
\bfpage{467}--\blpage{541}
(\byear{1964})
\doiurl{10.1029/RG002i003p00467}
\end{barticle}
\endbibitem

\bibitem[\protect\citeauthoryear{{Greenberg}}{2009}]{Greenberg2009}
\begin{barticle}
\bauthor{\bsnm{{Greenberg}}, \binits{R.}}:
\batitle{{Frequency Dependence of Tidal q}}.
\bjtitle{\apjl}
\bvolume{698},
\bfpage{42}--\blpage{45}
(\byear{2009})
\doiurl{10.1088/0004-637X/698/1/L42}
\end{barticle}
\endbibitem

\bibitem[\protect\citeauthoryear{{Hut}}{1980}]{Hut1980}
\begin{barticle}
\bauthor{\bsnm{{Hut}}, \binits{P.}}:
\batitle{{Stability of tidal equilibrium}}.
\bjtitle{\aap}
\bvolume{92},
\bfpage{167}--\blpage{170}
(\byear{1980})
\end{barticle}
\endbibitem

\bibitem[\protect\citeauthoryear{{Zahn} and {Bouchet}}{1989}]{ZahnBouchet1989}
\begin{barticle}
\bauthor{\bsnm{{Zahn}}, \binits{J.-P.}},
\bauthor{\bsnm{{Bouchet}}, \binits{L.}}:
\batitle{{Tidal evolution of close binary stars. II - Orbital circularization
  of late-type binaries}}.
\bjtitle{\aap}
\bvolume{223},
\bfpage{112}--\blpage{118}
(\byear{1989})
\end{barticle}
\endbibitem

\bibitem[\protect\citeauthoryear{{Witte} and
  {Savonije}}{2002}]{WitteSavonije2002}
\begin{barticle}
\bauthor{\bsnm{{Witte}}, \binits{M.G.}},
\bauthor{\bsnm{{Savonije}}, \binits{G.J.}}:
\batitle{{Orbital evolution by dynamical tides in solar type stars. Application
  to binary stars and planetary orbits}}.
\bjtitle{\aap}
\bvolume{386},
\bfpage{222}--\blpage{236}
(\byear{2002})
\doiurl{10.1051/0004-6361:20020155}
\end{barticle}
\endbibitem

\bibitem[\protect\citeauthoryear{{Efroimsky} and
  {Lainey}}{2007}]{EfroimskyLainey2007}
\begin{barticle}
\bauthor{\bsnm{{Efroimsky}}, \binits{M.}},
\bauthor{\bsnm{{Lainey}}, \binits{V.}}:
\batitle{{Physics of bodily tides in terrestrial planets and the appropriate
  scales of dynamical evolution}}.
\bjtitle{Journal of Geophysical Research (Planets)}
\bvolume{112}(\bissue{E11}),
\bfpage{12003}
(\byear{2007})
\doiurl{10.1029/2007JE002908}
{\href{https://arxiv.org/abs/0709.1995}{{arXiv:0709.1995}}}
\end{barticle}
\endbibitem

\bibitem[\protect\citeauthoryear{{Auclair-Desrotour} et~al.}{2014}]{ADLPM2014}
\begin{barticle}
\bauthor{\bsnm{{Auclair-Desrotour}}, \binits{P.}},
\bauthor{\bsnm{{Le Poncin-Lafitte}}, \binits{C.}},
\bauthor{\bsnm{{Mathis}}, \binits{S.}}:
\batitle{{Impact of the frequency dependence of tidal Q on the evolution of
  planetary systems}}.
\bjtitle{\aap}
\bvolume{561},
\bfpage{7}
(\byear{2014})
\doiurl{10.1051/0004-6361/201322782}
{\href{https://arxiv.org/abs/1311.4810}{{arXiv:1311.4810}}}
{[astro-ph.EP]}
\end{barticle}
\endbibitem

\bibitem[\protect\citeauthoryear{{Pollack} et~al.}{1996}]{Pollacketal1996}
\begin{barticle}
\bauthor{\bsnm{{Pollack}}, \binits{J.B.}},
\bauthor{\bsnm{{Hubickyj}}, \binits{O.}},
\bauthor{\bsnm{{Bodenheimer}}, \binits{P.}},
\bauthor{\bsnm{{Lissauer}}, \binits{J.J.}},
\bauthor{\bsnm{{Podolak}}, \binits{M.}},
\bauthor{\bsnm{{Greenzweig}}, \binits{Y.}}:
\batitle{{Formation of the Giant Planets by Concurrent Accretion of Solids and
  Gas}}.
\bjtitle{Icarus}
\bvolume{124},
\bfpage{62}--\blpage{85}
(\byear{1996})
\doiurl{10.1006/icar.1996.0190}
\end{barticle}
\endbibitem

\bibitem[\protect\citeauthoryear{{Efroimsky} and
  {Makarov}}{2013}]{Efroimsky2013}
\begin{barticle}
\bauthor{\bsnm{{Efroimsky}}, \binits{M.}},
\bauthor{\bsnm{{Makarov}}, \binits{V.V.}}:
\batitle{{Tidal Friction and Tidal Lagging. Applicability Limitations of a
  Popular Formula for the Tidal Torque}}.
\bjtitle{Astronomy \& Astrophysics}
\bvolume{764}(\bissue{1}),
\bfpage{26}
(\byear{2013})
\doiurl{10.1088/0004-637X/764/1/26}
{\href{https://arxiv.org/abs/1209.1615}{{1209.1615}}}
\end{barticle}
\endbibitem

\bibitem[\protect\citeauthoryear{{Tobie} et~al.}{2019}]{tobie2019}
\begin{barticle}
\bauthor{\bsnm{{Tobie}}, \binits{G.}},
\bauthor{\bsnm{{Grasset}}, \binits{O.}},
\bauthor{\bsnm{{Dumoulin}}, \binits{C.}},
\bauthor{\bsnm{{Mocquet}}, \binits{A.}}:
\batitle{{Tidal response of rocky and ice-rich exoplanets}}.
\bjtitle{Astronomy \& Astrophysics}
\bvolume{630},
\bfpage{70}
(\byear{2019})
\doiurl{10.1051/0004-6361/201935297}
\end{barticle}
\endbibitem

\bibitem[\protect\citeauthoryear{{Dermott}}{1979}]{Dermott1979}
\begin{barticle}
\bauthor{\bsnm{{Dermott}}, \binits{S.F.}}:
\batitle{{Tidal dissipation in the solid cores of the major planets}}.
\bjtitle{Icarus}
\bvolume{37},
\bfpage{310}--\blpage{321}
(\byear{1979})
\doiurl{10.1016/0019-1035(79)90137-4}
\end{barticle}
\endbibitem

\bibitem[\protect\citeauthoryear{{Remus} et~al.}{2015}]{Remusetal2015}
\begin{barticle}
\bauthor{\bsnm{{Remus}}, \binits{F.}},
\bauthor{\bsnm{{Mathis}}, \binits{S.}},
\bauthor{\bsnm{{Zahn}}, \binits{J.-P.}},
\bauthor{\bsnm{{Lainey}}, \binits{V.}}:
\batitle{{The surface signature of the tidal dissipation of the core in a
  two-layer planet}}.
\bjtitle{\aap}
\bvolume{573},
\bfpage{23}
(\byear{2015})
\doiurl{10.1051/0004-6361/201424472}
{\href{https://arxiv.org/abs/1409.8343}{{arXiv:1409.8343}}}
{[astro-ph.EP]}
\end{barticle}
\endbibitem

\bibitem[\protect\citeauthoryear{{Ogilvie} and {Lin}}{2004}]{Ogilvie2004}
\begin{barticle}
\bauthor{\bsnm{{Ogilvie}}, \binits{G.I.}},
\bauthor{\bsnm{{Lin}}, \binits{D.N.C.}}:
\batitle{{Tidal Dissipation in Rotating Giant Planets}}.
\bjtitle{Astrophysical Journal}
\bvolume{610}(\bissue{1}),
\bfpage{477}--\blpage{509}
(\byear{2004})
\doiurl{10.1086/421454}
\end{barticle}
\endbibitem

\bibitem[\protect\citeauthoryear{{Mathis} et~al.}{2016}]{Mathisetal2016}
\begin{barticle}
\bauthor{\bsnm{{Mathis}}, \binits{S.}},
\bauthor{\bsnm{{Auclair-Desrotour}}, \binits{P.}},
\bauthor{\bsnm{{Guenel}}, \binits{M.}},
\bauthor{\bsnm{{Gallet}}, \binits{F.}},
\bauthor{\bsnm{{Le Poncin-Lafitte}}, \binits{C.}}:
\batitle{{The impact of rotation on turbulent tidal friction in stellar and
  planetary convective regions}}.
\bjtitle{\aap}
\bvolume{592},
\bfpage{33}
(\byear{2016})
\doiurl{10.1051/0004-6361/201527545}
{\href{https://arxiv.org/abs/1604.08570}{{arXiv:1604.08570}}}
{[astro-ph.SR]}
\end{barticle}
\endbibitem

\bibitem[\protect\citeauthoryear{{Crida} and {Charnoz}}{2012}]{crida:12}
\begin{barticle}
\bauthor{\bsnm{{Crida}}, \binits{A.}},
\bauthor{\bsnm{{Charnoz}}, \binits{S.}}:
\batitle{{Formation of Regular Satellites from Ancient Massive Rings in the
  Solar System}}.
\bjtitle{Science}
\bvolume{338},
\bfpage{1196}
(\byear{2012})
\doiurl{10.1126/science.1226477}
{\href{https://arxiv.org/abs/1301.3808}{{arXiv:1301.3808}}}
{[astro-ph.EP]}
\end{barticle}
\endbibitem

\bibitem[\protect\citeauthoryear{{Shoji} and
  {Hussmann}}{2017}]{ShojiHussmann2017}
\begin{barticle}
\bauthor{\bsnm{{Shoji}}, \binits{D.}},
\bauthor{\bsnm{{Hussmann}}, \binits{H.}}:
\batitle{{Frequency-dependent tidal dissipation in a viscoelastic Saturnian
  core and expansion of Mimas' semi-major axis}}.
\bjtitle{\aap}
\bvolume{599},
\bfpage{10}
(\byear{2017})
\doiurl{10.1051/0004-6361/201630230}
{\href{https://arxiv.org/abs/1612.03664}{{arXiv:1612.03664}}}
{[astro-ph.EP]}
\end{barticle}
\endbibitem

\bibitem[\protect\citeauthoryear{{Storch} and {Lai}}{2014}]{StorchLai2014}
\begin{barticle}
\bauthor{\bsnm{{Storch}}, \binits{N.I.}},
\bauthor{\bsnm{{Lai}}, \binits{D.}}:
\batitle{{Viscoelastic tidal dissipation in giant planets and formation of hot
  Jupiters through high-eccentricity migration}}.
\bjtitle{\mnras}
\bvolume{438},
\bfpage{1526}--\blpage{1534}
(\byear{2014})
\doiurl{10.1093/mnras/stt2292}
{\href{https://arxiv.org/abs/1308.4968}{{arXiv:1308.4968}}}
{[astro-ph.EP]}
\end{barticle}
\endbibitem

\bibitem[\protect\citeauthoryear{{Storch} and {Lai}}{2015}]{StorchLai2015}
\begin{barticle}
\bauthor{\bsnm{{Storch}}, \binits{N.I.}},
\bauthor{\bsnm{{Lai}}, \binits{D.}}:
\batitle{{Analytical model of tidal distortion and dissipation for a giant
  planet with a viscoelastic core}}.
\bjtitle{\mnras}
\bvolume{450},
\bfpage{3952}--\blpage{3957}
(\byear{2015})
\doiurl{10.1093/mnras/stv904}
{\href{https://arxiv.org/abs/1502.06550}{{arXiv:1502.06550}}}
{[astro-ph.EP]}
\end{barticle}
\endbibitem

\bibitem[\protect\citeauthoryear{{Ogilvie} and {Lin}}{2004}]{OgilvieLin2004}
\begin{barticle}
\bauthor{\bsnm{{Ogilvie}}, \binits{G.I.}},
\bauthor{\bsnm{{Lin}}, \binits{D.N.C.}}:
\batitle{{Tidal Dissipation in Rotating Giant Planets}}.
\bjtitle{\apj}
\bvolume{610},
\bfpage{477}--\blpage{509}
(\byear{2004})
\doiurl{10.1086/421454}
{\href{https://arxiv.org/abs/astro-ph/0310218}{{astro-ph/0310218}}}
\end{barticle}
\endbibitem

\bibitem[\protect\citeauthoryear{{Ogilvie}}{2005}]{Ogilvie2005}
\begin{barticle}
\bauthor{\bsnm{{Ogilvie}}, \binits{G.I.}}:
\batitle{{Wave attractors and the asymptotic dissipation rate of tidal
  disturbances}}.
\bjtitle{Journal of Fluid Mechanics}
\bvolume{543},
\bfpage{19}--\blpage{44}
(\byear{2005})
\doiurl{10.1017/S0022112005006580}
{\href{https://arxiv.org/abs/astro-ph/0506450}{{astro-ph/0506450}}}
\end{barticle}
\endbibitem

\bibitem[\protect\citeauthoryear{{Goodman} and
  {Lackner}}{2009}]{GoodmanLackner2009}
\begin{barticle}
\bauthor{\bsnm{{Goodman}}, \binits{J.}},
\bauthor{\bsnm{{Lackner}}, \binits{C.}}:
\batitle{{Dynamical Tides in Rotating Planets and Stars}}.
\bjtitle{\apj}
\bvolume{696},
\bfpage{2054}--\blpage{2067}
(\byear{2009})
\doiurl{10.1088/0004-637X/696/2/2054}
{\href{https://arxiv.org/abs/0812.1028}{{arXiv:0812.1028}}}
\end{barticle}
\endbibitem

\bibitem[\protect\citeauthoryear{{Rieutord} and
  {Valdettaro}}{2010}]{rieutord:10}
\begin{barticle}
\bauthor{\bsnm{{Rieutord}}, \binits{M.}},
\bauthor{\bsnm{{Valdettaro}}, \binits{L.}}:
\batitle{{Viscous dissipation by tidally forced inertial modes in a rotating
  spherical shell}}.
\bjtitle{Journal of Fluid Mechanics}
\bvolume{643},
\bfpage{363}--\blpage{394}
(\byear{2010})
\doiurl{10.1017/S002211200999214X}
{\href{https://arxiv.org/abs/0909.3922}{{arXiv:0909.3922}}}
{[astro-ph.SR]}
\end{barticle}
\endbibitem

\bibitem[\protect\citeauthoryear{{Stevenson}}{1979}]{Stevenson1979}
\begin{barticle}
\bauthor{\bsnm{{Stevenson}}, \binits{D.J.}}:
\batitle{{Turbulent thermal convection in the presence of rotation and a
  magnetic field - A heuristic theory}}.
\bjtitle{Geophysical and Astrophysical Fluid Dynamics}
\bvolume{12},
\bfpage{139}--\blpage{169}
(\byear{1979})
\doiurl{10.1080/03091927908242681}
\end{barticle}
\endbibitem

\bibitem[\protect\citeauthoryear{{Barker} et~al.}{2014}]{Barkeretal2014}
\begin{barticle}
\bauthor{\bsnm{{Barker}}, \binits{A.J.}},
\bauthor{\bsnm{{Dempsey}}, \binits{A.M.}},
\bauthor{\bsnm{{Lithwick}}, \binits{Y.}}:
\batitle{{Theory and Simulations of Rotating Convection}}.
\bjtitle{\apj}
\bvolume{791},
\bfpage{13}
(\byear{2014})
\doiurl{10.1088/0004-637X/791/1/13}
{\href{https://arxiv.org/abs/1403.7207}{{arXiv:1403.7207}}}
{[astro-ph.SR]}
\end{barticle}
\endbibitem

\bibitem[\protect\citeauthoryear{{Currie} et~al.}{2020}]{Currieetal2020}
\begin{barticle}
\bauthor{\bsnm{{Currie}}, \binits{L.K.}},
\bauthor{\bsnm{{Barker}}, \binits{A.J.}},
\bauthor{\bsnm{{Lithwick}}, \binits{Y.}},
\bauthor{\bsnm{{Browning}}, \binits{M.K.}}:
\batitle{{Convection with misaligned gravity and rotation: simulations and
  rotating mixing length theory}}.
\bjtitle{\mnras}
\bvolume{493}(\bissue{4}),
\bfpage{5233}--\blpage{5256}
(\byear{2020})
\doiurl{10.1093/mnras/staa372}
{\href{https://arxiv.org/abs/2002.02461}{{arXiv:2002.02461}}}
{[astro-ph.SR]}
\end{barticle}
\endbibitem

\bibitem[\protect\citeauthoryear{{Vasil} et~al.}{2021}]{Vasiletal2021}
\begin{barticle}
\bauthor{\bsnm{{Vasil}}, \binits{G.M.}},
\bauthor{\bsnm{{Julien}}, \binits{K.}},
\bauthor{\bsnm{{Featherstone}}, \binits{N.A.}}:
\batitle{{Rotation suppresses giant-scale solar convection}}.
\bjtitle{Proceedings of the National Academy of Science}
\bvolume{118}(\bissue{31}),
\bfpage{2022518118}
(\byear{2021})
\doiurl{10.1073/pnas.2022518118}
\end{barticle}
\endbibitem

\bibitem[\protect\citeauthoryear{{Fuentes} et~al.}{2023}]{Fuentesetal2023}
\begin{barticle}
\bauthor{\bsnm{{Fuentes}}, \binits{J.R.}},
\bauthor{\bsnm{{Anders}}, \binits{E.H.}},
\bauthor{\bsnm{{Cumming}}, \binits{A.}},
\bauthor{\bsnm{{Hindman}}, \binits{B.W.}}:
\batitle{{Rotation Reduces Convective Mixing in Jupiter and Other Gas Giants}}.
\bjtitle{\apjl}
\bvolume{950}(\bissue{1}),
\bfpage{4}
(\byear{2023})
\doiurl{10.3847/2041-8213/acd774}
{\href{https://arxiv.org/abs/2305.09921}{{arXiv:2305.09921}}}
{[astro-ph.EP]}
\end{barticle}
\endbibitem

\bibitem[\protect\citeauthoryear{{de Vries} et~al.}{2023}]{devriesetal2023}
\begin{barticle}
\bauthor{\bsnm{{de Vries}}, \binits{N.B.}},
\bauthor{\bsnm{{Barker}}, \binits{A.J.}},
\bauthor{\bsnm{{Hollerbach}}, \binits{R.}}:
\batitle{{Tidal dissipation due to the elliptical instability and turbulent
  viscosity in convection zones in rotating giant planets and stars}}.
\bjtitle{\mnras}
\bvolume{524}(\bissue{2}),
\bfpage{2661}--\blpage{2683}
(\byear{2023})
\doiurl{10.1093/mnras/stad1990}
{\href{https://arxiv.org/abs/2306.17622}{{arXiv:2306.17622}}}
{[astro-ph.EP]}
\end{barticle}
\endbibitem

\bibitem[\protect\citeauthoryear{{Auclair Desrotour} et~al.}{2015}]{ADMLP2015}
\begin{barticle}
\bauthor{\bsnm{{Auclair Desrotour}}, \binits{P.}},
\bauthor{\bsnm{{Mathis}}, \binits{S.}},
\bauthor{\bsnm{{Le Poncin-Lafitte}}, \binits{C.}}:
\batitle{{Scaling laws to understand tidal dissipation in fluid planetary
  regions and stars I. Rotation, stratification and thermal diffusivity}}.
\bjtitle{\aap}
\bvolume{581},
\bfpage{118}
(\byear{2015})
\doiurl{10.1051/0004-6361/201526246}
{\href{https://arxiv.org/abs/1506.07705}{{arXiv:1506.07705}}}
{[astro-ph.EP]}
\end{barticle}
\endbibitem

\bibitem[\protect\citeauthoryear{{Ogilvie}}{2013}]{ogilvie:13}
\begin{barticle}
\bauthor{\bsnm{{Ogilvie}}, \binits{G.I.}}:
\batitle{{Tides in rotating barotropic fluid bodies: the contribution of
  inertial waves and the role of internal structure}}.
\bjtitle{\mnras}
\bvolume{429},
\bfpage{613}--\blpage{632}
(\byear{2013})
\doiurl{10.1093/mnras/sts362}
{\href{https://arxiv.org/abs/1211.0837}{{arXiv:1211.0837}}}
{[astro-ph.EP]}
\end{barticle}
\endbibitem

\bibitem[\protect\citeauthoryear{{Lazovik} et~al.}{2024}]{lazovik:24}
\begin{barticle}
\bauthor{\bsnm{{Lazovik}}, \binits{Y.A.}},
\bauthor{\bsnm{{Barker}}, \binits{A.J.}},
\bauthor{\bsnm{{de Vries}}, \binits{N.B.}},
\bauthor{\bsnm{{Astoul}}, \binits{A.}}:
\batitle{{Tidal dissipation in rotating and evolving giant planets with
  application to exoplanet systems}}.
\bjtitle{\mnras}
\bvolume{527}(\bissue{3}),
\bfpage{8245}--\blpage{8256}
(\byear{2024})
\doiurl{10.1093/mnras/stad3689}
{\href{https://arxiv.org/abs/2311.15815}{{arXiv:2311.15815}}}
{[astro-ph.EP]}
\end{barticle}
\endbibitem

\bibitem[\protect\citeauthoryear{{Ivanov} and
  {Papaloizou}}{2004}]{Papaloizou:04}
\begin{barticle}
\bauthor{\bsnm{{Ivanov}}, \binits{P.B.}},
\bauthor{\bsnm{{Papaloizou}}, \binits{J.C.B.}}:
\batitle{{On the tidal interaction of massive extrasolar planets on highly
  eccentric orbits}}.
\bjtitle{\mnras}
\bvolume{347}(\bissue{2}),
\bfpage{437}--\blpage{453}
(\byear{2004})
\doiurl{10.1111/j.1365-2966.2004.07238.x}
{\href{https://arxiv.org/abs/astro-ph/0303669}{{arXiv:astro-ph/0303669}}}
{[astro-ph]}
\end{barticle}
\endbibitem

\bibitem[\protect\citeauthoryear{{Ivanov} and {Papaloizou}}{2007}]{ivanov:07}
\begin{barticle}
\bauthor{\bsnm{{Ivanov}}, \binits{P.B.}},
\bauthor{\bsnm{{Papaloizou}}, \binits{J.C.B.}}:
\batitle{{Dynamic tides in rotating objects: orbital circularization of
  extrasolar planets for realistic planet models}}.
\bjtitle{\mnras}
\bvolume{376},
\bfpage{682}--\blpage{704}
(\byear{2007})
\doiurl{10.1111/j.1365-2966.2007.11463.x}
{\href{https://arxiv.org/abs/astro-ph/0512150}{{astro-ph/0512150}}}
\end{barticle}
\endbibitem

\bibitem[\protect\citeauthoryear{{Papaloizou} and
  {Ivanov}}{2010}]{papaloizou:10}
\begin{barticle}
\bauthor{\bsnm{{Papaloizou}}, \binits{J.C.B.}},
\bauthor{\bsnm{{Ivanov}}, \binits{P.B.}}:
\batitle{{Dynamic tides in rotating objects: a numerical investigation of
  inertial waves in fully convective or barotropic stars and planets}}.
\bjtitle{\mnras}
\bvolume{407},
\bfpage{1631}--\blpage{1656}
(\byear{2010})
\doiurl{10.1111/j.1365-2966.2010.17011.x}
{\href{https://arxiv.org/abs/1005.2397}{{arXiv:1005.2397}}}
{[astro-ph.SR]}
\end{barticle}
\endbibitem

\bibitem[\protect\citeauthoryear{{Guenel} et~al.}{2014}]{Gueneletal2014}
\begin{barticle}
\bauthor{\bsnm{{Guenel}}, \binits{M.}},
\bauthor{\bsnm{{Mathis}}, \binits{S.}},
\bauthor{\bsnm{{Remus}}, \binits{F.}}:
\batitle{{Unravelling tidal dissipation in gaseous giant planets}}.
\bjtitle{\aap}
\bvolume{566},
\bfpage{9}
(\byear{2014})
\doiurl{10.1051/0004-6361/201424010}
{\href{https://arxiv.org/abs/1406.1672}{{arXiv:1406.1672}}}
{[astro-ph.EP]}
\end{barticle}
\endbibitem

\bibitem[\protect\citeauthoryear{{Ogilvie}}{2013}]{Ogilvie2013}
\begin{barticle}
\bauthor{\bsnm{{Ogilvie}}, \binits{G.I.}}:
\batitle{{Tides in rotating barotropic fluid bodies: the contribution of
  inertial waves and the role of internal structure}}.
\bjtitle{\mnras}
\bvolume{429},
\bfpage{613}--\blpage{632}
(\byear{2013})
\doiurl{10.1093/mnras/sts362}
{\href{https://arxiv.org/abs/1211.0837}{{arXiv:1211.0837}}}
{[astro-ph.EP]}
\end{barticle}
\endbibitem

\bibitem[\protect\citeauthoryear{{Wu}}{2005a}]{Wu2005a}
\begin{barticle}
\bauthor{\bsnm{{Wu}}, \binits{Y.}}:
\batitle{{Origin of Tidal Dissipation in Jupiter. I. Properties of Inertial
  Modes}}.
\bjtitle{\apj}
\bvolume{635},
\bfpage{674}--\blpage{687}
(\byear{2005})
\doiurl{10.1086/497354}
{\href{https://arxiv.org/abs/astro-ph/0407627}{{astro-ph/0407627}}}
\end{barticle}
\endbibitem

\bibitem[\protect\citeauthoryear{{Wu}}{2005b}]{Wu2005b}
\begin{barticle}
\bauthor{\bsnm{{Wu}}, \binits{Y.}}:
\batitle{{Origin of Tidal Dissipation in Jupiter. II. The Value of Q}}.
\bjtitle{\apj}
\bvolume{635},
\bfpage{688}--\blpage{710}
(\byear{2005})
\doiurl{10.1086/497355}
{\href{https://arxiv.org/abs/astro-ph/0407628}{{astro-ph/0407628}}}
\end{barticle}
\endbibitem

\bibitem[\protect\citeauthoryear{{Mathis}}{2015}]{Mathis2015}
\begin{barticle}
\bauthor{\bsnm{{Mathis}}, \binits{S.}}:
\batitle{{Variation of tidal dissipation in the convective envelope of low-mass
  stars along their evolution}}.
\bjtitle{\aap}
\bvolume{580},
\bfpage{3}
(\byear{2015})
\doiurl{10.1051/0004-6361/201526472}
{\href{https://arxiv.org/abs/1507.00165}{{arXiv:1507.00165}}}
{[astro-ph.SR]}
\end{barticle}
\endbibitem

\bibitem[\protect\citeauthoryear{{Ioannou} and {Lindzen}}{1993a}]{ioannou:93a}
\begin{barticle}
\bauthor{\bsnm{{Ioannou}}, \binits{P.J.}},
\bauthor{\bsnm{{Lindzen}}, \binits{R.S.}}:
\batitle{{Gravitational tides in the outer planets. I - Implications of
  classical tidal theory. II - Interior calculations and estimation of the
  tidal dissipation factor}}.
\bjtitle{\apj}
\bvolume{406},
\bfpage{252}--\blpage{278}
(\byear{1993})
\doiurl{10.1086/172437}
\end{barticle}
\endbibitem

\bibitem[\protect\citeauthoryear{{Ioannou} and {Lindzen}}{1993b}]{ioannou:93b}
\begin{barticle}
\bauthor{\bsnm{{Ioannou}}, \binits{P.J.}},
\bauthor{\bsnm{{Lindzen}}, \binits{R.S.}}:
\batitle{{Gravitational Tides in the Outer Planets. II. Interior Calculations
  and Estimation of the Tidal Dissipation Factor}}.
\bjtitle{\apj}
\bvolume{406},
\bfpage{266}
(\byear{1993})
\doiurl{10.1086/172438}
\end{barticle}
\endbibitem

\bibitem[\protect\citeauthoryear{{Andr{\'e}} et~al.}{2017}]{Andreetal2017}
\begin{botherref}
\oauthor{\bsnm{{Andr{\'e}}}, \binits{Q.}},
\oauthor{\bsnm{{Barker}}, \binits{A.J.}},
\oauthor{\bsnm{{Mathis}}, \binits{S.}}:
{Layered semi-convection and tides in giant planet interiors - I. Propagation
  of internal waves}.
ArXiv e-prints
(2017)
{\href{https://arxiv.org/abs/1704.08974}{{arXiv:1704.08974}}}
{[astro-ph.EP]}
\end{botherref}
\endbibitem

\bibitem[\protect\citeauthoryear{{Andr{\'e}} et~al.}{2019}]{Andreetal2019}
\begin{barticle}
\bauthor{\bsnm{{Andr{\'e}}}, \binits{Q.}},
\bauthor{\bsnm{{Mathis}}, \binits{S.}},
\bauthor{\bsnm{{Barker}}, \binits{A.J.}}:
\batitle{{Layered semi-convection and tides in giant planet interiors. II.
  Tidal dissipation}}.
\bjtitle{\aap}
\bvolume{626},
\bfpage{82}
(\byear{2019})
\doiurl{10.1051/0004-6361/201833674}
{\href{https://arxiv.org/abs/1902.04848}{{arXiv:1902.04848}}}
{[astro-ph.EP]}
\end{barticle}
\endbibitem

\bibitem[\protect\citeauthoryear{{Pontin} et~al.}{2020}]{Pontinetal2020}
\begin{barticle}
\bauthor{\bsnm{{Pontin}}, \binits{C.M.}},
\bauthor{\bsnm{{Barker}}, \binits{A.J.}},
\bauthor{\bsnm{{Hollerbach}}, \binits{R.}},
\bauthor{\bsnm{{Andr{\'e}}}, \binits{Q.}},
\bauthor{\bsnm{{Mathis}}, \binits{S.}}:
\batitle{{Wave propagation in semiconvective regions of giant planets}}.
\bjtitle{\mnras}
\bvolume{493}(\bissue{4}),
\bfpage{5788}--\blpage{5806}
(\byear{2020})
\doiurl{10.1093/mnras/staa664}
{\href{https://arxiv.org/abs/2003.02595}{{arXiv:2003.02595}}}
{[astro-ph.EP]}
\end{barticle}
\endbibitem

\bibitem[\protect\citeauthoryear{{Pontin} et~al.}{2023}]{Pontinetal2023a}
\begin{barticle}
\bauthor{\bsnm{{Pontin}}, \binits{C.M.}},
\bauthor{\bsnm{{Barker}}, \binits{A.J.}},
\bauthor{\bsnm{{Hollerbach}}, \binits{R.}}:
\batitle{{Tidal Dissipation in Stratified and Semi-convective Regions of Giant
  Planets}}.
\bjtitle{\apj}
\bvolume{950}(\bissue{2}),
\bfpage{176}
(\byear{2023})
\doiurl{10.3847/1538-4357/accd67}
{\href{https://arxiv.org/abs/2304.11898}{{arXiv:2304.11898}}}
{[astro-ph.EP]}
\end{barticle}
\endbibitem

\bibitem[\protect\citeauthoryear{{Lin}}{2023}]{Lin2023}
\begin{barticle}
\bauthor{\bsnm{{Lin}}, \binits{Y.}}:
\batitle{{Dynamical tides in Jupiter and the role of interior structure}}.
\bjtitle{\aap}
\bvolume{671},
\bfpage{37}
(\byear{2023})
\doiurl{10.1051/0004-6361/202245112}
{\href{https://arxiv.org/abs/2301.02418}{{arXiv:2301.02418}}}
{[astro-ph.EP]}
\end{barticle}
\endbibitem

\bibitem[\protect\citeauthoryear{{Dewberry}}{2023}]{Dewberry2023}
\begin{barticle}
\bauthor{\bsnm{{Dewberry}}, \binits{J.W.}}:
\batitle{{Dynamical tides in Jupiter and other rotationally flattened planets
  and stars with stable stratification}}.
\bjtitle{\mnras}
\bvolume{521}(\bissue{4}),
\bfpage{5991}--\blpage{6004}
(\byear{2023})
\doiurl{10.1093/mnras/stad546}
{\href{https://arxiv.org/abs/2301.07097}{{arXiv:2301.07097}}}
{[astro-ph.EP]}
\end{barticle}
\endbibitem

\bibitem[\protect\citeauthoryear{{Fuller} et~al.}{2016}]{Fuller2016}
\begin{barticle}
\bauthor{\bsnm{{Fuller}}, \binits{J.}},
\bauthor{\bsnm{{Luan}}, \binits{J.}},
\bauthor{\bsnm{{Quataert}}, \binits{E.}}:
\batitle{{Resonance locking as the source of rapid tidal migration in the
  Jupiter and Saturn moon systems}}.
\bjtitle{Monthly Notices of the Royal Astronomical Society}
\bvolume{458}(\bissue{4}),
\bfpage{3867}--\blpage{3879}
(\byear{2016})
\doiurl{10.1093/mnras/stw609}
{\href{https://arxiv.org/abs/1601.05804}{{arXiv:1601.05804}}}
{[astro-ph.EP]}
\end{barticle}
\endbibitem

\bibitem[\protect\citeauthoryear{{Luan} et~al.}{2018}]{Luanetal2018}
\begin{barticle}
\bauthor{\bsnm{{Luan}}, \binits{J.}},
\bauthor{\bsnm{{Fuller}}, \binits{J.}},
\bauthor{\bsnm{{Quataert}}, \binits{E.}}:
\batitle{{How Cassini can constrain tidal dissipation in Saturn}}.
\bjtitle{\mnras}
\bvolume{473},
\bfpage{5002}--\blpage{5014}
(\byear{2018})
\doiurl{10.1093/mnras/stx2714}
{\href{https://arxiv.org/abs/1707.02519}{{arXiv:1707.02519}}}
{[astro-ph.EP]}
\end{barticle}
\endbibitem

\bibitem[\protect\citeauthoryear{{Barker} and
  {Ogilvie}}{2010}]{BarkerOgilvie2010}
\begin{barticle}
\bauthor{\bsnm{{Barker}}, \binits{A.J.}},
\bauthor{\bsnm{{Ogilvie}}, \binits{G.I.}}:
\batitle{{On internal wave breaking and tidal dissipation near the centre of a
  solar-type star}}.
\bjtitle{\mnras}
\bvolume{404},
\bfpage{1849}--\blpage{1868}
(\byear{2010})
\doiurl{10.1111/j.1365-2966.2010.16400.x}
{\href{https://arxiv.org/abs/1001.4009}{{arXiv:1001.4009}}}
{[astro-ph.EP]}
\end{barticle}
\endbibitem

\bibitem[\protect\citeauthoryear{{Press}}{1981}]{Press1981}
\begin{barticle}
\bauthor{\bsnm{{Press}}, \binits{W.H.}}:
\batitle{{Radiative and other effects from internal waves in solar and stellar
  interiors}}.
\bjtitle{\apj}
\bvolume{245},
\bfpage{286}--\blpage{303}
(\byear{1981})
\doiurl{10.1086/158809}
\end{barticle}
\endbibitem

\bibitem[\protect\citeauthoryear{{Rogers} et~al.}{2013}]{Rogersetal2013}
\begin{barticle}
\bauthor{\bsnm{{Rogers}}, \binits{T.M.}},
\bauthor{\bsnm{{Lin}}, \binits{D.N.C.}},
\bauthor{\bsnm{{McElwaine}}, \binits{J.N.}},
\bauthor{\bsnm{{Lau}}, \binits{H.H.B.}}:
\batitle{{Internal Gravity Waves in Massive Stars: Angular Momentum
  Transport}}.
\bjtitle{\apj}
\bvolume{772}(\bissue{1}),
\bfpage{21}
(\byear{2013})
\doiurl{10.1088/0004-637X/772/1/21}
{\href{https://arxiv.org/abs/1306.3262}{{arXiv:1306.3262}}}
{[astro-ph.SR]}
\end{barticle}
\endbibitem

\bibitem[\protect\citeauthoryear{{Weinberg} et~al.}{2012}]{Weinberg2012}
\begin{barticle}
\bauthor{\bsnm{{Weinberg}}, \binits{N.N.}},
\bauthor{\bsnm{{Arras}}, \binits{P.}},
\bauthor{\bsnm{{Quataert}}, \binits{E.}},
\bauthor{\bsnm{{Burkart}}, \binits{J.}}:
\batitle{{Nonlinear Tides in Close Binary Systems}}.
\bjtitle{\apj}
\bvolume{751},
\bfpage{136}
(\byear{2012})
\doiurl{10.1088/0004-637X/751/2/136}
{\href{https://arxiv.org/abs/1107.0946}{{arXiv:1107.0946}}}
{[astro-ph.SR]}
\end{barticle}
\endbibitem

\bibitem[\protect\citeauthoryear{{Essick} and {Weinberg}}{2016}]{Essick2016}
\begin{barticle}
\bauthor{\bsnm{{Essick}}, \binits{R.}},
\bauthor{\bsnm{{Weinberg}}, \binits{N.N.}}:
\batitle{{Orbital Decay of Hot Jupiters Due to Nonlinear Tidal Dissipation
  within Solar-type Hosts}}.
\bjtitle{\apj}
\bvolume{816},
\bfpage{18}
(\byear{2016})
\doiurl{10.3847/0004-637X/816/1/18}
{\href{https://arxiv.org/abs/1508.02763}{{arXiv:1508.02763}}}
{[astro-ph.EP]}
\end{barticle}
\endbibitem

\bibitem[\protect\citeauthoryear{{Terquem}}{2021}]{Terquem2021}
\begin{barticle}
\bauthor{\bsnm{{Terquem}}, \binits{C.}}:
\batitle{{On a new formulation for energy transfer between convection and fast
  tides with application to giant planets and solar type stars}}.
\bjtitle{\mnras}
\bvolume{503}(\bissue{4}),
\bfpage{5789}--\blpage{5806}
(\byear{2021})
\doiurl{10.1093/mnras/stab224}
{\href{https://arxiv.org/abs/2101.10047}{{arXiv:2101.10047}}}
{[astro-ph.SR]}
\end{barticle}
\endbibitem

\bibitem[\protect\citeauthoryear{{Barker} and {Astoul}}{2021}]{Barker2021}
\begin{barticle}
\bauthor{\bsnm{{Barker}}, \binits{A.J.}},
\bauthor{\bsnm{{Astoul}}, \binits{A.A.V.}}:
\batitle{{On the interaction between fast tides and convection}}.
\bjtitle{\mnras}
\bvolume{506}(\bissue{1}),
\bfpage{69}--\blpage{73}
(\byear{2021})
\doiurl{10.1093/mnrasl/slab077}
{\href{https://arxiv.org/abs/2105.00757}{{arXiv:2105.00757}}}
{[astro-ph.SR]}
\end{barticle}
\endbibitem

\bibitem[\protect\citeauthoryear{{Jouve} and {Ogilvie}}{2014}]{JO2014}
\begin{barticle}
\bauthor{\bsnm{{Jouve}}, \binits{L.}},
\bauthor{\bsnm{{Ogilvie}}, \binits{G.I.}}:
\batitle{{Direct numerical simulations of an inertial wave attractor in linear
  and nonlinear regimes}}.
\bjtitle{Journal of Fluid Mechanics}
\bvolume{745},
\bfpage{223}--\blpage{250}
(\byear{2014})
\doiurl{10.1017/jfm.2014.63}
{\href{https://arxiv.org/abs/1402.2769}{{arXiv:1402.2769}}}
{[astro-ph.SR]}
\end{barticle}
\endbibitem

\bibitem[\protect\citeauthoryear{{Le Reun} et~al.}{2017}]{LeReunetal2017}
\begin{barticle}
\bauthor{\bsnm{{Le Reun}}, \binits{T.}},
\bauthor{\bsnm{{Favier}}, \binits{B.}},
\bauthor{\bsnm{{Barker}}, \binits{A.J.}},
\bauthor{\bsnm{{Le Bars}}, \binits{M.}}:
\batitle{{Inertial Wave Turbulence Driven by Elliptical Instability}}.
\bjtitle{\prl}
\bvolume{119}(\bissue{3}),
\bfpage{034502}
(\byear{2017})
\doiurl{10.1103/PhysRevLett.119.034502}
{\href{https://arxiv.org/abs/1706.07378}{{arXiv:1706.07378}}}
{[physics.flu-dyn]}
\end{barticle}
\endbibitem

\bibitem[\protect\citeauthoryear{{Astoul} and
  {Barker}}{2022}]{AstoulBarker2022}
\begin{barticle}
\bauthor{\bsnm{{Astoul}}, \binits{A.}},
\bauthor{\bsnm{{Barker}}, \binits{A.J.}}:
\batitle{{The effects of non-linearities on tidal flows in the convective
  envelopes of rotating stars and planets in exoplanetary systems}}.
\bjtitle{\mnras}
\bvolume{516}(\bissue{2}),
\bfpage{2913}--\blpage{2935}
(\byear{2022})
\doiurl{10.1093/mnras/stac2117}
{\href{https://arxiv.org/abs/2207.12780}{{arXiv:2207.12780}}}
{[astro-ph.SR]}
\end{barticle}
\endbibitem

\bibitem[\protect\citeauthoryear{{Dandoy} et~al.}{2023}]{Dandoyetal2023}
\begin{barticle}
\bauthor{\bsnm{{Dandoy}}, \binits{V.}},
\bauthor{\bsnm{{Park}}, \binits{J.}},
\bauthor{\bsnm{{Augustson}}, \binits{K.}},
\bauthor{\bsnm{{Astoul}}, \binits{A.}},
\bauthor{\bsnm{{Mathis}}, \binits{S.}}:
\batitle{{How tidal waves interact with convective vortices in rapidly rotating
  planets and stars}}.
\bjtitle{\aap}
\bvolume{673},
\bfpage{6}
(\byear{2023})
\doiurl{10.1051/0004-6361/202243586}
{\href{https://arxiv.org/abs/2211.05900}{{arXiv:2211.05900}}}
{[astro-ph.EP]}
\end{barticle}
\endbibitem

\bibitem[\protect\citeauthoryear{{Astoul} and
  {Barker}}{2023}]{AstoulBarker2023}
\begin{botherref}
\oauthor{\bsnm{{Astoul}}, \binits{A.}},
\oauthor{\bsnm{{Barker}}, \binits{A.J.}}:
{Tidally-excited inertial waves in stars and planets: exploring the
  frequency-dependent and averaged dissipation with nonlinear simulations}.
arXiv e-prints,
2309--02520
(2023)
\doiurl{10.48550/arXiv.2309.02520}
{\href{https://arxiv.org/abs/2309.02520}{{arXiv:2309.02520}}}
{[astro-ph.SR]}
\end{botherref}
\endbibitem

\bibitem[\protect\citeauthoryear{{Terquem}}{2023}]{Terquem2023}
\begin{barticle}
\bauthor{\bsnm{{Terquem}}, \binits{C.}}:
\batitle{{On the energetics of a tidally oscillating convective flow}}.
\bjtitle{\mnras}
\bvolume{525}(\bissue{1}),
\bfpage{508}--\blpage{526}
(\byear{2023})
\doiurl{10.1093/mnras/stad2163}
{\href{https://arxiv.org/abs/2309.01450}{{arXiv:2309.01450}}}
{[astro-ph.SR]}
\end{barticle}
\endbibitem

\bibitem[\protect\citeauthoryear{{Duguid} et~al.}{2020a}]{Duguidetal2020a}
\begin{barticle}
\bauthor{\bsnm{{Duguid}}, \binits{C.D.}},
\bauthor{\bsnm{{Barker}}, \binits{A.J.}},
\bauthor{\bsnm{{Jones}}, \binits{C.A.}}:
\batitle{{Tidal flows with convection: frequency dependence of the effective
  viscosity and evidence for antidissipation}}.
\bjtitle{\mnras}
\bvolume{491}(\bissue{1}),
\bfpage{923}--\blpage{943}
(\byear{2020})
\doiurl{10.1093/mnras/stz2899}
{\href{https://arxiv.org/abs/1910.06034}{{arXiv:1910.06034}}}
{[astro-ph.SR]}
\end{barticle}
\endbibitem

\bibitem[\protect\citeauthoryear{{Duguid} et~al.}{2020b}]{Duguidetal2020b}
\begin{barticle}
\bauthor{\bsnm{{Duguid}}, \binits{C.D.}},
\bauthor{\bsnm{{Barker}}, \binits{A.J.}},
\bauthor{\bsnm{{Jones}}, \binits{C.A.}}:
\batitle{{Convective turbulent viscosity acting on equilibrium tidal flows: new
  frequency scaling of the effective viscosity}}.
\bjtitle{\mnras}
\bvolume{497}(\bissue{3}),
\bfpage{3400}--\blpage{3417}
(\byear{2020})
\doiurl{10.1093/mnras/staa2216}
{\href{https://arxiv.org/abs/2007.12624}{{arXiv:2007.12624}}}
{[astro-ph.EP]}
\end{barticle}
\endbibitem

\bibitem[\protect\citeauthoryear{{Vidal} and
  {Barker}}{2020a}]{VidalBarker2020a}
\begin{barticle}
\bauthor{\bsnm{{Vidal}}, \binits{J.}},
\bauthor{\bsnm{{Barker}}, \binits{A.J.}}:
\batitle{{Turbulent Viscosity Acting on the Equilibrium Tidal Flow in
  Convective Stars}}.
\bjtitle{\apjl}
\bvolume{888}(\bissue{2}),
\bfpage{31}
(\byear{2020})
\doiurl{10.3847/2041-8213/ab6219}
{\href{https://arxiv.org/abs/1912.07910}{{arXiv:1912.07910}}}
{[astro-ph.SR]}
\end{barticle}
\endbibitem

\bibitem[\protect\citeauthoryear{{Vidal} and
  {Barker}}{2020b}]{VidalBarker2020b}
\begin{barticle}
\bauthor{\bsnm{{Vidal}}, \binits{J.}},
\bauthor{\bsnm{{Barker}}, \binits{A.J.}}:
\batitle{{Efficiency of tidal dissipation in slowly rotating fully convective
  stars or planets}}.
\bjtitle{\mnras}
\bvolume{497}(\bissue{4}),
\bfpage{4472}--\blpage{4485}
(\byear{2020})
\doiurl{10.1093/mnras/staa2239}
{\href{https://arxiv.org/abs/2007.13392}{{arXiv:2007.13392}}}
{[astro-ph.SR]}
\end{barticle}
\endbibitem

\bibitem[\protect\citeauthoryear{{Barker} and
  {Astoul}}{2021}]{BarkerAstoul2021}
\begin{barticle}
\bauthor{\bsnm{{Barker}}, \binits{A.J.}},
\bauthor{\bsnm{{Astoul}}, \binits{A.A.V.}}:
\batitle{{On the interaction between fast tides and convection}}.
\bjtitle{\mnras}
\bvolume{506}(\bissue{1}),
\bfpage{69}--\blpage{73}
(\byear{2021})
\doiurl{10.1093/mnrasl/slab077}
{\href{https://arxiv.org/abs/2105.00757}{{arXiv:2105.00757}}}
{[astro-ph.SR]}
\end{barticle}
\endbibitem

\bibitem[\protect\citeauthoryear{{Kerswell}}{2002}]{Kerswell2002}
\begin{barticle}
\bauthor{\bsnm{{Kerswell}}, \binits{R.R.}}:
\batitle{{Elliptical instability}}.
\bjtitle{Annual Review of Fluid Mechanics}
\bvolume{34},
\bfpage{83}--\blpage{113}
(\byear{2002})
\doiurl{10.1146/annurev.fluid.34.081701.171829}
\end{barticle}
\endbibitem

\bibitem[\protect\citeauthoryear{{Le Bars} et~al.}{2015}]{LeBarsetal2015}
\begin{barticle}
\bauthor{\bsnm{{Le Bars}}, \binits{M.}},
\bauthor{\bsnm{{C{\'e}bron}}, \binits{D.}},
\bauthor{\bsnm{{Le Gal}}, \binits{P.}}:
\batitle{{Flows Driven by Libration, Precession, and Tides}}.
\bjtitle{Annual Review of Fluid Mechanics}
\bvolume{47},
\bfpage{163}--\blpage{193}
(\byear{2015})
\doiurl{10.1146/annurev-fluid-010814-014556}
\end{barticle}
\endbibitem

\bibitem[\protect\citeauthoryear{{Barker} and
  {Lithwick}}{2013}]{BarkerLithwick2013}
\begin{barticle}
\bauthor{\bsnm{{Barker}}, \binits{A.J.}},
\bauthor{\bsnm{{Lithwick}}, \binits{Y.}}:
\batitle{{Non-linear evolution of the tidal elliptical instability in gaseous
  planets and stars}}.
\bjtitle{\mnras}
\bvolume{435},
\bfpage{3614}--\blpage{3626}
(\byear{2013})
\doiurl{10.1093/mnras/stt1561}
{\href{https://arxiv.org/abs/1309.0107}{{arXiv:1309.0107}}}
{[astro-ph.EP]}
\end{barticle}
\endbibitem

\bibitem[\protect\citeauthoryear{{Barker}}{2016}]{Barker2016}
\begin{barticle}
\bauthor{\bsnm{{Barker}}, \binits{A.J.}}:
\batitle{{Non-linear tides in a homogeneous rotating planet or star: global
  simulations of the elliptical instability}}.
\bjtitle{\mnras}
\bvolume{459},
\bfpage{939}--\blpage{956}
(\byear{2016})
\doiurl{10.1093/mnras/stw702}
{\href{https://arxiv.org/abs/1603.06840}{{arXiv:1603.06840}}}
{[astro-ph.EP]}
\end{barticle}
\endbibitem

\bibitem[\protect\citeauthoryear{{Barker} and
  {Lithwick}}{2014}]{BarkerLithwick2014}
\begin{barticle}
\bauthor{\bsnm{{Barker}}, \binits{A.J.}},
\bauthor{\bsnm{{Lithwick}}, \binits{Y.}}:
\batitle{{Non-linear evolution of the elliptical instability in the presence of
  weak magnetic fields}}.
\bjtitle{\mnras}
\bvolume{437},
\bfpage{305}--\blpage{315}
(\byear{2014})
\doiurl{10.1093/mnras/stt1884}
{\href{https://arxiv.org/abs/1309.0108}{{arXiv:1309.0108}}}
{[astro-ph.EP]}
\end{barticle}
\endbibitem

\bibitem[\protect\citeauthoryear{{Guillot} et~al.}{2018}]{Guillotetal2018}
\begin{barticle}
\bauthor{\bsnm{{Guillot}}, \binits{T.}},
\bauthor{\bsnm{{Miguel}}, \binits{Y.}},
\bauthor{\bsnm{{Militzer}}, \binits{B.}},
\bauthor{\bsnm{{Hubbard}}, \binits{W.B.}},
\bauthor{\bsnm{{Kaspi}}, \binits{Y.}},
\bauthor{\bsnm{{Galanti}}, \binits{E.}},
\bauthor{\bsnm{{Cao}}, \binits{H.}},
\bauthor{\bsnm{{Helled}}, \binits{R.}},
\bauthor{\bsnm{{Wahl}}, \binits{S.M.}},
\bauthor{\bsnm{{Iess}}, \binits{L.}},
\bauthor{\bsnm{{Folkner}}, \binits{W.M.}},
\bauthor{\bsnm{{Stevenson}}, \binits{D.J.}},
\bauthor{\bsnm{{Lunine}}, \binits{J.I.}},
\bauthor{\bsnm{{Reese}}, \binits{D.R.}},
\bauthor{\bsnm{{Biekman}}, \binits{A.}},
\bauthor{\bsnm{{Parisi}}, \binits{M.}},
\bauthor{\bsnm{{Durante}}, \binits{D.}},
\bauthor{\bsnm{{Connerney}}, \binits{J.E.P.}},
\bauthor{\bsnm{{Levin}}, \binits{S.M.}},
\bauthor{\bsnm{{Bolton}}, \binits{S.J.}}:
\batitle{{A suppression of differential rotation in Jupiter's deep interior}}.
\bjtitle{\nat}
\bvolume{555},
\bfpage{227}--\blpage{230}
(\byear{2018})
\doiurl{10.1038/nature25775}
\end{barticle}
\endbibitem

\bibitem[\protect\citeauthoryear{{Galanti} et~al.}{2019}]{Galantietal2019}
\begin{barticle}
\bauthor{\bsnm{{Galanti}}, \binits{E.}},
\bauthor{\bsnm{{Kaspi}}, \binits{Y.}},
\bauthor{\bsnm{{Miguel}}, \binits{Y.}},
\bauthor{\bsnm{{Guillot}}, \binits{T.}},
\bauthor{\bsnm{{Durante}}, \binits{D.}},
\bauthor{\bsnm{{Racioppa}}, \binits{P.}},
\bauthor{\bsnm{{Iess}}, \binits{L.}}:
\batitle{{Saturn's Deep Atmospheric Flows Revealed by the Cassini Grand Finale
  Gravity Measurements}}.
\bjtitle{\grl}
\bvolume{46}(\bissue{2}),
\bfpage{616}--\blpage{624}
(\byear{2019})
\doiurl{10.1029/2018GL078087}
{\href{https://arxiv.org/abs/1902.04268}{{arXiv:1902.04268}}}
{[astro-ph.EP]}
\end{barticle}
\endbibitem

\bibitem[\protect\citeauthoryear{{Mathis}}{2009}]{Mathis2009}
\begin{barticle}
\bauthor{\bsnm{{Mathis}}, \binits{S.}}:
\batitle{{Transport by gravito-inertial waves in differentially rotating
  stellar radiation zones. I - Theoretical formulation}}.
\bjtitle{\aap}
\bvolume{506},
\bfpage{811}--\blpage{828}
(\byear{2009})
\doiurl{10.1051/0004-6361/200810544}
\end{barticle}
\endbibitem

\bibitem[\protect\citeauthoryear{{Baruteau} and
  {Rieutord}}{2013}]{BaruteauRieutord2013}
\begin{barticle}
\bauthor{\bsnm{{Baruteau}}, \binits{C.}},
\bauthor{\bsnm{{Rieutord}}, \binits{M.}}:
\batitle{{Inertial waves in a differentially rotating spherical shell}}.
\bjtitle{Journal of Fluid Mechanics}
\bvolume{719},
\bfpage{47}--\blpage{81}
(\byear{2013})
\doiurl{10.1017/jfm.2012.605}
{\href{https://arxiv.org/abs/1203.4347}{{arXiv:1203.4347}}}
{[astro-ph.SR]}
\end{barticle}
\endbibitem

\bibitem[\protect\citeauthoryear{{Guenel} et~al.}{2016}]{Gueneletal2016}
\begin{barticle}
\bauthor{\bsnm{{Guenel}}, \binits{M.}},
\bauthor{\bsnm{{Baruteau}}, \binits{C.}},
\bauthor{\bsnm{{Mathis}}, \binits{S.}},
\bauthor{\bsnm{{Rieutord}}, \binits{M.}}:
\batitle{{Tidal inertial waves in differentially rotating convective envelopes
  of low-mass stars. I. Free oscillation modes}}.
\bjtitle{\aap}
\bvolume{589},
\bfpage{22}
(\byear{2016})
\doiurl{10.1051/0004-6361/201527621}
{\href{https://arxiv.org/abs/1601.04617}{{arXiv:1601.04617}}}
{[astro-ph.SR]}
\end{barticle}
\endbibitem

\bibitem[\protect\citeauthoryear{{Mirouh} et~al.}{2016}]{Mirouhetal2016}
\begin{barticle}
\bauthor{\bsnm{{Mirouh}}, \binits{G.M.}},
\bauthor{\bsnm{{Baruteau}}, \binits{C.}},
\bauthor{\bsnm{{Rieutord}}, \binits{M.}},
\bauthor{\bsnm{{Ballot}}, \binits{J.}}:
\batitle{{Gravito-inertial waves in a differentially rotating spherical
  shell}}.
\bjtitle{Journal of Fluid Mechanics}
\bvolume{800},
\bfpage{213}--\blpage{247}
(\byear{2016})
\doiurl{10.1017/jfm.2016.382}
{\href{https://arxiv.org/abs/1511.05832}{{arXiv:1511.05832}}}
{[astro-ph.SR]}
\end{barticle}
\endbibitem

\bibitem[\protect\citeauthoryear{{Goldreich} and
  {Nicholson}}{1989}]{GoldreichNicholson1989}
\begin{barticle}
\bauthor{\bsnm{{Goldreich}}, \binits{P.}},
\bauthor{\bsnm{{Nicholson}}, \binits{P.D.}}:
\batitle{{Tidal friction in early-type stars}}.
\bjtitle{\apj}
\bvolume{342},
\bfpage{1079}--\blpage{1084}
(\byear{1989})
\doiurl{10.1086/167665}
\end{barticle}
\endbibitem

\bibitem[\protect\citeauthoryear{{Favier} et~al.}{2014}]{Favieretal2014}
\begin{barticle}
\bauthor{\bsnm{{Favier}}, \binits{B.}},
\bauthor{\bsnm{{Barker}}, \binits{A.J.}},
\bauthor{\bsnm{{Baruteau}}, \binits{C.}},
\bauthor{\bsnm{{Ogilvie}}, \binits{G.I.}}:
\batitle{{Non-linear evolution of tidally forced inertial waves in rotating
  fluid bodies}}.
\bjtitle{\mnras}
\bvolume{439},
\bfpage{845}--\blpage{860}
(\byear{2014})
\doiurl{10.1093/mnras/stu003}
{\href{https://arxiv.org/abs/1401.0643}{{arXiv:1401.0643}}}
{[astro-ph.EP]}
\end{barticle}
\endbibitem

\bibitem[\protect\citeauthoryear{{Astoul} et~al.}{2021}]{Astouletal2021}
\begin{barticle}
\bauthor{\bsnm{{Astoul}}, \binits{A.}},
\bauthor{\bsnm{{Park}}, \binits{J.}},
\bauthor{\bsnm{{Mathis}}, \binits{S.}},
\bauthor{\bsnm{{Baruteau}}, \binits{C.}},
\bauthor{\bsnm{{Gallet}}, \binits{F.}}:
\batitle{{The complex interplay between tidal inertial waves and zonal flows in
  differentially rotating stellar and planetary convective regions. I. Free
  waves}}.
\bjtitle{\aap}
\bvolume{647},
\bfpage{144}
(\byear{2021})
\doiurl{10.1051/0004-6361/202039148}
{\href{https://arxiv.org/abs/2101.04656}{{arXiv:2101.04656}}}
{[astro-ph.SR]}
\end{barticle}
\endbibitem

\bibitem[\protect\citeauthoryear{{Morize} et~al.}{2010}]{Morizeetal2010}
\begin{barticle}
\bauthor{\bsnm{{Morize}}, \binits{C.}},
\bauthor{\bsnm{{Le Bars}}, \binits{M.}},
\bauthor{\bsnm{{Le Gal}}, \binits{P.}},
\bauthor{\bsnm{{Tilgner}}, \binits{A.}}:
\batitle{{Experimental Determination of Zonal Winds Driven by Tides}}.
\bjtitle{\prl}
\bvolume{104}(\bissue{21}),
\bfpage{214501}
(\byear{2010})
\doiurl{10.1103/PhysRevLett.104.214501}
\end{barticle}
\endbibitem

\bibitem[\protect\citeauthoryear{{C{\'e}bron} et~al.}{2021}]{Cebronetal2021}
\begin{barticle}
\bauthor{\bsnm{{C{\'e}bron}}, \binits{D.}},
\bauthor{\bsnm{{Vidal}}, \binits{J.}},
\bauthor{\bsnm{{Schaeffer}}, \binits{N.}},
\bauthor{\bsnm{{Borderies}}, \binits{A.}},
\bauthor{\bsnm{{Sauret}}, \binits{A.}}:
\batitle{{Mean zonal flows induced by weak mechanical forcings in rotating
  spheroids}}.
\bjtitle{Journal of Fluid Mechanics}
\bvolume{916},
\bfpage{39}
(\byear{2021})
\doiurl{10.1017/jfm.2021.220}
{\href{https://arxiv.org/abs/2103.10260}{{arXiv:2103.10260}}}
{[physics.flu-dyn]}
\end{barticle}
\endbibitem

\bibitem[\protect\citeauthoryear{{Connerney} et~al.}{2018}]{Connerneyetal2018}
\begin{barticle}
\bauthor{\bsnm{{Connerney}}, \binits{J.E.P.}},
\bauthor{\bsnm{{Kotsiaros}}, \binits{S.}},
\bauthor{\bsnm{{Oliversen}}, \binits{R.J.}},
\bauthor{\bsnm{{Espley}}, \binits{J.R.}},
\bauthor{\bsnm{{Joergensen}}, \binits{J.L.}},
\bauthor{\bsnm{{Joergensen}}, \binits{P.S.}},
\bauthor{\bsnm{{Merayo}}, \binits{J.M.G.}},
\bauthor{\bsnm{{Herceg}}, \binits{M.}},
\bauthor{\bsnm{{Bloxham}}, \binits{J.}},
\bauthor{\bsnm{{Moore}}, \binits{K.M.}},
\bauthor{\bsnm{{Bolton}}, \binits{S.J.}},
\bauthor{\bsnm{{Levin}}, \binits{S.M.}}:
\batitle{{A New Model of Jupiter's Magnetic Field From Juno's First Nine
  Orbits}}.
\bjtitle{\grl}
\bvolume{45}(\bissue{6}),
\bfpage{2590}--\blpage{2596}
(\byear{2018})
\doiurl{10.1002/2018GL077312}
\end{barticle}
\endbibitem

\bibitem[\protect\citeauthoryear{{Duarte} et~al.}{2018}]{Duarteetal2018}
\begin{barticle}
\bauthor{\bsnm{{Duarte}}, \binits{L.D.V.}},
\bauthor{\bsnm{{Wicht}}, \binits{J.}},
\bauthor{\bsnm{{Gastine}}, \binits{T.}}:
\batitle{{Physical conditions for Jupiter-like dynamo models}}.
\bjtitle{Icarus}
\bvolume{299},
\bfpage{206}--\blpage{221}
(\byear{2018})
\doiurl{10.1016/j.icarus.2017.07.016}
\end{barticle}
\endbibitem

\bibitem[\protect\citeauthoryear{{Dougherty} et~al.}{2018}]{Doughertyetal2018}
\begin{barticle}
\bauthor{\bsnm{{Dougherty}}, \binits{M.K.}},
\bauthor{\bsnm{{Cao}}, \binits{H.}},
\bauthor{\bsnm{{Khurana}}, \binits{K.K.}},
\bauthor{\bsnm{{Hunt}}, \binits{G.J.}},
\bauthor{\bsnm{{Provan}}, \binits{G.}},
\bauthor{\bsnm{{Kellock}}, \binits{S.}},
\bauthor{\bsnm{{Burton}}, \binits{M.E.}},
\bauthor{\bsnm{{Burk}}, \binits{T.A.}},
\bauthor{\bsnm{{Bunce}}, \binits{E.J.}},
\bauthor{\bsnm{{Cowley}}, \binits{S.W.H.}},
\bauthor{\bsnm{{Kivelson}}, \binits{M.G.}},
\bauthor{\bsnm{{Russell}}, \binits{C.T.}},
\bauthor{\bsnm{{Southwood}}, \binits{D.J.}}:
\batitle{{Saturn's magnetic field revealed by the Cassini Grand Finale}}.
\bjtitle{Science}
\bvolume{362}(\bissue{6410}),
\bfpage{5434}
(\byear{2018})
\doiurl{10.1126/science.aat5434}
\end{barticle}
\endbibitem

\bibitem[\protect\citeauthoryear{{Yadav} et~al.}{2022}]{Yadavetal2022}
\begin{barticle}
\bauthor{\bsnm{{Yadav}}, \binits{R.K.}},
\bauthor{\bsnm{{Cao}}, \binits{H.}},
\bauthor{\bsnm{{Bloxham}}, \binits{J.}}:
\batitle{{A Global Simulation of the Dynamo, Zonal Jets, and Vortices on
  Saturn}}.
\bjtitle{\apj}
\bvolume{940}(\bissue{2}),
\bfpage{185}
(\byear{2022})
\doiurl{10.3847/1538-4357/ac9d94}
{\href{https://arxiv.org/abs/2212.10617}{{arXiv:2212.10617}}}
{[astro-ph.EP]}
\end{barticle}
\endbibitem

\bibitem[\protect\citeauthoryear{{Wei}}{2016}]{Wei2016}
\begin{barticle}
\bauthor{\bsnm{{Wei}}, \binits{X.}}:
\batitle{{Calculating Rotating Hydrodynamic and Magnetohydrodynamic Waves to
  Understand Magnetic Effects on Dynamical Tides}}.
\bjtitle{\apj}
\bvolume{828},
\bfpage{30}
(\byear{2016})
\doiurl{10.3847/0004-637X/828/1/30}
{\href{https://arxiv.org/abs/1606.06232}{{arXiv:1606.06232}}}
{[astro-ph.SR]}
\end{barticle}
\endbibitem

\bibitem[\protect\citeauthoryear{{Lin} and {Ogilvie}}{2018}]{LinOgilvie2018}
\begin{barticle}
\bauthor{\bsnm{{Lin}}, \binits{Y.}},
\bauthor{\bsnm{{Ogilvie}}, \binits{G.I.}}:
\batitle{{Tidal dissipation in rotating fluid bodies: the presence of a
  magnetic field}}.
\bjtitle{\mnras}
\bvolume{474},
\bfpage{1644}--\blpage{1656}
(\byear{2018})
\doiurl{10.1093/mnras/stx2764}
{\href{https://arxiv.org/abs/1710.07690}{{arXiv:1710.07690}}}
{[astro-ph.EP]}
\end{barticle}
\endbibitem

\bibitem[\protect\citeauthoryear{{Wei}}{2018}]{Wei2018}
\begin{barticle}
\bauthor{\bsnm{{Wei}}, \binits{X.}}:
\batitle{{The Magnetic Effect on Dynamical Tide in Rapidly Rotating
  Astronomical Objects}}.
\bjtitle{\apj}
\bvolume{854},
\bfpage{34}
(\byear{2018})
\doiurl{10.3847/1538-4357/aaa54d}
{\href{https://arxiv.org/abs/1801.05552}{{arXiv:1801.05552}}}
{[astro-ph.SR]}
\end{barticle}
\endbibitem

\bibitem[\protect\citeauthoryear{{C{\'e}bron} and
  {Hollerbach}}{2014}]{CebronHollerbach2014}
\begin{barticle}
\bauthor{\bsnm{{C{\'e}bron}}, \binits{D.}},
\bauthor{\bsnm{{Hollerbach}}, \binits{R.}}:
\batitle{{Tidally Driven Dynamos in a Rotating Sphere}}.
\bjtitle{\apjl}
\bvolume{789},
\bfpage{25}
(\byear{2014})
\doiurl{10.1088/2041-8205/789/1/L25}
{\href{https://arxiv.org/abs/1406.3431}{{arXiv:1406.3431}}}
{[astro-ph.SR]}
\end{barticle}
\endbibitem

\bibitem[\protect\citeauthoryear{{Witte} and {Savonije}}{1999}]{witte:99}
\begin{barticle}
\bauthor{\bsnm{{Witte}}, \binits{M.G.}},
\bauthor{\bsnm{{Savonije}}, \binits{G.J.}}:
\batitle{{Tidal evolution of eccentric orbits in massive binary systems. A
  study of resonance locking}}.
\bjtitle{\aap}
\bvolume{350},
\bfpage{129}--\blpage{147}
(\byear{1999})
{\href{https://arxiv.org/abs/astro-ph/9909073}{{astro-ph/9909073}}}
\end{barticle}
\endbibitem

\bibitem[\protect\citeauthoryear{{Ma} and {Fuller}}{2021}]{Ma2021}
\begin{barticle}
\bauthor{\bsnm{{Ma}}, \binits{L.}},
\bauthor{\bsnm{{Fuller}}, \binits{J.}}:
\batitle{{Orbital Decay of Short-period Exoplanets via Tidal Resonance
  Locking}}.
\bjtitle{\apj}
\bvolume{918}(\bissue{1}),
\bfpage{16}
(\byear{2021})
\doiurl{10.3847/1538-4357/ac088e}
{\href{https://arxiv.org/abs/2105.09335}{{arXiv:2105.09335}}}
{[astro-ph.EP]}
\end{barticle}
\endbibitem

\bibitem[\protect\citeauthoryear{{Fuller} et~al.}{2017}]{fuller:17}
\begin{barticle}
\bauthor{\bsnm{{Fuller}}, \binits{J.}},
\bauthor{\bsnm{{Hambleton}}, \binits{K.}},
\bauthor{\bsnm{{Shporer}}, \binits{A.}},
\bauthor{\bsnm{{Isaacson}}, \binits{H.}},
\bauthor{\bsnm{{Thompson}}, \binits{S.}}:
\batitle{{Accelerated tidal circularization via resonance locking in KIC
  8164262}}.
\bjtitle{\mnras}
\bvolume{472}(\bissue{1}),
\bfpage{25}--\blpage{29}
(\byear{2017})
\doiurl{10.1093/mnrasl/slx130}
{\href{https://arxiv.org/abs/1706.05053}{{arXiv:1706.05053}}}
{[astro-ph.SR]}
\end{barticle}
\endbibitem

\bibitem[\protect\citeauthoryear{{Lainey} et~al.}{2020}]{Lainey2020}
\begin{barticle}
\bauthor{\bsnm{{Lainey}}, \binits{V.}},
\bauthor{\bsnm{{Casajus}}, \binits{L.G.}},
\bauthor{\bsnm{{Fuller}}, \binits{J.}},
\bauthor{\bsnm{{Zannoni}}, \binits{M.}},
\bauthor{\bsnm{{Tortora}}, \binits{P.}},
\bauthor{\bsnm{{Cooper}}, \binits{N.}},
\bauthor{\bsnm{{Murray}}, \binits{C.}},
\bauthor{\bsnm{{Modenini}}, \binits{D.}},
\bauthor{\bsnm{{Park}}, \binits{R.S.}},
\bauthor{\bsnm{{Robert}}, \binits{V.}},
\bauthor{\bsnm{{Zhang}}, \binits{Q.}}:
\batitle{{Resonance locking in giant planets indicated by the rapid orbital
  expansion of Titan}}.
\bjtitle{Nature Astronomy}
\bvolume{4},
\bfpage{1053}--\blpage{1058}
(\byear{2020})
\doiurl{10.1038/s41550-020-1120-5}
{\href{https://arxiv.org/abs/2006.06854}{{arXiv:2006.06854}}}
{[astro-ph.EP]}
\end{barticle}
\endbibitem

\bibitem[\protect\citeauthoryear{{Ogilvie} and {Lin}}{2004}]{ogilvie:04}
\begin{barticle}
\bauthor{\bsnm{{Ogilvie}}, \binits{G.I.}},
\bauthor{\bsnm{{Lin}}, \binits{D.N.C.}}:
\batitle{{Tidal Dissipation in Rotating Giant Planets}}.
\bjtitle{\apj}
\bvolume{610},
\bfpage{477}--\blpage{509}
(\byear{2004})
\doiurl{10.1086/421454}
{\href{https://arxiv.org/abs/astro-ph/0310218}{{astro-ph/0310218}}}
\end{barticle}
\endbibitem

\bibitem[\protect\citeauthoryear{{Lin} and {Ogilvie}}{2021}]{lin:21}
\begin{barticle}
\bauthor{\bsnm{{Lin}}, \binits{Y.}},
\bauthor{\bsnm{{Ogilvie}}, \binits{G.I.}}:
\batitle{{Resonant Tidal Responses in Rotating Fluid Bodies: Global Modes
  Hidden beneath Localized Wave Beams}}.
\bjtitle{\apjl}
\bvolume{918}(\bissue{1}),
\bfpage{21}
(\byear{2021})
\doiurl{10.3847/2041-8213/ac1f23}
{\href{https://arxiv.org/abs/2108.08515}{{arXiv:2108.08515}}}
{[physics.flu-dyn]}
\end{barticle}
\endbibitem

\end{thebibliography}

\end{document}